\documentclass[a4paper,twocolumn,11pt,accepted=2025-04-23]{quantumarticle}
\pdfoutput=1
\usepackage{colortbl}
\usepackage[table]{xcolor}
\usepackage{enumitem}
\setitemize{noitemsep,topsep=2pt,parsep=0pt,partopsep=0pt}
\setenumerate{noitemsep,topsep=2pt,parsep=0pt,partopsep=0pt}

\usepackage{color}
\definecolor{deepblue}{HTML}{002AA0}
\definecolor{deepred}{rgb}{0.6,0,0}
\definecolor{deepgreen}{rgb}{0,0.5,0}
\definecolor{lightblue}{HTML}{F0F8FF}

\usepackage{listings}
\definecolor{codegreen}{rgb}{0,0.6,0}
\definecolor{codegray}{rgb}{0.5,0.5,0.5}
\definecolor{backcolor}{HTML}{ECEFF1}

\lstdefinestyle{codeblock}{
	backgroundcolor=\color{backcolor},   
	commentstyle=\color{codegreen},
	keywordstyle=\color{deepblue},
	numberstyle=\tiny\color{codegray},
	stringstyle=\color{deepgreen},
	basicstyle=\footnotesize,
	escapechar=\¢,escapebegin=\color{purple}, % highlighting for decorator
	otherkeywords={with},
	breakatwhitespace=false,         
	breaklines=true,      
	lineskip=1pt,
	captionpos=b,                    
	keepspaces=true,
	language=Python,
	numbers=left,                    
	numbersep=6pt,                  
	showspaces=false,                
	showstringspaces=false,
	showtabs=false,                  
	tabsize=2,
	frame=single,
	framerule=0pt,
	basicstyle=\fontencoding{T1}\ttfamily\footnotesize
}
\lstdefinestyle{output}{
	backgroundcolor=\color{white},   
	commentstyle=\color{codegreen},
	keywordstyle=\color{deepblue},
	numberstyle=\tiny\color{white},
	stringstyle=\color{deepgreen},
	basicstyle=\footnotesize,
	escapechar=\¢,escapebegin=\color{purple}, % highlighting for decorator
	otherkeywords={with},
	breakatwhitespace=false,         
	breaklines=true,      
	lineskip=1pt,
	captionpos=b,                    
	keepspaces=true,
	language=Python,
	numbers=none,                    
	numbersep=6pt,                  
	showspaces=false,                
	showstringspaces=false,
	showtabs=false,                  
	tabsize=2,
	frame=single,
	framerule=0pt,
	basicstyle=\fontencoding{T1}\ttfamily\footnotesize
}
\newcommand{\pyl}[1]{\lstinline!#1!}

\usepackage{graphicx}
\usepackage{subcaption}
\usepackage[T1]{fontenc}%
\usepackage[utf8]{inputenc}%
\usepackage{textcomp}%
\usepackage{lastpage}%
\usepackage{geometry}%
\usepackage{amsmath}
\usepackage{amssymb}
\usepackage{pifont}
\usepackage{diagbox}
\usepackage{longtable}
\usepackage{xfrac}
\usepackage{tensor}
\usepackage{multirow}
\usepackage{rotating}
\usepackage[numbers]{natbib}
\usepackage{algorithm}
\usepackage{algorithmicx}
\usepackage{algpseudocode}
\floatname{algorithm}{Algorithm}

\usepackage{xcolor}
\usepackage[colorlinks,allcolors=blue]{hyperref}

\newcommand\ket[1]{\ensuremath{|#1\rangle}}

\newcommand{\hamparam}{\vec{\theta}_\mathrm{ham}}

\newcommand{\opn}{\hat{n}}
\newcommand{\opv}{\hat{\varphi}}

\newcommand{\aref}[1]{\hyperref[#1]{Appendix~\ref*{#1}}}
\newcommand{\sref}[1]{\hyperref[#1]{Sec.~\ref*{#1}}}
\usepackage[normalem]{ulem}

\def\<{\langle}
\def\>{\rangle}

\begin{document}
\captionsetup[lstlisting]{position=bottom}

\title{SuperGrad: a differentiable simulator for superconducting processors}

\author{Ziang Wang}
\affiliation{Zhejiang Institute of Modern Physics and Zhejiang Key Laboratory of Micro-nano Quantum Chips and Quantum Control, Zhejiang University, Hangzhou 310027, China}

\author{Feng Wu}
\affiliation{Zhongguancun Laboratory, Beijing, China}

\author{Hui-Hai Zhao}
\email{zhaohuihai@iqubit.org}
\affiliation{Zhongguancun Laboratory, Beijing, China}

\author{Xin Wan}
\affiliation{Zhejiang Institute of Modern Physics and Zhejiang Key Laboratory of Micro-nano Quantum Chips and Quantum Control, Zhejiang University, Hangzhou 310027, China}

\author{Xiaotong Ni}
\email{xiaotong.ni@gmail.com}
\affiliation{Quantum Science Center of Guangdong-Hong Kong-Macao, Shenzhen 518045, China}

\begin{abstract}

One significant advantage of superconducting processors is their extensive design flexibility, which encompasses various types of qubits and interactions. Given the large number of tunable parameters of a processor, the ability to perform gradient optimization would be highly beneficial. Efficient backpropagation for gradient computation requires a tightly integrated software library, for which no open-source implementation is currently available.
In this work, we introduce SuperGrad, a simulator that accelerates the design of superconducting quantum processors by incorporating gradient computation capabilities. SuperGrad offers a user-friendly interface for constructing Hamiltonians and computing both static and dynamic properties of composite systems.
This differentiable simulation is valuable for a range of applications, including optimal control, design optimization, and experimental data fitting. In this paper, we demonstrate these applications through examples and code snippets. The code is available at
\url{https://github.com/iqubit-org/supergrad}.
\end{abstract}

\maketitle

\tableofcontents

\section{Introduction}

    Superconducting quantum processors have made significant strides recently, with their performance steadily improving and approaching the fault-tolerance threshold~\cite{PhysRevLett.129.010502, PhysRevX.13.031035, sung_realization_2021, PhysRevLett.127.080505, zhang2023tunable,acharyaSuppressingQuantumErrors2023, zhaoRealizationErrorCorrectingSurface2022, krinnerRealizingRepeatedQuantum2022}. These processors have not only demonstrated successful simulations of various quantum systems~\cite{georgescu_quantum_2014, zhu_observation_2022, google_quantum_ai_and_collaborators_non-abelian_2023, shi_quantum_2023, xiang_simulating_2023, xu_digital_2023} but have also shown promise in running other applications~\cite{riste_demonstration_2017, gong_quantum_2021}. As we inch closer to realizing the full potential of these quantum devices, it is natural to seek ways to optimize their parameters to maximize performance across various tasks. However, as discussed in~\cite{ni_superconducting_2023,acharyaSuppressingQuantumErrors2023}, simulating and optimizing superconducting quantum processors for error correction is a challenging endeavor. These processors are intrinsically quantum many-body systems with numerous optimizable parameters, making the optimization process computationally demanding and complex.

To address these challenges, we introduce SuperGrad~\cite{supergrad2024github}, a differentiable simulator for superconducting quantum processors. It is written in Python and offers a simple and intuitive interface for describing various types of superconducting qubits and their interactions. By also providing information about qubit control and the desired objective function, the library can efficiently compute the gradients with respect to the processor and control parameters through backpropagation.
Intuitively, the gradients contain information about how each parameter should be changed to improve the processors, and this is the idea behind various optimizers based on gradient descent.
In general, the multiplicative speedup achieved by using backpropagation to compute the gradients is proportional to the number of parameters. Therefore, for a time-consuming simulation with dozens of parameters, backpropagation can offer a decent multiplicative speedup and potentially a very large absolute speedup.
After the gradients are computed, they can help us perform joint optimization of these parameters~\cite{ni_integrating_2022}, which is beneficial because these parameters can be heavily intertwined in the Hamiltonian.
The efficient gradient computation is a result of considering the simulation from constructing superconducting qubits to dynamics as a whole. Our library can be viewed as a combination of previous libraries such as SCQubits~\cite{groszkowski_scqubits_2021} and Qiskit Dynamics~\cite{puzzuoli_qiskit_2023}. However, since they each focus on a specific part of the simulation, they cannot perform backpropagation of gradients between them.

    SuperGrad can be used to optimize gate fidelities and logical error rates of quantum processors~\cite{ni_superconducting_2023}. Furthermore, we foresee that our library will be useful for optimizing bosonic code processors~\cite{regentHighperformanceRepetitionCat2023, chamberlandBuildingFaultTolerantQuantum2022}, where complex control schemes and scaling to more modes are ongoing challenges.  
    Additionally, a very common task in experiments is extracting parameters from the Hamiltonian (see, for example~\cite{qfit}). This task can also be viewed as an optimization problem, where the objective function is the difference between experimental data and simulation data. The gradients computed using SuperGrad can again make the optimization process faster and more stable. This extends previous studies~\cite{krastanov_stochastic_2019} by the ability to fit parameters using both time dynamics data and energy spectra data. We provide an example of fitting qubit parameters from experimental data in~\sref{sec:fitting_exp_data}.

    SuperGrad is based on the JAX library~\cite{jax2018github}, leveraging its ability to compute gradients of NumPy-like programs and accelerate computations with GPUs. The main challenge in designing SuperGrad was managing the various parameters in a scalable way while maintaining compatibility with JAX's automatic differentiation (auto-diff) functionality. We employed the Haiku library~\cite{haiku2020github} to achieve this task, as discussed in~\sref{sec:software_arch}.

    In the current SuperGrad library, we do not focus extensively on the efficiency of numerical solvers, as simulating many-body processors to an accuracy suitable for predicting quantum error correction performance remains an open problem. The main time evolution solver is based on Trotterization~\cite{Trotter1959, Suzuki1976GeneralizedTF, suzuki_fractal_1990}. It is known that naively computing the gradients through backpropagation will lead to a large memory cost proportional to the number of steps for discretizing the time interval. The common countermeasure to save memory costs is using the continuous adjoint method~\cite{pontryagin_mathematical_1985, chen_neural_nodate, blondel_elements_2024}. In the library, we implement an improved variant of the continuous adjoint method, which utilizes the locality of the Hamiltonians.

    The remainder of this paper is structured as follows. ~\sref{sec:implementation} presents background information on superconducting quantum processors and introduces our approach to Hamiltonian simulation and optimization. ~\sref{sec:tutorial} provides a hands-on tutorial for designing a quantum processor and showcases examples of SuperGrad's capabilities. ~\sref{sec:benchmark} benchmarks SuperGrad's performance for differentiable simulation. ~\sref{sec:discussion} shows the extensibility of SuperGrad and future work.

\section{Implementation}
\label{sec:implementation}

\subsection{Hamiltonians of superconducting processors}

    In general, a multi-qubit superconducting circuit comprises numerous qubit modes that should thus be treated as a many-body system. Within the realm of superconducting quantum processors, the dynamics of this system are efficiently captured by 1-body and 2-body Hamiltonians. External control pulses can be applied to individual qubit operators, preserving the structure of the time-dependent Hamiltonian as outlined in~\autoref{eq:hamiltonian_decomposation}. The Hamiltonian of the composited system $\mathcal{H}(t)$ that acts on $N$ qubits can be expressed as a sum of local Hamiltonians, denoted as
\begin{equation}
    \mathcal{H}(t) = \sum_i H_i(t) + \sum_{e(G)} H_{ij},
    \label{eq:hamiltonian_decomposation}
\end{equation}
where the 1-body Hamiltonian $H_i(t)$ and 2-body Hamiltonian $H_{ij}$ act non-trivially only on the corresponding qubits $i$ and $j$. Here, $e(G)$ represents the edges of the graph $G$, which indicate the interacting pairs within the quantum processor. Notably, these couplings can be arbitrarily long-ranged in the graph.

    The initial step in simulating a quantum processor involves calculating the low-energy basis for each qubit. Following this, we derive the Hamiltonian for the composite quantum system. In circuit quantum electrodynamics (circuit-QED), the Hamiltonian of superconducting qubits involves device parameters such as Josephson energy ($E_J$), charging energy ($E_C$), inductive energy ($E_L$), and the coupling strengths ($J_C$, $J_L$). 
    
    The next step of simulating a quantum processor’s dynamics necessitates solving the time-dependent Schrödinger equation~\cite{efthymiou_qibo_2022,puzzuoli_qiskit_2023, JOHANSSON20121760}. In scenarios where the quantum processor is influenced by external control pulses, the driving frequency ($\omega_d$) and the pulse amplitude ($\epsilon_d$) become relevant control parameters. These parameters are crucial for characterizing quantum processors in both simulation and experimental setups. Furthermore, our library supports the computation of gradients with respect to these parameters.

\subsection{Library architecture}
\label{sec:library_architecture}

We suggest incorporating the above steps to compute the dynamics within a unified and differentiable framework to streamline the simulation process. This approach offers two primary advantages:
\begin{enumerate}
    \item Enable the computation of gradients with respect to both control parameters and device parameters.
    \item Defining a unified data structure to bridge the quantized Hamiltonian and the ODE solver which facilitates the adoption of advanced solvers.
\end{enumerate}

\begin{figure}
        \includegraphics[width=1\textwidth]{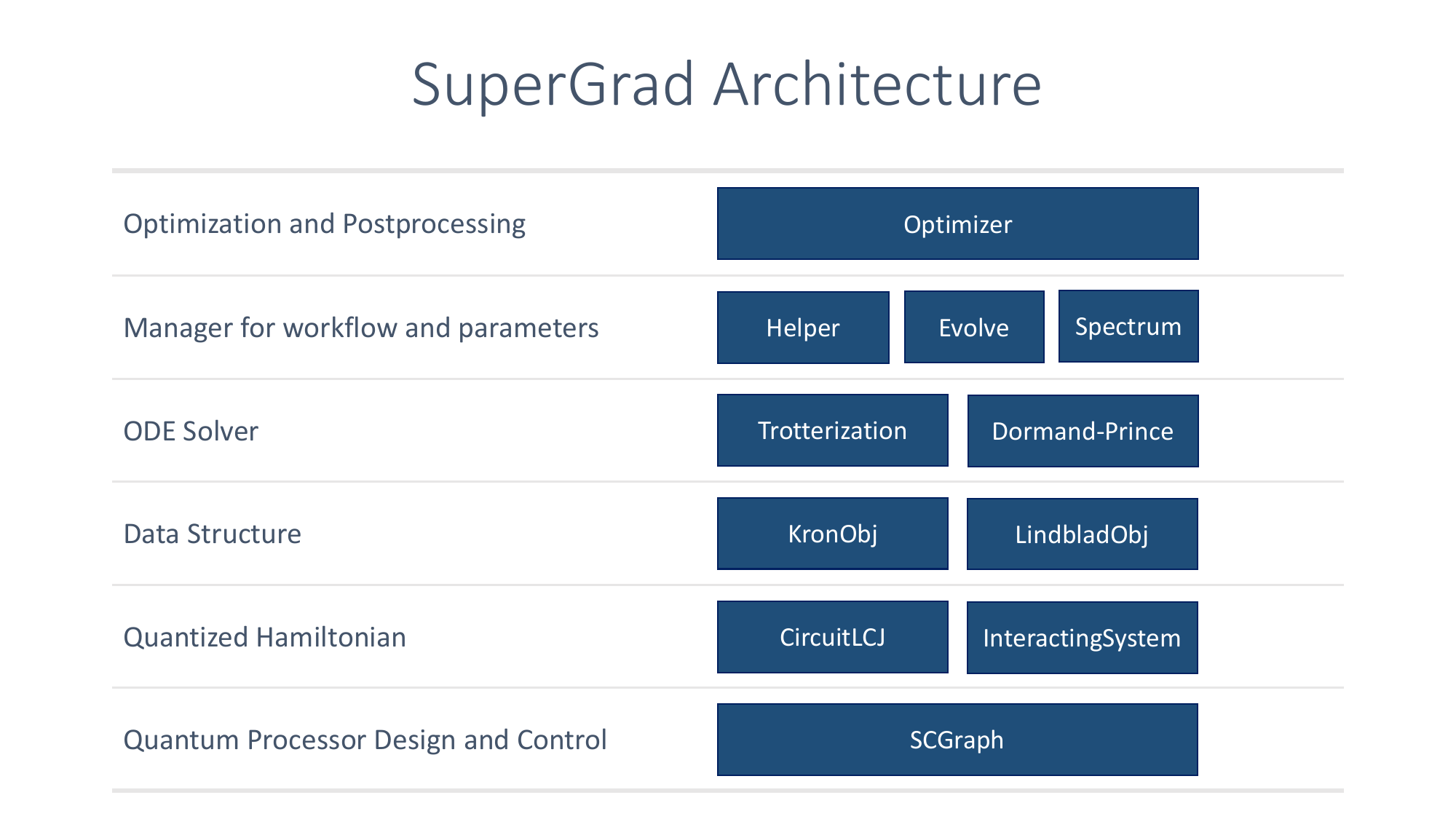}
        \caption{The overview of the SuperGrad library with the components described in~\sref{sec:library_architecture}.}
        \label{fig:supergrad_ecosystem}
    \end{figure}

    Consequently, this method allows for the optimization of quantum processor designs through direct time evolution simulations, offering a comprehensive solution for differentiable quantum processor simulation. An overview of the SuperGrad library is illustrated in~\autoref{fig:supergrad_ecosystem}. A brief description of each component is as follows:

\begin{itemize}
    \item \textbf{SCGraph} provides a comprehensive representation of the superconducting quantum processor structure, incorporating time-dependent control pulses.
    \item \textbf{CircuitLCJ and InteractingSystem}: $\textbf{CircuitLCJ}$ computes the Hamiltonians and operators for each qubit, whereas $\textbf{InteractingSystem}$ computes the Hamiltonian for the quantum processor.
    \item \textbf{KronObj and LindbladObj} store the Hamiltonian and operators in their local forms and are designed to solve the Schrödinger equation and the Lindblad master equation, respectively.
    \item \textbf{ODE solvers} such as Trotterization and Dormand-Prince can compute gradients by the continuous adjoint method.
    \item \textbf{Helper} processes $\textbf{SCGraph}$ to generate differentiable functions for specific computations. This component includes Evolve (for evaluating time evolution) and Spectrum (for assessing static properties).
\end{itemize}

\onecolumn\newpage

\twocolumn

\subsection{Trotterization solver}
\label{sec:trotterization_solver}

    An efficient method for simulating time evolution utilizes Trotterization. This technique decomposes the time evolution operator into a sequence of local operators, facilitating the efficient computation of time evolution using 1-body and 2-body operators. Consequently, memory costs are primarily influenced by quantum states, with the memory cost scaling as $O(d^N)$ where $N$ is the number of qudits, and $d$ represents the dimension of each qudit subsystem. From an alternative perspective, the quantum state may be regarded as a high-rank state tensor. Wherein the product of a local operator and state is calculated through tensor contraction, with the time cost scaling as $O(d^{N+2})$.

    We define the time evolution operator by solving the time-dependent Schrödinger equation:
\begin{equation}
        U(t_0,t_g) = \mathcal{T} \exp \left\{-i \int_{t_0}^{t_g} \mathcal{H}(t) dt \right\},
\end{equation}
    where $\mathcal{T} \exp$ denotes the time-ordered exponential. This operator $U(t_0,t_g)$ describes the time evolution from the initial time $t_0$ to the final time $t_g$. 

    We utilize the Suzuki-Trotter decomposition (STD) to restrict the matrix exponential to each local Hamiltonian separately. Initially, we employ the product formula to approximate the time evolution operator, dividing it into a sequence of operators over $\delta t$ time intervals, ensuring that
\begin{multline}
U(t_{0},t_{g})=U(t_{g}-\delta t,t_{g})U(t_{g}-2\delta t,t_{g}-\delta t) \cdots \\
U(t_{0}+\delta t,t_{0}+2\delta t)U(t_{0},t_{0}+\delta t).
\label{eq:discrete_time_evo_unitary}
\end{multline}
Additionally, we apply the $n$-th Suzuki-Trotter decomposition as a further approximation of the time evolution operator, which is denoted as
\begin{equation}
        U(t, t+\delta t) = Q_n(\delta t) + O(\delta t^{n+1}),
\end{equation}
    where $Q_n(\delta t)$ is recursively constructed as
\begin{equation}
    Q_n(\delta t) = \prod_{j=1}^r Q_{n-1}(p_{n,j}\delta t).
\end{equation}
    The parameters ${p_{n,j}}$ satisfy the decomposition conditions $\sum^{r}_{j=1} (p_{n,j})^n=0$ with $\sum^{r}_{j=1}p_{n,j}=1$. For a Hamiltonian of the form $\mathcal{H}=\sum H_k$, the first-order Suzuki-Trotter decomposition is simply
\begin{equation}
    Q_1(\delta t) = \prod \exp \left\{-i H_k \delta t \right\}.
\end{equation}
    We derive the formulations for higher-order STD in~\aref{appen:higher_order_std}. Notably, the product of the STD operator with the quantum state corresponds to the contraction of a tensor network. Details of our implementation can be found in~\aref{appen:tn_contraction}.

\subsection{Efficiency of computing gradients}
To compare different gradient computation methods, we can look at the time overhead $O_t$ and memory overhead $O_m$
\begin{align*}
    O_t &= \frac{T_{\text{grad}}}{T_{\text{forward}}} \\
    O_m &= \frac{M_{\text{grad}}}{M_{\text{forward}}},
\end{align*}
where $T_{\text{forward}}$, $T_{\text{grad}}$ is the time needed for computing $f$ and $\nabla f$, and similarly $M$ is the memory needed for the computation.
We care about the overhead instead of the time or memory costs because $T_{\text{forward}}$ and $M_{\text{forward}}$ often have exponentially scaling themselves in quantum simulators.
$O_t$ and $O_m$ depend heavily on the concrete implementation of software and hardware, but we can still discuss their scaling with respect to some parameters.
For example, $O_t$ is proven to be smaller than 5 for a given set of elementary operations~\cite{baurComplexityPartialDerivatives1983}. 
In practice, for the reverse-mode auto-diff done by libraries such as JAX or PyTorch, $O_t$ is indeed around 5 for algorithms which are mostly matrix multiplications.
However, $O_m$ is unbounded when using the reverse-mode auto-diff.
For example, $O_m$ is proportional to the depth of feedforward neural networks with auto-diff, because all values of intermediate hidden neurons need to be stored.
The same issue exists if we try to use auto-diff for Schr\"odinger's equation. Moreover, the overhead has a larger impact when simulating quantum systems because of the exponential memory cost of storing a $n$-qubit state.

\subsection{Continuous adjoint method}

    To address the large memory overhead $O_m$, we explore the continuous adjoint method, a technique that significantly reduces memory costs associated with differentiating through ODEs. 
    % Notably, this method has also been recently adopted in the machine learning community for learning neural ODE~\cite{chen_neural_nodate, ijcai2019p103}.
    This approach treats the backpropagation problem as the task of solving an augmented ODE for the adjoint state.
    We also propose the local continuous adjoint method, which in addition utilizes the locality of terms in the Hamiltonian.

\begin{figure}
    \includegraphics[width=1.1\linewidth]{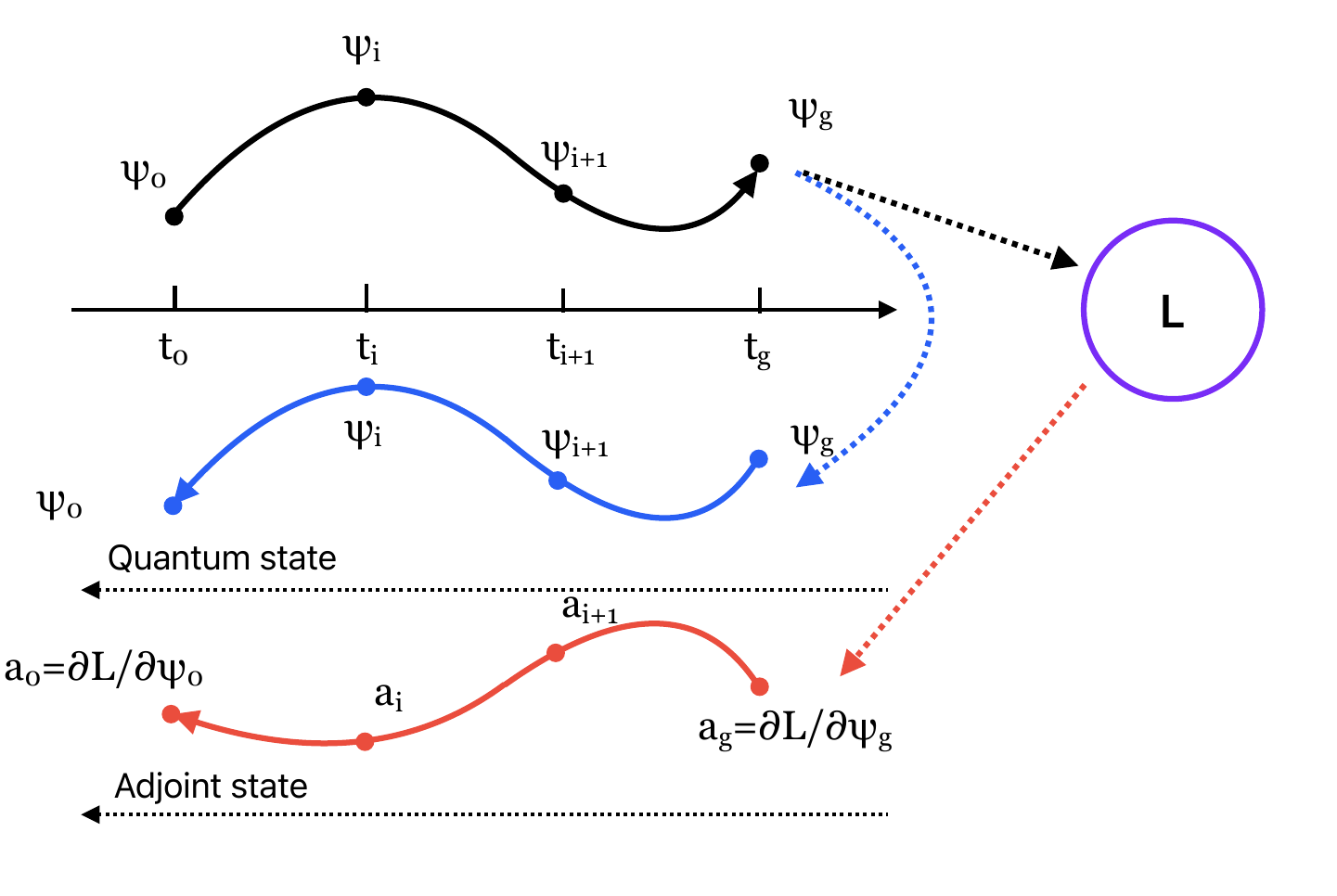}
    \caption{Outline of the continuous adjoint method. The black lines represent time evolution in the forward pass. After the final state $\psi_g$ and the objective function $\mathcal{L}$ are computed, we can use the continuous adjoint method to compute the gradient. To do that, we can evolve $\psi_g$ and $a_g$ backward in time, which means we do not need to store them during the forward pass. During the backward evolution, we can accumulate the gradient according to~\autoref{eq:u_partial_derivative_sum} and~\autoref{eq:gradient_ith_term}.} 
    \label{fig:adjoint_evolve}
\end{figure}

Here, we will briefly explain these methods.
For the more rigorous derivation of the continuous adjoint method, we refer the reader to~\cite{blondel_elements_2024}.
The main idea of the method is that we can simply do a reverse time evolution on the quantum states to avoid the need to store them.
This is illustrated in~\autoref{fig:adjoint_evolve}.
On the other hand, to understand this, it suffices to look at the time evolution after discretization.
We can rewrite~\autoref{eq:discrete_time_evo_unitary} as
\begin{equation}
    \ket{\psi (t_g, \theta)} = U_g(\theta) U_{g-1}(\theta) \cdots U_0(\theta) \ket{\psi (t_0)}.
\end{equation}
Here,  $\theta$ are the tunable parameters from the Hamiltonian $\mathcal{H}(\theta)$.
For brevity, let us assume all quantities in the above equation are real numbers\footnote{In general, the expression $\frac{\partial \mathcal{L}}{\partial z}$ is not well-defined for a complex number $z$. Therefore, we need to convert $z$ to a pair of real numbers. See~\cite{jaxautodiff} for more details about how JAX handles complex numbers.}.
Then, we have 
\begin{equation}
    \frac{\partial \ket{\psi (t_g, \theta)}}{\partial \theta} = \sum_i U_g(\theta) \cdots \frac{\partial U_i( \theta)}{\partial \theta} \cdots U_0(\theta) \ket{\psi (t_0)}.
    \label{eq:u_partial_derivative_sum}
\end{equation}
We note that the above equation already provides one way to compute the gradient with a memory cost independent of $g$, since we only need to keep track of the products of unitaries before and after $\frac{\partial U_i( \theta)}{\partial \theta}$.
But with more careful treatment, we will show that we only need to store states with size $d^N$ instead of matrices with size $d^N\times d^N$, where $N$ is the number of qudits.

To do this, let us examine the $i$-th term in the summation of the above equation.
We will use the simplified notation
\begin{equation}
    \ket{\psi_i} = U_{i}(\theta) \cdots U_0(\theta) \ket{\psi (t_0)}.
\end{equation}
We will also consider an objective function $\mathcal{L}(\ket{\psi (t_g, \theta)})$ which only depends on the final state.
By rewriting $\mathcal{L}$ as a function of $ \ket{\psi_i}$
\begin{equation}
    \mathcal{L}(\ket{\psi (t_g, \theta)}) = \mathcal{L}(U_{g}(\theta) \cdots U_{i+1}(\theta)  \ket{\psi_i}),
\end{equation}
we can consider the partial derivative $a_i = \frac{\partial\mathcal{L}}{\partial \psi_i}$ which is a column vector of size $d^N$.
One can verify that the $i$-th term is
\begin{equation}
    a_i^{T}  \frac{\partial U_i( \theta)}{\partial \theta} \ket{\psi_{i-1}},
    \label{eq:gradient_ith_term}
\end{equation}
where we use the transpose since we already assume all vectors here are real.
Therefore, we only need to keep track of $a_i$ and $\psi_i$ during backpropagation.
By using the chain rule, we have
\begin{equation}    
    a_i = U_{i+1}^T a_{i+1}.
\end{equation}
To compute $\psi_i$ and $a_i$ from $\psi_{i+1}$ and $a_{i+1}$, we can use an approximation such as Suzuki-Trotter decomposition to write $U_{i+1}$ as a product of 2-body terms.
Then, we can compute $ \frac{\partial U_i( \theta)}{\partial \theta} \ket{\psi_{i-1}}$ using an expansion similar to the R.H.S of~\autoref{eq:u_partial_derivative_sum}.
This way, we can also avoid handling full $d^N\times d^N$ matrices during the computation of gradient $\frac{\partial \mathcal{L}}{\partial \theta}$.
This is what we call the local continuous adjoint method in the following text.

\section{Using the Library: Hands-On Examples}
\label{sec:tutorial}

    In this section, we provide a detailed guide on the use of SuperGrad, by using the multipath coupling scheme~\cite{nguyen_blueprint_2022} in a fluxonium qubit chain as the example. The two-fluxonium cross-resonance gate has been successfully realized in the laboratory~\cite{dogan_two-fluxonium_2023, PRXQuantum.6.010349}, which uses a similar architecture. Our initial task involves optimizing the coupling strengths in a fluxonium 3-qubit system. Subsequently, we show how to construct a longer chain of fluxonium qubits based on a frequency pattern. Finally, we optimize this frequency pattern to enhance the fidelity of the simultaneous $\mathrm{CNOT}^{\otimes 5}$ gate.

    The tutorials are based on the current version of SuperGrad 0.2.3. Future releases may introduce breaking changes, and we will ensure that the documentation is kept up to date on GitHub~\cite{supergrad2024github}.

    \begin{figure}
        \includegraphics[width=1.0\linewidth]{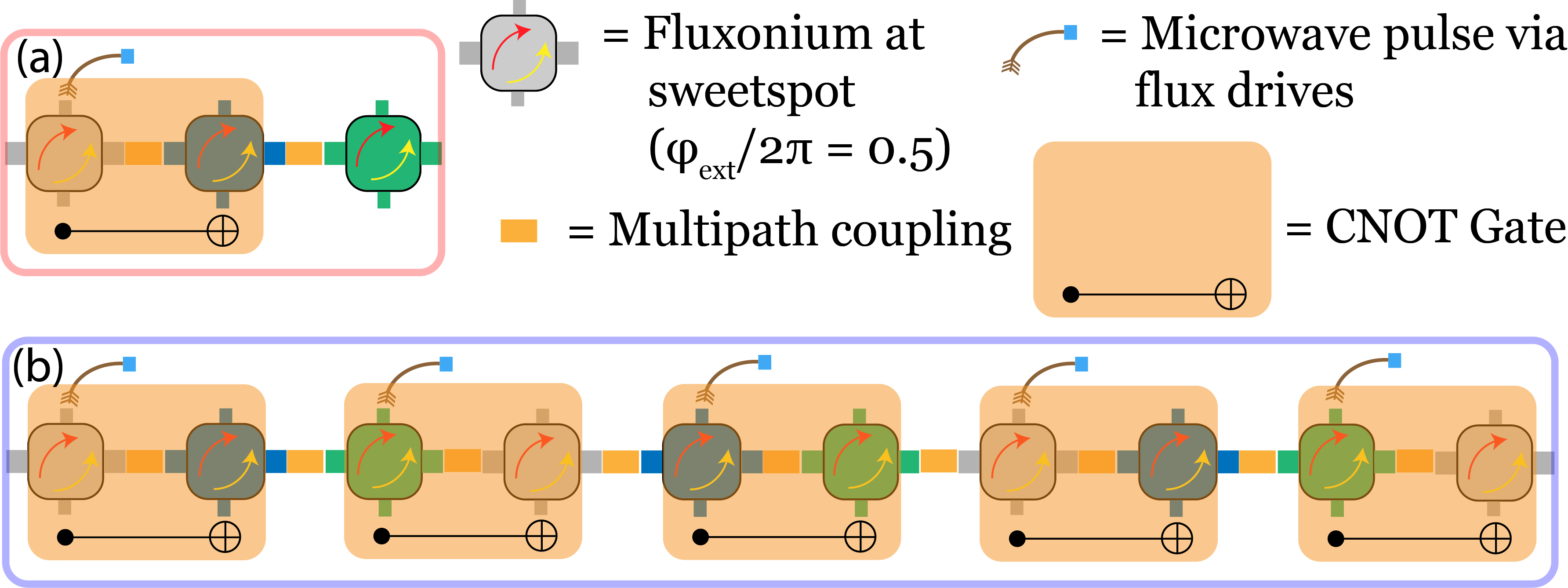}
        \caption{Sketch of a chain-like fluxonium quantum processor with fluxonium qubits biased at the sweetspot, and multipath coupling is used between neighboring qubit pairs. (a) The fluxonium 3-qubit chain arranged in a frequency pattern is highlighted in the red rectangle. Microwave pulses manipulate the gray qubit via its flux drive in~\sref{ssec:workflow_time_evo}. (b) 10-qubit chain-like quantum processor with simultaneous gate control. The fluxonium qubits are periodically arranged in a 3-qubit chain pattern, while their parameters have additional small deviations (see~\sref{ssec:device_and_control_optimization}).}
        \label{fig:quantum_processor}
    \end{figure}

\subsection{Create a 3-qubit chain}
\label{sec:create_multipath_fluxonium_model}
    This section includes a hands-on example using SuperGrad. We explore a 3-qubit chain consisting of three varied fluxonium qubits (highlighted by the red-lined rectangle with label (a) in~\autoref{fig:quantum_processor}). Both capacitive and inductive couplings exist between qubits, configured such that the left and right pairs exhibit identical coupling strengths. The device parameters for each fluxonium are sourced from multipath coupling reference designs~\cite{nguyen_blueprint_2022}. The time-independent part of the 3-qubit chain Hamiltonian is
\begin{equation}
    \label{eq:hamiltonian_3qubit}
    \mathcal{H}(0)=\sum_{1\leq i\leq 3} H_{f,i} + H_{12} + H_{23},
\end{equation}
    where $H_{f,i}$ is the Hamiltonian for the individual fluxonium qubit. It takes the form as
\begin{equation}
    H_{\mathrm{f},i} = 4E_{\mathrm{C},i} \opn_i^2 + \frac{1}{2} E_{\mathrm{L},i} (\opv_i +\varphi_{\mathrm{ext},i})^2
    -E_{\mathrm{J},i}\cos \left( \opv_i \right),
    \label{eq:hamiltonian_single_fluxonium}
\end{equation}
    where $\opv_i$ is the phase operator, and $\opn_i$ is the conjugate charge operator. $\varphi_{\mathrm{ext},i}$ represents the external flux. We will keep $\varphi_{\mathrm{ext},i}=\pi$ to set the fluxonium qubits at their sweetspot. The multipath coupling terms satisfy
\begin{equation}
    \label{eq:multipath_coupling_term}
    H_{ij} = J_{\mathrm{C}} \opn_i \opn_j - J_{\mathrm{L}} \opv_i \opv_j,
\end{equation}
    where $J_{\mathrm{C}}$ and $J_{\mathrm{L}}$ are the strengths of capacitive and inductive coupling.

\begin{lstlisting}[caption={Python Dictionaries for parameters of a 3-qubit chain example. Each parameter is defined using specific keywords.}, label={code:params_dict}]
import numpy as np

fluxonium_1 = {
    "ec": 1.0 * 2 * np.pi,
    "ej": 4.0 * 2 * np.pi,
    "el": 0.9 * 2 * np.pi,
    "shared_param_mark": "grey",
    "phiext": np.pi,
    "system_type": "fluxonium",
}

fluxonium_2 = {
    "ec": 1.0 * 2 * np.pi,
    "ej": 4.0 * 2 * np.pi,
    "el": 1.0 * 2 * np.pi,
    "shared_param_mark": "blue",
    "phiext": np.pi,
    "system_type": "fluxonium",
}

fluxonium_3 = {
    "ec": 1.0 * 2 * np.pi,
    "ej": 4.0 * 2 * np.pi,
    "el": 1.1 * 2 * np.pi,
    "shared_param_mark": "green",
    "phiext": np.pi,
    "system_type": "fluxonium",
}

coupling = {
    "capacitive_coupling": {
        "strength": 0.02 * 2 * np.pi},
    "inductive_coupling": {
        "strength": -0.002 * 2 * np.pi},
}
\end{lstlisting}
    % The Hamiltonian of a single qubit is written in the phase basis over a finite phase range, specifically chosen as $\varphi \in [-5\pi,5\pi]$.

    The parameters of each qubit and their couplings are managed via Python Dictionaries and can be exported as JSON files for subsequent utilization. We now explore the structure of the dictionary used in our implementation, detailed in~\autoref{code:params_dict}. The dictionary defines each qubit type as fluxonium and sets the external flux to bias the fluxonium at its sweetspot. In this work, we will employ the specified fluxonium parameters in~\autoref{code:params_dict} that form a frequency pattern. Each parameter set is uniquely labeled with distinct colors, such as $\textbf{grey}$, $\textbf{blue}$ and $\textbf{green}$. Those colors should be declared in $\textbf{shared\_param\_mark}$ for parameter sharing.

    In addition, we depict the topology of the quantum processor using a graph-based approach. In this representation, nodes correspond to qubits while edges represent the two-qubit interactions. This graph-based methodology facilitates the adaptation of various quantum processor configurations into a graph, leveraging the data structures and algorithms provided by NetworkX~\cite{SciPyProceedings_11}.
    We develop a class $\textbf{SCGraph}$ based on NetworkX's graph. In~\autoref{code:params_graph}, we show how to define a 3-qubit chain using $\textbf{SCGraph}$ class. This class effectively encapsulates all relevant details about our quantum system, including the Hamiltonian parameters and the couplings.

\begin{lstlisting}[caption={The $\textbf{SCGraph}$ stores the structure and parameters of a quantum processor. We attach parameter dictionaries to nodes and edges of the graph.}, label={code:params_graph}]
from supergrad.scgraph.graph import SCGraph

class MultipathThreeQubit(SCGraph):
    def __init__(self):
        super().__init__()

        # nodes represent qubits
        self.add_node("q1", **fluxonium_1)
        self.add_node("q2", **fluxonium_2)
        self.add_node("q3", **fluxonium_3)
        # edges represent two-qubit interactions
        self.add_edge("q1", "q2",**coupling)
        self.add_edge("q2", "q3",**coupling)
\end{lstlisting}

    We provide tools in the $\textbf{Helper}$ subpackage to compute common quantities such as static spectra and the time evolution unitary matrices based on the provided graphs.
    More concretely, $\textbf{Helper}$ can turn these computations into pure functions, which are functions without any side effects. This is a prerequisite for function transformation in JAX (see~\sref{sec:software_arch} for more discussion). We will sometimes refer to such functions as models. 
    To demonstrate the $\textbf{Helper}$ utility in converting $\textbf{SCGraph}$ into models, we engage it to compute the eigenenergies of the composite quantum system and identify the dressed states. The output of this eigenenergy computation is a tensor with $N$ indices, where each index is associated with a label of the corresponding bare state. The Python code snippet~\autoref{code:spectrum} exemplifies this process.
% \newpage
\begin{lstlisting}[escapeinside={(*@}{@*)}, caption={The code block about computing eigenenergies of the composite quantum system and identifying the dressed states.}, label={code:spectrum}]
from supergrad.helper import Spectrum

chain_3q = MultipathThreeQubit()
spec = Spectrum(chain_3q, share_params=True, unify_coupling=True)
params = spec.all_params
energy = spec.energy_tensor(params)
dressed_freq_q1 = (energy[1, 0, 0] - energy[0, 0, 0])
dressed_freq_q2 = (energy[0, 1, 0] - energy[0, 0, 0])
print(dressed_freq_q1 / 2 / np.pi)
# (*@\textcolor{codegreen}{$\bar\omega_{01}/2\pi$}@*) = 0.499 GHz
print(dressed_freq_q2 / 2 / np.pi)
# (*@\textcolor{codegreen}{$\bar\omega_{01}/2\pi$}@*) = 0.582 GHz
\end{lstlisting}

    In the code~\autoref{code:spectrum}, the class property $\textbf{spec.all\_params}$ is responsible for extracting all relevant parameters from the graph. These parameters are subsequently passed as arguments to the pure function \textbf{spec.energy\_tensor}. The pure function can be transformed by JAX to compute gradients or other functions if needed. Users can further customize or expand the functionality by writing their own $\textbf{Helper}$.

    Following~\cite{nguyen_blueprint_2022}, let us analyze the static longitudinal coupling rate. The rate of the left qubit pair is
    \begin{equation}
        \label{eq:static_zz}
        \zeta_{\mathrm{ZZI}} = \omega_{\ket{000}} + \omega_{\ket{110}} - \omega_{\ket{100}} - \omega_{\ket{010}}.
    \end{equation}
    Here, $\omega_{ijk}$ is the dressed state's eigenenergy where the dressed state is close to the bare state $|ijk\rangle$.
    Ideally, we want $\zeta_{\mathrm{ZZI}}$ to be $0$.
    Let us keep the rightmost qubit in its ground state, and optimize the capacitive coupling parameter $J_C$ by sweeping it. This is done in~\autoref{code:compute_coupling_zzi}. We adopt a simplified approach named unify-coupling, which shares parameters between all coupling configurations. This is enabled by setting $\textbf{share\_params}$ and $\textbf{unify\_coupling}$ to True.

    \begin{figure}
        \includegraphics[width=\linewidth]{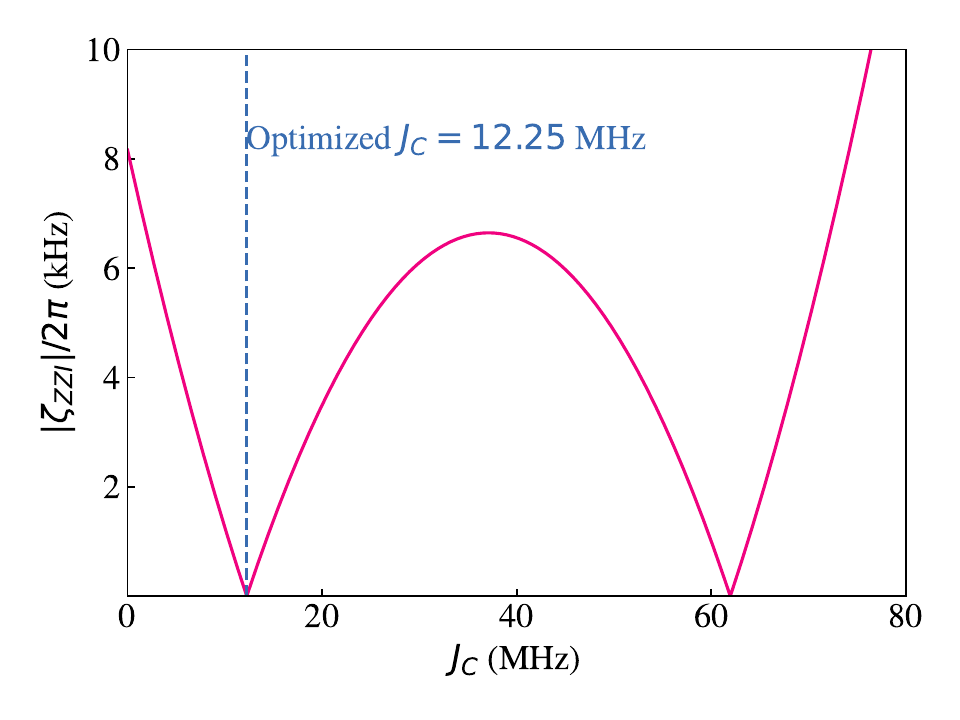}
        \caption{Static longitudinal coupling rate $\zeta_{\mathrm{ZZI}}$ as a function of the capacitive coupling strength $J_C$. The dotted blue line labels optimized $J_C$ found in~\autoref{code:minimize_zz}.} \label{fig:static_zzi}
    \end{figure}

\begin{lstlisting}[caption={The code block about computing static longitudinal coupling rate while sweeping the capacitive coupling parameters.}, label={code:compute_coupling_zzi}]
import jax
import jax.numpy as jnp

def compute_static_longitudinal_coupling(jc):
    params = spec.all_params
    params["capacitive_coupling_all_unify"].update({"strength": jnp.array(jc)})
    energy = spec.energy_tensor(params)
    zeta_zzi = (
        energy[0, 0, 0] + energy[1, 1, 0] - energy[0, 1, 0] - energy[1, 0, 0]
    )
    return (zeta_zzi / 2 / np.pi * 1e6)**2

jc_sweep = jnp.linspace(0.0, 80e-3, 1000)
zeta_sweep = jax.vmap(compute_static_longitudinal_coupling)(jc_sweep * 2 * jnp.pi)
\end{lstlisting}

\subsection{Optimization via JAX Framework}
\label{ssec:multipath_coupling_optimize}

    The JAX framework offers efficient computation of objective function values and gradients through automatic differentiation, which can be used to speed up optimization. 

    In~\autoref{code:minimize_zz}, we perform the minimization of $\zeta_{\mathrm{ZZI}}$ over $J_C$ when $J_L$ is fixed at 2 MHz. We use the BFGS optimizer provided by SciPy. In this example, we use \textbf{jax.grad} to transform \textbf{compute\_static\_longitudinal\_coupling} function to its gradient function by performing automatic differentiation. Then, the gradient function is passed to the minimizer.

\newpage
\begin{lstlisting}[escapeinside={(*@}{@*)}, caption={Code block demonstrating the optimization of the static longitudinal coupling rate using a gradient-based algorithm.}, label={code:minimize_zz}]
from scipy.optimize import minimize

func = lambda x0: compute_static_longitudinal_coupling(*x0)
res = minimize(func, 0.1, method='bfgs', jac=jax.grad(func))
jc_opt = res.x / 2 / np.pi * 1000
print(jc_opt)
# (*@\textcolor{codegreen}{$J_C$}@*) = 12.25 MHz
\end{lstlisting}

    The multipath coupling incorporates both an inductive coupling term $-J_L\opv_A \opv_B$ and a capacitive coupling term $J_C\opn_A \opn_B$ to suppress residual static $ZZ$ interactions. When the inductive coupling constant $J_L$ is fixed at 2 MHz, our goal is to fine-tune the capacitive coupling to minimize the longitudinal coupling rate $\zeta_{\mathrm{ZZI}}$. We define $\zeta_{\mathrm{ZZI}}$ in~\autoref{eq:static_zz} as the objective function as detailed in~\autoref{code:minimize_zz}. The optimization yields $J_C=12.25$ MHz, delineated by the blue line in \autoref{fig:static_zzi}, which corroborates the optimal value identified through a parametric sweep of $J_C$.

\subsection{Workflow for Time Evolution}
    \label{ssec:workflow_time_evo}

    This subsection describes the procedure for simulating the time evolution of a quantum system using the $\textbf{Evolve}$ class within the \textbf{Helper} package, specifically focusing on the implementation of a Cross-Resonance ($\mathrm{CR}$) pulse in a 3-qubit chain.
    First, we initialize pulse parameters for the time evolution, as detailed in the code block~\autoref{code:compute_initial_guess}. This step involves setting the length of the $\mathrm{CR}$ pulse and computing the effective coupling strength $J_{\mathrm{eff}}$ that is extracted from the multipath coupling term~\autoref{eq:multipath_coupling_term}, as well as the detuning between the dressed $0-1$ transition frequencies of qubits 1 and 2. The driving amplitude $\epsilon_d$ of the pulse is then computed based on the detuning and the effective coupling strength.

\begin{lstlisting}[caption={Implementing a $\mathrm{CNOT}$ gate via a cross-resonance pulse. The computation of the initial guess is based on the dressed states' eigenenergies and effective coupling rate.}, label={code:compute_initial_guess}]
from supergrad.helper import Evolve
from supergrad.utils import tensor, compute_fidelity_with_1q_rotation_axis
from supergrad.utils.gates import cnot

length = 100.0
detuning = jnp.abs(dressed_freq_q1 - dressed_freq_q2)
j_eff = 0.01 * 2 * np.pi
tau_eps_drive = np.pi / 2.0 * detuning / j_eff

cr_pulse = {
    "pulse": {
        "amp": tau_eps_drive / length,
        "omega_d": dressed_freq_q2,
        "phase": 0.0,
        "length": length,
        "pulse_type": "cos",
        "operator_type": "phi_operator",
        "delay": 0.0,
    }
}

cr_chain_3q = MultipathThreeQubit()
cr_chain_3q.add_node("q1", **cr_pulse)
\end{lstlisting}

    Subsequent to setting the initial parameters, the $\textbf{Evolve}$ class is employed to simulate the time evolution of the system, detailed in the code block~\autoref{code:evolve}. Hyperparameters are declared for the simulation, such as truncated dimension (the number of qudit's energy levels), parameter sharing (compute gradient with respect to the frequency pattern), and compensation options (the type of errorless single-qubit rotations before and after the time evolution operator). The unitary evolution of the $\mathrm{CR}$ pulse is computed by utilizing the $\textbf{eigen\_basis}$ method, which performs the time evolution simulation within the multi-qubit eigenbasis. This method employs a basis transform technique that computes the evolution within the eigenbasis while benefiting from the local Hamiltonian in the product basis. The detail is shown in~\aref{appen:basis_transform}.

\begin{lstlisting}[caption={Performing time evolution via the Evolve class and calculating the fidelity of the simulated result against the ideal $\mathrm{CNOT}$ gate. Arbitrary single-qubit rotations are added before and after the time evolution to implement the $\mathrm{CNOT}\otimes \mathrm{I}$ gate.}, label={code:evolve}]
target_unitary = tensor(cnot(), jnp.eye(2))
evo = Evolve(cr_chain_3q, truncated_dim=3, share_params=True, unify_coupling=True, compensation_option='no_comp')
params = evo.pulse_params
cr_unitary = evo.eigen_basis(evo.all_params)
fidelity, res_unitary = compute_fidelity_with_1q_rotation_axis(target_unitary, cr_unitary, compensation_option='arbit_single')
print(fidelity)  # 0.98824
\end{lstlisting}

    The $\mathrm{CR}$ pulse introduces an effective $\mathrm{ZX}$ term, which can lead to an implementation of a $\mathrm{CNOT}$ gate when augmented by single-qubit gates. Given the typically lower error rates of single-qubit gates, here we assume they are perfect. These single-qubit compensations are optimized inside the $\textbf{compute\_fidelity\_with\_1q\_rotation\_axis}$ method. The fidelity of the resultant unitary is measured against the target unitary and achieves a high value (0.98824), and the high fidelity affirms the successful implementation of the $\mathrm{CNOT}$ gate with the initial guess of pulse parameters.
    Additionally, we optimize pulse parameters, as illustrated in the accompanying code block~\autoref{code:minimize_fidelity}. This process employs backpropagation to calculate the gradients effectively.
    The function $\textbf{scipy\_optimize}$ is a wrapper to achieve the same functionality shown in~\autoref{code:minimize_zz}.
    After optimization, the fidelity of the $\mathrm{CNOT} \otimes \mathrm{I}$ gate is substantially improved (achieving 99.99\%).

\begin{lstlisting}[caption={Code block showing the objective function and performing pulse parameter optimization.}, label={code:minimize_fidelity}]
import haiku as hk
from supergrad.utils.optimize import scipy_minimize

def infidelity(params):
    params = hk.data_structures.merge(evo.all_params, params)
    cr_unitary = evo.eigen_basis(params)
    fidelity, res_unitary = compute_fidelity_with_1q_rotation_axis(target_unitary, cr_unitary, compensation_option='arbit_single')
    return jnp.abs(1 - fidelity)

scipy_minimize(infidelity, params, method='L-BFGS-B', logging=True)

#  message: CONVERGENCE: REL_REDUCTION_OF_F_<=_FACTR*EPSMCH
#  success: True
#   status: 0
#      fun: 1.716181357702684e-05
\end{lstlisting}

\subsection{Device and control gradient optimization}
\label{ssec:device_and_control_optimization}

    In this section, we explore a one-dimensional chain of fluxonium qubits with the optimized multipath coupling designs in~\sref{ssec:multipath_coupling_optimize}. The qubits are characterized by nine frequency pattern parameters in~\autoref{table:frequency_pattern} that form a frequency pattern along the chain. For our example, we employ a chain-like quantum processor comprising ten qubits. This setup is illustrated in~\autoref{fig:quantum_processor}. The initial values of the frequency pattern parameters are provided in~\autoref{table:frequency_pattern}. We vary $E_L$ by $\pm 0.1\text{GHz}$ to create differences in the qubit spectrums. Notably, the precision fabrication of superconducting qubits presents challenges, leading to deviations in the final device parameters of the quantum processor post-fabrication. Interestingly, as mentioned in~\cite{berke_transmon_2022}, these deviations can enhance the localization and addressability of the qubits. We model these deviations by assuming that the device parameters follow a normal distribution. The mean of this distribution corresponds to the values $E_{C/J/L}$ in~\autoref{table:frequency_pattern}, and the standard deviation is $0.01\times E_{C/J/L}$.

\begin{table}[h]
    \centering
    \begin{tabular}{l c c c}
    \hline
    \hline
    Qubit & $E_J$ (GHz) & $E_C$ (GHz) & $E_L$ (GHz) \\
    \hline
    A (Grey) & 4.0 & 1.0 & 0.9 \\
    B (Blue) & 4.0 & 1.0 & 1.0 \\
    C (Green) & 4.0 & 1.0 & 1.1 \\
    \hline
    \hline
    \end{tabular}
    \caption{The frequency pattern for the fluxonium chain. The colors of qubits are referring to the colors in~\autoref{fig:quantum_processor} and in~\autoref{code:params_dict}.}
    \label{table:frequency_pattern}
\end{table}

    We extend the time evolution workflow as described in~\sref{ssec:workflow_time_evo}, to simulate systems of ten fluxonium qubits. This is achieved by applying simultaneous control pulses to execute $\mathrm{CNOT}^{\otimes 5}$ quantum gates. SuperGrad provides a multitude of options for configuring time evolution, each affecting the simulation's accuracy and the performance of the solver. These options include the number of time steps, the Trotter order of the Trotterization solver, and the basis for the unitary matrix. The code template for implementing the $\mathrm{CNOT}^{\otimes 5}$ gate is presented in~\autoref{code:10qubit_cnot}. The class $\textbf{MPCFluxonium}$ is a $\textbf{SCGraph}$ with many custom methods. For example, we use $\textbf{create\_cr\_pulse}$ to add $\mathrm{CR}$ pulses in a way similar to~\autoref{code:compute_initial_guess}. We add random deviations to the device parameters by setting $\textbf{add\_random}$ to True. We enable the parameter sharing so that the gradient computed in~\autoref{table:gradient_origin} is with respect to the parameters in the frequency pattern. 

\begin{lstlisting}[caption={The code template for constructing a 10-qubit chain and performing simultaneous $\mathrm{CNOT}^{\otimes 5}$ gates. The initial values for the pulses are set inside the \textbf{MPCFluxonium1D} class, in the same way as in~\autoref{code:evolve}. The argument $\textbf{astep}$ is the number of time steps during time evolution, which is equal to $t_g/\delta_t$.}, label={code:10qubit_cnot}]
from supergrad.scgraph.graph_mpc_fluxonium_1d import MPCFluxonium1D

chain = MPCFluxonium1D(n_qubit=10, periodic=False, seed=0)
chain.create_cr_pulse(
    ix_control_list=[0, 2, 4, 6, 8],
    ix_target_list=[1, 3, 5, 7, 9],
    tg_list=[100.0, 100.0, 100.0, 100.0, 100.0], add_random=True)
evo = Evolve(chain, truncated_dim=2,
    share_params=True, unify_coupling=True,
    options={'astep': 2000, 
             'trotter_order': 4})
# Compute the time evolution unitary in the eigenbasis.
sim_u = evo.eigen_basis(evo.all_params)
\end{lstlisting}

\begin{figure}
    \includegraphics[width=\linewidth]{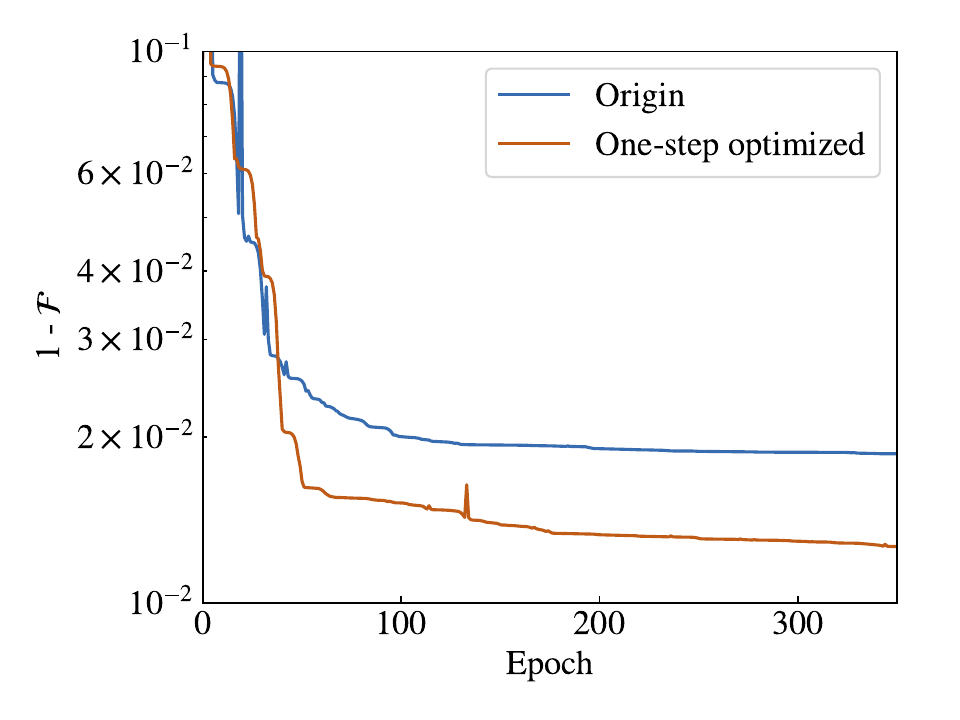}
    \caption{The optimization processes for the control parameters for both the original and one-step optimized frequency patterns are recorded. The x-axis represents the number of minimization steps (Epochs). The fidelity of the simultaneous $\mathrm{CNOT}^{\otimes 5}$ gate exhibits an increase following the one-step frequency pattern optimization.} \label{fig:control_optimizing}
\end{figure}

Given the substantial differences in optimal learning rates for each variable~\cite{ni_superconducting_2023}, here we will only demonstrate a proof of principle optimization loop. We first optimize the control parameters, then perform a single-step update of the fluxonium qubit parameters $E_{C/J/L}$. Finally, we execute another control optimization with the newly updated qubit parameters and reinitialized control parameters.  The multipath coupling strengths remain unchanged in this optimization. We use $1 - \mathcal{F}$ as the objective function to optimize the performance of the simultaneous $\mathrm{CNOT}^{\otimes 5}$ gate. This optimization is carried out using the gradient-based algorithm L-BFGS-B~\cite{doi:10.1137/0916069}.
The objective function during the optimization process is shown in~\autoref{fig:control_optimizing}. Following the optimization of the control parameters within the original frequency pattern, the gradients with respect to the device and control parameters are detailed in~\autoref{table:gradient_origin}.

\begin{table}[h]
    \centering
    \begin{tabular}{ l | c r r}%
    \hline%
    \hline%
    \multicolumn{4}{c}{Parameters and Gradients for Frequency Patterns}\\%
    \hline%
    Type&Parameter&Value (GHz)&Gradient\\%
    \hline%
    \multirow{3}{*}{A (Grey)}&$E_{C, A}$&1.000e+00&{-}4.280e{-}01\\%
    &$E_{J, A}$&4.000e+00&1.323e{-}01\\%
    &$E_{L, A}$&9.000e{-}01&{-}3.866e{-}01\\%
    \hline%
    \multirow{3}{*}{B (Blue)}&$E_{C, B}$&1.000e+00&8.286e{-}01\\%
    &$E_{J, B}$&4.000e+00&{-}2.582e{-}01\\%
    &$E_{L, B}$&1.000e+00&7.027e{-}01\\%
    \hline%
    \multirow{3}{*}{C (Green)}&$E_{C, C}$&1.000e+00&{-}9.257e{-}01\\%
    &$E_{J, C}$&4.000e+00&3.076e{-}01\\%
    &$E_{L, C}$&1.100e+00&{-}8.193e{-}01\\%
    \hline%
    \hline%
    \end{tabular}%
    \caption{Gradients table for the objective function with the original frequency pattern and optimized control parameters.}
    \label{table:gradient_origin}
\end{table}

    Our ultimate goal is to facilitate simultaneous gate operations within the quantum processor. To achieve this, we modify the frequency pattern based on the gradient table detailed in~\autoref{table:gradient_origin}. Selecting appropriate learning rates for various parameters with disparate units can pose a challenge (for an in-depth discussion, refer to the appendix of~\cite{ni_superconducting_2023}). In an ideal scenario, effective learning rates could be determined by calculating the Hessian matrix. However, in this tutorial, the coupling parameters remain constant, and $E_{C/J/L}$ are optimized according to the following learning rate:
\begin{equation}
    \label{eq:one_step_optimizing}
    E_{\mathrm{C/J/L},i} \rightarrow E_{\mathrm{C/J/L},i} - 0.01 \mathrm{GHz^{2}}\times \frac{\partial \mathcal{L}}{\partial E_{\mathrm{C/J/L},i}}.
\end{equation}

    This process results in a more favorable frequency pattern. The optimization processes for the control parameters for both the original and one-step optimized frequency patterns are recorded in~\autoref{fig:control_optimizing}. As the number of minimization steps (epochs) increases, we observe the objective function gradually converging, indicating an improvement in gate performance with the optimized frequency pattern compared to the original one. The fidelity of the simultaneous $\mathrm{CNOT}^{\otimes 5}$ gate exhibits an increase following the one-step frequency pattern optimization. 
    % Accomplishing this task would be difficult without SuperGrad, especially considering the prohibitive memory cost of the Trotterization auto-diff.

\subsection{Fitting experimental data}
\label{sec:fitting_exp_data}

    In this section, we give an example of using SuperGrad to characterize quantum processors through fitting experimental data.
    % This task can be viewed as a Hamiltonian parameterized inverse eigenvalue problem (HamPIEP)~\cite{willsch_observation_nodate}. 
    Our example is based on the spectrum measurement of a single fluxonium qubit.
    Our objective is to deduce the parameters in~\autoref{eq:hamiltonian_single_fluxonium} by measuring the energy eigenvalues when the fluxonium is at different external flux $\varphi_{\mathrm{ext}}$ in experiments. For example, we can measure the energy difference between the two lowest eigenvalues $\varepsilon_{01} = E_1(\varphi_{\mathrm{ext}})-E_0(\varphi_{\mathrm{ext}})$ at varying $\varphi_{\mathrm{ext}}$. We could also calculate the theoretical energy difference $f_{01}(\varphi_{\mathrm{ext}})$ by diagonalizing the Hamiltonian that contains learnable parameters $E_C,E_J,E_L,\varphi_{\mathrm{ext}}$. The caveat is that we cannot directly manipulate $\varphi_{\mathrm{ext}}$ in the experiment. Instead, we apply a voltage $x (\text{V})$ that is approximately linear to $\varphi_{\mathrm{ext}}$. 
    
\begin{figure}
    \includegraphics[width=\linewidth]{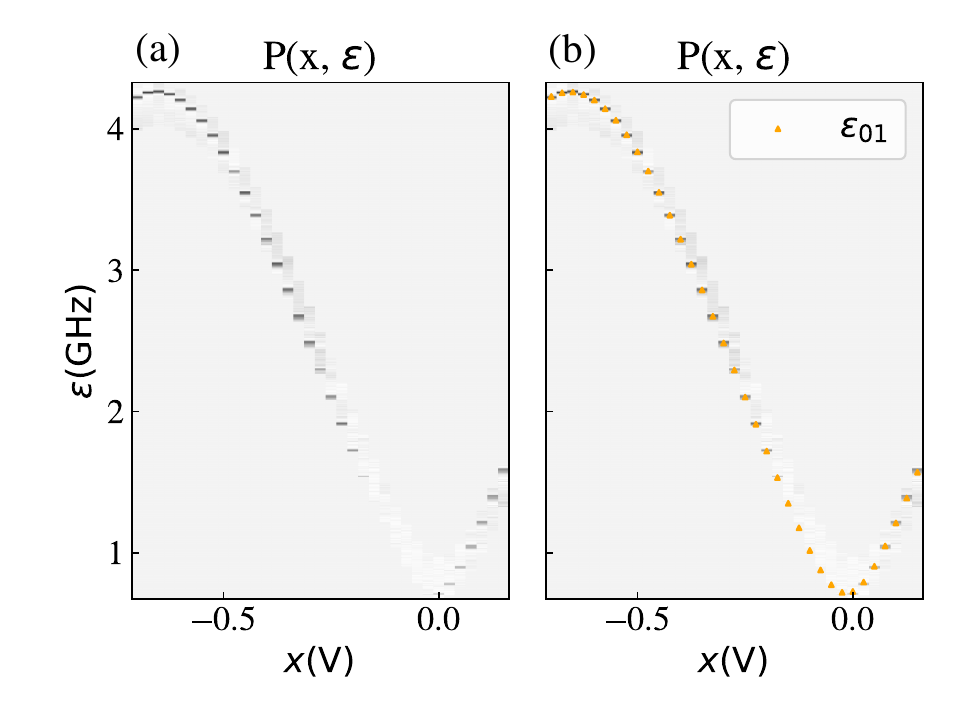}
    \caption{The experimental data and the fitting for the spectra measurement of a fluxonium qubit. The data are normalized as a probability distribution and then plotted as a color mesh from white to gray. The label of the x-axis is voltage, which is approximately linear with external flux $\varphi_{\text{ext}}$. After optimizing the KL-DIV objective function, we plot the result $f_{01}$ as orange dots. (a) The experiment spectra measurement with proper normalization. (b) The fitted spectra dots together with the original measured data.} 
    \label{fig:exp_spectra}
\end{figure}

    However, extracting $\varepsilon_{01}$ from the experimental data is not always straightforward, particularly when the signal-to-noise ratio is low. Moreover, we might miss some signals if we are scanning a small part of the parameter space. For example, \autoref{fig:exp_spectra} shows the experimental data and a fitting that we will obtain at the end of this tutorial. If we simply locate peaks in every column, the columns without a signal might produce wrong $\varepsilon_{01}$ values.

    Here, we propose a novel way~\cite{private_communication_lei_wang} to improve the robustness of the fitting. We will first normalize the data to be positive and sum to 1, and then treat it as a probability distribution $P(x, \varepsilon)$ on 2D.  We construct the target probability distribution $Q(x, \varepsilon)$ based on numerical simulation of the fluxonium Hamiltonian. $Q$ has learnable parameters including ``linewidth'' parameter $\lambda$, x-axis linear transformation parameters $a,b$, and a background noise parameter $c$. The $Q(x, \varepsilon)$ satisfies
\begin{equation}
Q(x, \varepsilon) = \frac{1}{Z}([\frac{\lambda}{(\varepsilon-f_{01}(ax+b))^2+\lambda^2}] + c),
\end{equation}
    where $Z$ is the normalization factor. We choose the objective function to be the Kullback-Leibler divergence (KL-DIV)
\begin{equation}
\mathcal{C}(P, Q) = \sum_{x, \varepsilon} P\ln (\frac{P}{Q}).
\end{equation}
    With SuperGrad, we can then compute the gradient and Hessian matrix of the objective function $\mathcal{C}$ utilizing higher-order automatic differentiation provided by the JAX framework. We employ the Newton-CG optimizer to minimize the objective function. The fitting result is depicted in~\autoref{fig:exp_spectra}.

\section{Benchmarks}
\label{sec:benchmark}
    In this section, we benchmark SuperGrad performances for differentiable simulation, which comprises cost function evaluation and the corresponding backpropagation for gradient computation. We examine how the time and memory costs scale with the dimension of the Hilbert space by performing time evolution on $N$ fluxonium qubits with multipath coupling, which is an extension of the Hamiltonian in~\autoref{eq:hamiltonian_3qubit}. In the time evolution, each qubit simultaneously undergoes an $\mathrm{X}$ gate operation. The detailed descriptions of the gate scheme are provided in~\aref{appen:gate_simulation}. 
    
\onecolumn\newpage
    \begin{figure}
        \begin{minipage}{0.5\textwidth}
            \begin{subfigure}{\textwidth}
                \includegraphics[width=\textwidth]{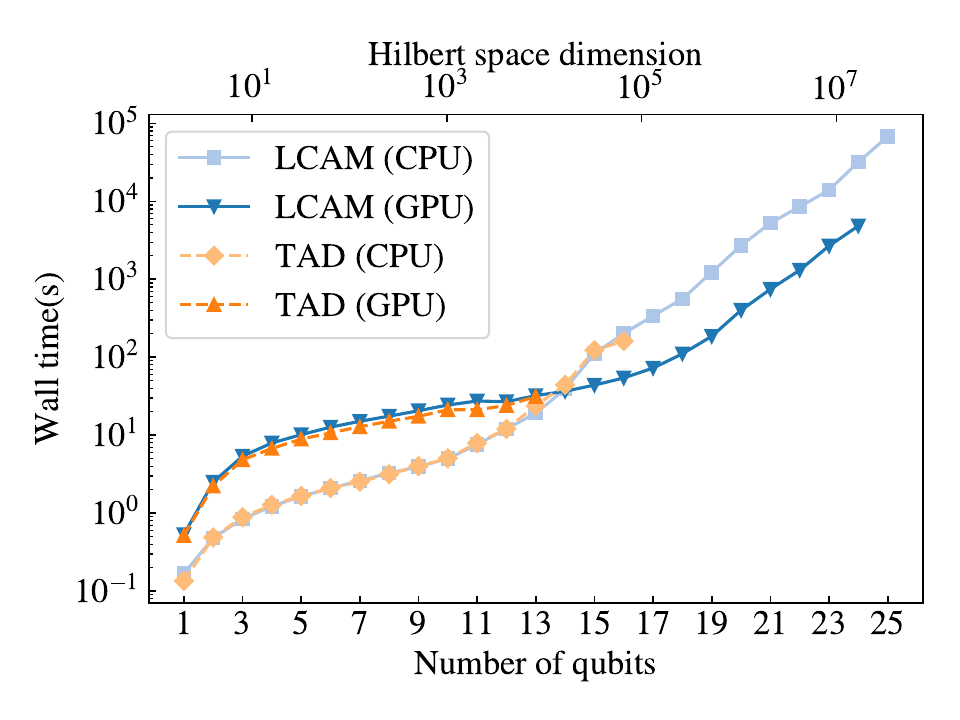}
                \caption{Computing final state and gradients ($\delta t=0.01, T=50$)}
                \label{fig:state_benchmark_grad_method}
            \end{subfigure}
        \end{minipage}\hfill
        \begin{minipage}{0.5\textwidth}
            \begin{subfigure}{\textwidth}
                \includegraphics[width=\textwidth]{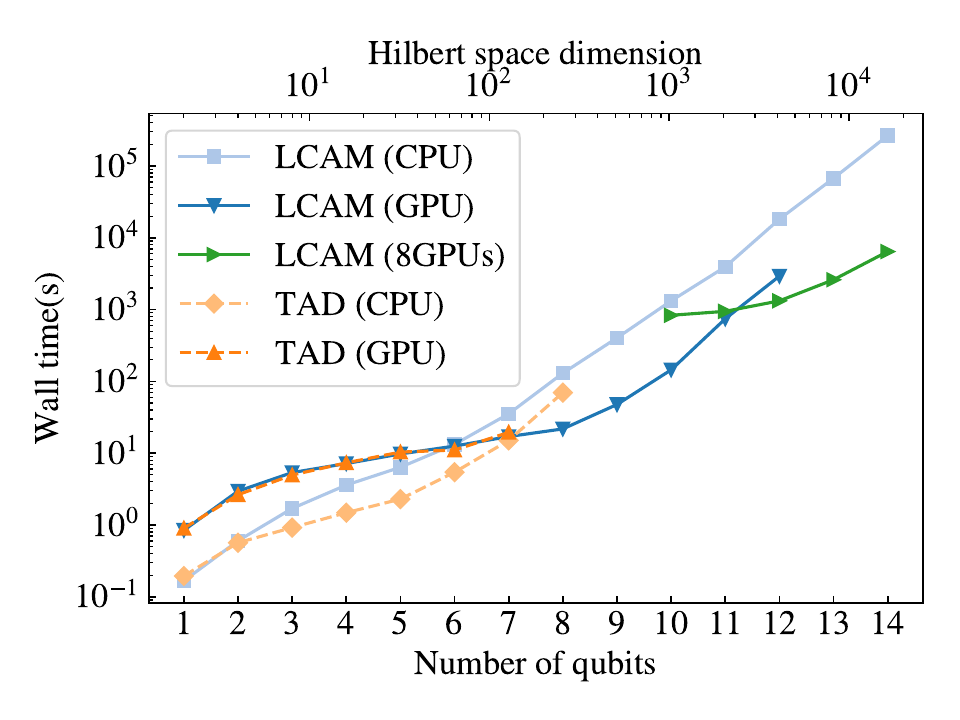}
                \caption{Computing unitary and gradients ($\delta t=0.01, T=50$)}
                \label{fig:unitary_state_benchmark_grad_method}
            \end{subfigure}
        \end{minipage}
        \caption{Benchmarking comparison of the auto-diff of Trotterization (TAD) and our local continuous adjoint method (LCAM) focused on the wall time required to evaluate (a) final state fidelity and (b) gate fidelity, as well as gradients with respect to control and device parameters, respectively. The process wall time was measured against the number of qubits, and the results were minimized across three runs. For state differentiable simulation (a), the TAD incurs a significantly higher memory cost, approximately 38 times greater than LCAM for a chain of 16 qubits (in CPU). The TAD exceeds our GPU's memory limitation (32GB) for qubit chains larger than 13 qubits, while the LCAM method remains feasible until qubit chains exceed 24 qubits, at which point it also surpasses GPU memory limits. For unitary differentiable simulation (b), the memory cost of TAD is 30 times greater than LCAM for an 8-qubit chain (in CPU). TAD's memory cost is out of GPU memory for qubit chains larger than 7 qubits; in contrast, the LCAM method remains within GPU memory up to a qubit chain size of 12. By performing the unitary differentiable simulation in an 8GPUs configuration, the LCAM method remains available until qubit chains exceed 14 qubits. We note that this memory overhead of TAD is dependent on the time step size $\delta t$, which is chosen to be $0.01$ ns.}
        \label{fig:benchmark_grad_method}
    \end{figure}
\twocolumn

Our experimental setup consists of an AMD\textsuperscript{\textregistered} EPYC\textsuperscript{\textregistered} 7513 CPU running at 3.2 GHz (32-core/64-threads), paired with 256 GB of random-access memory (RAM), and eight NVIDIA\textsuperscript{\textregistered} Tesla\textsuperscript{\textregistered} V100 GPUs, each equipped with 32 GB of memory. The operating system is Ubuntu 22.04 LTS, with CUDA 12.3 and JAX 0.4.33. The SuperGrad version is 0.2.3, and the source code for the benchmarks presented in this section is accessible at~\cite{supergrad2024github}.

\subsection{Benchmark backpropagation algorithm}
\label{sec:benchmark_grad_method}
    For computing the time evolution unitaries, we will always use the Trotterization solver detailed in~\sref{sec:trotterization_solver}.
    Only the gradient computation method will be changed and compared.

    We present benchmarks for the wall time across various numbers of qubits. For the differentiable simulation of the state as shown in~\autoref{fig:state_benchmark_grad_method}, we employ the target-state infidelity $\mathcal{L}(|\psi_g\rangle)=1-|\langle\psi_t|\psi_g\rangle|^2$ as the cost function, where $|\psi_t\rangle$ is the target final state. We measure the wall time including evaluating the final state and computing the gradients with respect to control and device parameters. 

    In~\autoref{fig:unitary_state_benchmark_grad_method}, we benchmark for optimizing the evolution unitary. It's achieved by computing the time evolution of a set of initial states $\{|\psi_0^s\rangle\} (s=1,2,\dots, S)$. For this task, we use the composite state-transfer cost function~\cite{PhysRevA.95.042318}, which takes the form as
    \begin{equation}
        \label{eq:composite_state_cost}
        \mathcal{L}(U(t_0,t_g)) = 1-|\frac{1}{S}\sum_s\langle\psi_t^s|\psi_g^s\rangle|^2,
    \end{equation}
    where the set of target states $\{|\psi_t^s\rangle=U_t|\psi_0^s\rangle\} (s=1,2,\dots, S)$ represents the column vectors of the target unitary $U_t$. This composite state-transfer cost function is equivalent to the gate fidelity when used over a complete basis. Note that the cost function could be computed using distributed methods to support multi-GPU configurations.

    We computed gradients by the auto-diff of Trotterization (TAD) and the local continuous adjoint method (LCAM), denoted as $\mathbf{g}^{\mathrm{TAD}}$ and $\mathbf{g}^{\mathrm{LCAM}}$ respectively.
    To ensure the time evolution computation is accurate, we employ a time step size of $\delta t = 0.01$ ns and utilize a second-order Trotter decomposition to achieve a converged time evolution, where LCAM uses the same step size and Trotter order.
    Convergence is confirmed through comparisons between the time evolution unitary produced by the Trotterization solver with the unitary computed by Qiskit Dynamics using a very small ODE solver tolerance (dataset is shown in~\cite{ziang_2024_11192761}). We evaluate the accuracy of LCAM by computing the relative error for each component in the gradient, which is listed in~\autoref{table:parameters_relative_error}. We can see that the relative errors are all smaller than $10^{-7}$.
    The LCAM offers significant memory efficiency (e.g., one-thirtieth of memory usage compared to TAD for computing eight qubits evolution unitary) as it only needs to store a single copy of the quantum state and the adjoint state, and it does not need to represent the Hamiltonian as a dense matrix for evaluating $\partial \mathcal{H}(t,\theta)/\partial \theta$. 

\subsection{Leveraging multi-GPU configuration}
\label{ssec:hpc_extensibility}
\begin{figure}[t]
    \includegraphics[width=\linewidth]{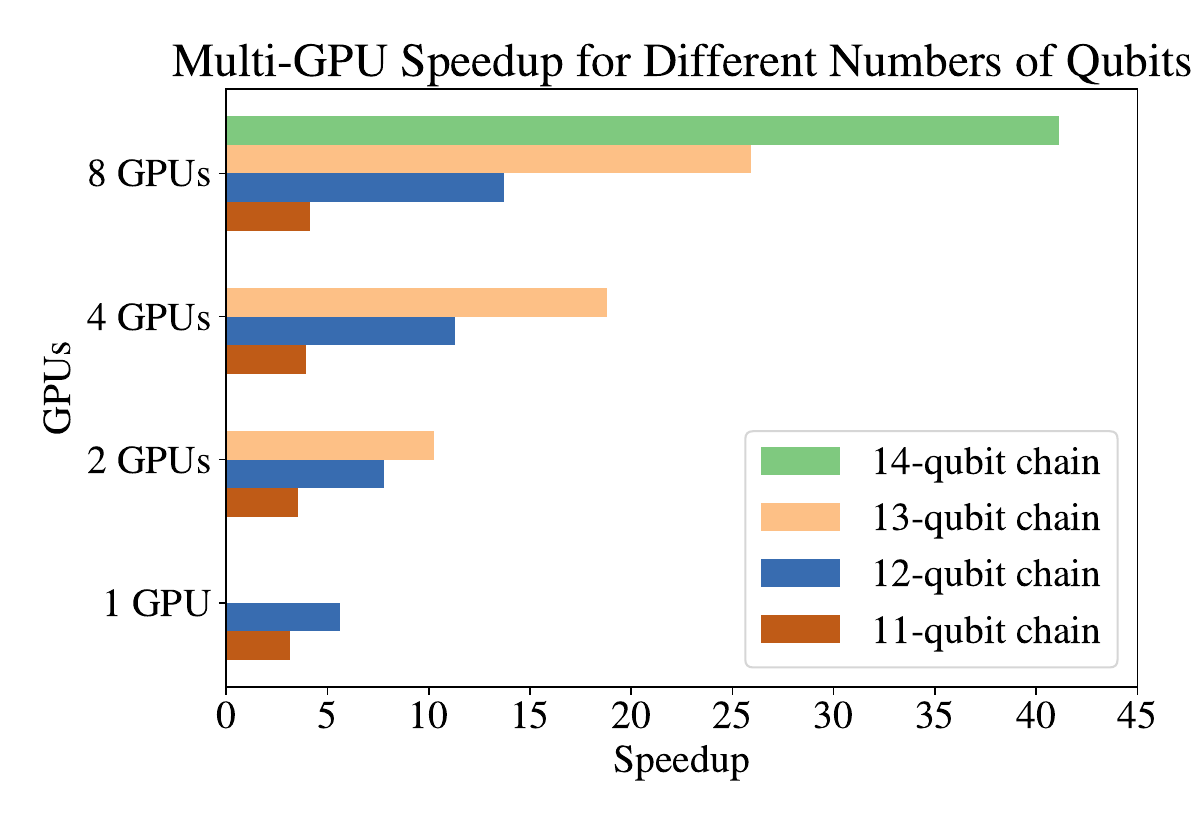}
    \caption{Benchmark of the multi-GPU speedup when computing gradients using LCAM with NVIDIA\textsuperscript{\textregistered} Tesla\textsuperscript{\textregistered} V100 32GB GPUs over CPU implementations on AMD\textsuperscript{\textregistered} EPYC\textsuperscript{\textregistered} 7513 CPU (32-core/64-threads) at 3.2GHz. The multi-GPU implementation could surpass the memory limitation of a single GPU by distributing unitary across multiple GPU instances. During our benchmark, 2 GPUs were able to simulate up to 13 qubits, while 8 GPUs could handle 14 qubits, both of them exceeding the memory capacity of a single GPU.}
    \label{fig:speed_multi_gpu}
\end{figure}

    % High-performance computing (HPC) clusters connect many separate nodes (CPUs, GPUs) via fast interconnects, serving as powerful tools for scientific computations and machine learning tasks. Their ability to process vast amounts of data at high speed and in parallel makes them particularly suitable for simulating time evolution unitary that exceeds the memory capabilities of a single node.

    We have implemented multi-GPU support for time evolution unitary simulations.
    In our time evolution solvers, the largest memory cost is for storing the quantum states array, which scales as $O(d^{2N})$. Where the quantum processor has $N$ qudits, and each qudit subsystem has a dimension $d$. By distributing the unitary across multiple devices, we can simulate larger quantum processors in parallel.

    We benchmark the speedup of the multi-GPU backend over the CPU backend by performing the same time evolution for unitary as described in \sref{sec:benchmark_grad_method} on an $N$-qubit chain, where gradients are computed using the local continuous adjoint method. The speedup results shown in \autoref{fig:speed_multi_gpu} are based on the minimal wall time across three runs.
    This benchmark indicates that larger Hilbert space sizes better harness the floating-point computational capabilities of the GPU. Additionally, employing multi-GPU configurations allows us to mitigate the memory limitations of individual GPUs. However, the number of qubits that can be simulated remains constrained by the exponential growth of the memory cost associated with a single quantum state. To address this limitation, we plan to explore the use of tensor networks in future work.

\subsection{Comparison with other simulation software}
\label{ssec:benchmark_comparision}

    In this subsection, we compare SuperGrad against similar software packages, such as SCQubits, QuTiP, and Qiskit Dynamics. We list their features in~\autoref{table:benchmark_comp_setup}. SuperGrad stands out as the only library that integrates Hamiltonian construction and time evolution for superconducting processors. Therefore, it is capable of computing the gradients of control and device parameters in a single backpropagation. Qiskit Dynamics is the only one that supports perturbation-theory-based solvers, such as the Dyson series and Magnus expansion~\cite{PUZZUOLI2023112262}. Qibo stands out for its advanced hardware accelerator support, which takes full advantage of GPU and multi-GPU devices~\cite{efthymiou_qibo_2022}.
    
    To get a benchmark for the time and memory cost of these libraries, we will need to use a combination of SCQubits and QuTiP / Qiskit Dynamics to be near feature parity with SuperGrad.
    We again run the time evolution of $X^{\otimes 6}$ as the benchmark. The result is listed in~\autoref{table:benchmark_comparison}.

\onecolumn\newpage
\begin{table}[h]
\centering
\begin{tabular}{ l | l | l | l | l | l }%
\hline%
\hline%
& SuperGrad & SCQubits & QuTiP & Qiskit Dynamics & Qibo \\
\hline%
\hline
Version & 0.2.3 & 4.3.0 & 5.1.1 & 0.5.1 & 0.2.16 \\
\hline%
Hamiltonian construction & \checkmark & \checkmark & \ding{55} & \ding{55} & \ding{55} \\
\hline%
Time evolution & \checkmark & \ding{55} & \checkmark & \checkmark & \checkmark \\
\hline%
Gradients of device parameters & \checkmark & \ding{55} & \ding{55} & \ding{55} & \ding{55} \\
\hline%
Gradients of control parameters & \checkmark & \ding{55} & Pre-Alpha & \checkmark & \checkmark \\
\hline%
GPU acceleration & \checkmark & \checkmark & Pre-Alpha & \checkmark & \checkmark \\
\hline%
Multi-GPU support & \checkmark & \ding{55} & \ding{55} & \ding{55} & \checkmark \\

\hline%
\hline%
\end{tabular}%
    \caption{A comparison of the capabilities of SuperGrad and other packages reveals notable differences. All tools were benchmarked using the version listing in the table. SuperGrad stands out by providing the full simulation pipeline, from superconducting processor Hamiltonian construction to time evolution, whereas SCQubits is limited to Hamiltonian construction. To use the time evolution solver of QuTiP and Qiskit Dynamics, one needs to supply the Hamiltonians, e.g., from SCQubits. This entire process simulation support in SuperGrad enables efficient differentiable simulation, while Qiskit Dynamics and Qibo only support backpropagation for the time evolution part. QuTiP provides partial support for differentiable simulations and GPU acceleration via optional plug-ins (QuTiP-JAX and QuTiP-CuPy), but these features are in the pre-alpha stage. Both Qibo and SuperGrad support distributed time evolution across multiple GPUs, which is crucial for simulating larger quantum systems.
    Additionally, Qiskit Dynamics offers perturbation-theory-based solvers, which deliver a significant speed advantage over traditional solvers, though at the cost of increased initial compilation time~\cite{PUZZUOLI2023112262}.
}
    \label{table:benchmark_comp_setup}
\end{table}

\begin{table}[h]
    \centering
    \begin{tabular}{ l | c | c | c | c | c }%
    \hline%
    \hline%
    \multicolumn{2}{l|}{} & SuperGrad & \multicolumn{2}{c|}{SCQubits + Qiskit Dynamics} & SCQubits + QuTiP\\%
    \hline%
    \hline
    \multicolumn{2}{l|}{ODE solver} & Trotterization & \multicolumn{3}{c}{Dormand-Prince} \\%
    \hline
    \multicolumn{2}{l|}{Differential method} & LCAM & CAM & \multicolumn{2}{c}{FDM}\\%
    \hline%
    \multicolumn{2}{l|}{Gradient computation scope} & Device and control & Control & \multicolumn{2}{c}{Device and control} \\%
    \hline%
    \multirow{3}{*}{Time} & Forward $f$ & 2.02 s  & 9.28 s & 9.28 s & 173.81 s\\%
    \cline{2-6}
    & Gradients $\nabla f$ & 8.55 s & 14.73 s & 554.66 s & 10151.47 s \\
    \cline{2-6}
    & Overhead $O_t$ & 4.23 & 1.59 & 59.77 & 58.41 \\
    \hline%
    \multirow{3}{*}{Memory} & Forward $f$ & 2.01 GB & 1.96 GB & 1.96 GB  & 1.83 GB \\%
    \cline{2-6}
    & Gradients $\nabla f$ & 2.74 GB & 2.97 GB & 8.54 GB & 6.31 GB \\
    \cline{2-6}
    & Overhead $O_m$ & 1.36 & 1.52 & 4.36 & 3.45 \\
    \hline%
    \hline%
\end{tabular}%
    \caption{Benchmark of the performance of SuperGrad and other packages, focusing on the wall time and memory usage required to compute the simultaneous $\mathrm{X}^{\otimes 6}$ gate fidelity and its gradient. For the case that the gradient computation scope is "control", gradients are computed only for control parameters (24 parameters in total). In contrast, in the cases where the gradient computation scope is "device and control", gradients are computed for both control and device parameters (58 parameters in total). Time and memory overheads are calculated relative to the corresponding forward simulation. The Dormand-Prince solver in Qiskit Dynamics supports the continuous adjoint method, which allows for efficient backpropagation of time evolution with a small time overhead, as it bypasses gradient computation for the device Hamiltonian. The case is labeled as FDM using the standard finite difference method (FDM), where time overheads are approximated to the number of parameters. Due to Python’s garbage collection strategy, the memory usage is likely overestimated.
    }
    \label{table:benchmark_comparison}
\end{table}
\twocolumn

    For fairness, the ODE solver and its error control parameters (atol = $1 \times 10^{-8}$, rtol = $1 \times 10^{-8}$) were kept the same for both QuTiP and Qiskit Dynamics. We used the JAX backend of Qiskit Dynamics to perform computations.
    For SuperGrad, we employ a time step size of $\delta t = 0.01$ ns and utilize second-order Trotter decomposition. We observed the unitary infidelity $1 - \mathcal{F}$ between SuperGrad and Qiskit Dynamics is smaller than $1 \times 10^{-6}$.
    Note that this does not imply that the results they computed have the same accuracies.
    Therefore, the time overheads for computing gradients are more important than the absolute times in~\autoref{table:benchmark_comparison}.
    
    All benchmarks were run on the CPU since some of these packages do not have GPU backends. The minimal wall time across three runs was recorded, and memory usage was defined by the maximum memory resident set size, as reported by the psutil library during computation. Due to Python’s lazy garbage collection strategy, unused memory may not be released immediately, which affects the accuracy of memory usage benchmarks. This can lead to an overestimation of memory consumption, particularly in FDM computations. In theory, the memory usage for FDM computation should be equivalent to that of the corresponding forward computation.

    One of SuperGrad’s key contributions is its support for the differentiable simulation of both control and device parameters.
On the other hand, Qiskit Dynamics can only compute gradients with respect to control parameters.
We can still estimate the gradients with respect to the device parameters using the finite difference method (FDM). For FDM, we used forward differences, which require one additional computation per parameter, with a step size of $1 \times 10^{-4}$ for finite differences. The results of the computation wall time and memory usage are shown in~\autoref{table:benchmark_comparison}. These benchmarks demonstrate that SuperGrad computes gradients more efficiently than similar software, particularly when dealing with a large number of parameters.

\section{Discussion}
\label{sec:discussion}

\subsection{Library design}
\label{sec:software_arch}

The first question we need to think about when designing the library is how to be compatible with JAX.
The main requirement is that the functions which we want to compute the gradients are pure.
A pure function is a function that, given the same input, will always return the same output and does not produce any side effects.
On the other hand, it is natural to construct the library with some elements of object-oriented programming (OOP) in most modern languages such as Python, C++, etc.
    We have also adopted OOP for SuperGrad. The fundamental features of OOP, such as inheritance and object composition, provide a suitable framework for describing superconducting quantum processors. In our software architecture, we conceptualize the local Hamiltonian terms, such as qubits, couplings, and control fields, as objects in OOP. Here are some examples for which OOP is useful:
\begin{enumerate}
    \item Qubits can be categorized into specific types based on the structure of their Hamiltonians. This mirrors the OOP requirement for treating an object as an instance of a particular class.
    \item As a quantum system, each qubit's eigenvalues could be computed in the same way (diagonalization). This is similar to the OOP class, which could inherit methods from other classes.
    \item The quantum processor can be viewed as a complex composition of various objects, analogous to the concept of object composition in OOP, which allows for the assembly of complex entities from simpler ones.
\end{enumerate}

    In the OOP paradigm, objects have methods that define procedures for the object's behavior, allowing us to compute outputs by running object methods. However, object methods may not be pure functions that satisfy JAX's transformable requirements. For example, consider a fluxonium qubit biased at the sweetspot while external flux $\varphi_{\mathrm{ext}}$ is an inherited parameter. JAX's auto-diff engine will fail to compute the gradient with respect to $\varphi_{\mathrm{ext}}$ when given the object method. This occurs because JAX does not trace inherited parameters. This incompatibility between the OOP paradigm and JAX's pure function model is something that we should always keep in mind.

    Neural network libraries such as Flax~\cite{flax2020github} and Haiku~\cite{haiku2020github} ensure flexible usage of the JAX framework, mitigating the conflict between OOP and JAX's requirement for pure functions. In SuperGrad, we use Haiku to transform the quantum processor composed of objects into pure functions. Haiku can help us collect all parameters declared in a computation into a Python dictionary and pass the dictionary through a transformed pure function. In some quantum processor optimization tasks, it is reasonable to partition the parameters into trainable and non-trainable sets, and Haiku provides utilities for manipulating the parameters dictionary.
    % While Haiku is widely used in SuperGrad, it will soon enter maintenance mode without active development,

However, we are considering replacing Haiku with customized utilities.
This is because Haiku is designed primarily for neural network tasks, e.g., sequentially transforming a tensor by neural network layers. This is not the case for superconducting processor simulation.
For example, there are few intermediate results we want to extract or modify when computing time evolution from \textbf{SCGraph}.
Therefore, we only need to make sure such functions starting from SCGraph can be auto-differentiated.

\subsection{SuperGrad Enhancement Proposals}

The usability of our current version SuperGrad 0.2.3 can be enhanced in several respects, particularly concerning the user's interaction and the system's flexibility:

\begin{enumerate}
    \item Users are required to construct a "graph-like" structure to input all necessary data. This process could be streamlined with the development of a graphical user interface (GUI), which would also help in detecting typos.
    \item Gradient computation and optimization processes require users to specify variable sharing and fix certain parameters. This selection process can be cumbersome. A well-designed GUI could include features to facilitate these selections, enhancing user interaction and system usability.
    \item We are considering making $\textbf{SCGraph}$  a JAX-differentiable data structure. The transformed pure functions will use $\textbf{SCGraph}$ as input data, and the gradient of the objective function will take the same data structure as $\textbf{SCGraph}$. Moreover, the enhanced $\textbf{SCGraph}$ will support functions for handling random deviations and parameter sharing, which further expands its functionality.
    \item Beyond standard operations like computing eigenvalues or simulating time evolution, users may need to perform unconventional calculations. The current object-oriented programming (OOP) structure aims to understand the codebase; however, it still needs developer experience.
\end{enumerate}

\subsection{Role of SuperGrad in experiments}

SuperGrad, like other simulation software, can be an invaluable tool in design optimization, cost-efficient development acceleration, pre-experimental planning, and post-experimental validation for fabricating superconducting processors, as is the case in other fields, such as aircraft, robotics, and semiconductor chip designs. While the exact simulation of a complex processor is generally unfeasible, the software complements expensive design iteration and device testing, and helps plan and understand the experimental measurements of device behavior.
To showcase the multifaceted capabilities of SuperGrad, we provide the following two scenarios.

First, we often need to prototype novel processor designs with new types of qubits and couplings. For this purpose, we need to estimate important quantities such as qubit frequencies and 2-qubit gate fidelities.
We then need to optimize the device design for fabrication.
The experimental data obtained by measuring the fabricated devices can be employed to refine the simulation, progressively enhancing the predictive power of simulation models. As a result, the trial-and-error iterations can be substantially accelerated through this synergy between simulations and experiments.

Secondly, most of the time, we are looking for robust features in simulations and experiments.
For example, in the experimental spectra data in~\autoref{fig:exp_spectra}, the transition frequencies $f_{01}$ can have small shifts due to two-level systems and other sources in the processors.
However, for a working qubit, the overall shape of the spectra is close to the theoretical curve $f_{01}(\varphi_{\text{ext}})$, and therefore we can extract a good approximation of qubit parameters.
Another example is the inaccuracy from control pulse distortions, which means that the qubits never experience the exact same control pulses as the simulation.
However, we can use the simulation software to verify that for a gate scheme, the pulse parameters can be calibrated to compensate for pulse distortions.
The average fidelities after calibration are thus a more robust quantity that can be extracted from simulation.

\subsection{Realization of other superconducting qubit architectures}
    As demonstrated in the previous section, SuperGrad is capable of simulating and optimizing a multipath coupling fluxonium processor. Naturally, SuperGrad also supports transmon qubits. For example, we also reproduced schemes such as transmon qubits with tunable couplers~\cite{xu_high-fidelity_2020} and fluxonium qubits mediated by transmon couplers (FTF)~\cite{PhysRevX.13.031035}. These codes could be found in Zenodo~\cite{ziang_2024_11192761}. The extensible design of SuperGrad allows the integration of novel gate schemes, enabling users to implement their quantum processors accordingly.

    One class of processors that are not currently supported by SuperGrad are qubits with tunable frequencies. More specifically, it is difficult to simulate the performance when idling and interaction frequencies are optimized directly in experiments (e.g., calibration done in~\cite{acharyaSuppressingQuantumErrors2023}). One potential solution for these processors is simulating the optimization process in numerical simulation.

\subsection{Adding a Tensor Network Backend}
    Quantum many-body systems pose a numerical challenge due to the large Hilbert space, which grows exponentially with system size. Various numerical methods such as Matrix Product States (MPS) have been developed to navigate this challenge. Typically, high-rank state tensors can be represented as MPS, which consist of chains of rank-3 tensors truncated by a bond dimension. The time evolution simulations of quantum processors could be improved by employing tensor network algorithms (e.g., the algorithm in~\cite{PhysRevLett.98.070201}). We should be careful that a large bond dimension may be required if we want to accurately represent a state with highly entangled qubits as a tensor network. 
    
    At times, it's necessary to compute correlated errors with the highest precision~\cite{ni_superconducting_2023}, and we wish to compute exact time evolution in this case. With careful design, many superconducting qubit systems can exhibit predominantly local couplings, meaning that simulations in a subset of quantum processors are sufficient to capture the correlated errors. Analyzing these errors is useful for optimizing both the fabrication and control of superconducting qubits, leading to improved quantum processor performance and reduced error rates.
    
    However, for many common operations, the entanglement is very localized. Therefore, it is expected that tensor network algorithms can accurately approximate these kinds of time evolutions. We believe that efficiently simulating larger quantum processors has the potential to significantly advance the device design and control methods, pushing them to the next stage. Hence, we can gain deeper insights into the behavior of the scalable quantum processors.

    Our Trotterization solver has already incorporated some tensor network features. It can transform the time evolution computation into a tensor network format and contract the tensor network through an optimized path. However, we have yet to implement tensor network-based time evolution algorithms in the SuperGrad, which we aim to incorporate in future releases.

\section{Conclusion}

    As a versatile library, SuperGrad is specifically designed for the simulation and optimization of quantum processors. SuperGrad seamlessly integrates the quantum processor device and control optimization within a gradient-based framework. In this work, we have shown that the Trotterization solver serves as an efficient differentiable time evolution engine. With the help of SuperGrad's components, users can easily simulate their quantum processors, even with novel gate schemes. We want to emphasize that SuperGrad supports a wide range of gate schemes and is not limited to the specific showcases presented in this article.

    We acknowledge that efficiently solving both forward and backward problems for time evolution in the many-body quantum system remains an open and challenging question. We are committed to continuing our exploration of viable solutions and improvements in this area. Our SuperGrad library is open source on GitHub~\cite{supergrad2024github}. The SuperGrad library is continually evolving to meet the escalating demands for simulating larger quantum processors, and future releases may introduce significant changes.

\section{Acknowledgement}

The authors would like to thank Wangwei Lan for constructive discussions. ZAW and XW acknowledge the partial support from the National Key Research and Development Program of China (2021YFA1401902). XTN acknowledges the partial support by Guangdong Provincial Quantum Science Strategic Initiative (GDZX2203001, GDZX2403001) and NSFC (No.92476206). FW and HHZ are supported by Zhongguancun Laboratory.

\bibliography{main_cite}

\begin{thebibliography}{60}%
\makeatletter
\providecommand \@ifxundefined [1]{%
 \@ifx{#1\undefined}
}%
\providecommand \@ifnum [1]{%
 \ifnum #1\expandafter \@firstoftwo
 \else \expandafter \@secondoftwo
 \fi
}%
\providecommand \@ifx [1]{%
 \ifx #1\expandafter \@firstoftwo
 \else \expandafter \@secondoftwo
 \fi
}%
\providecommand \natexlab [1]{#1}%
\providecommand \enquote  [1]{``#1''}%
\providecommand \bibnamefont  [1]{#1}%
\providecommand \bibfnamefont [1]{#1}%
\providecommand \citenamefont [1]{#1}%
\providecommand \href@noop [0]{\@secondoftwo}%
\providecommand \href [0]{\begingroup \@sanitize@url \@href}%
\providecommand \@href[1]{\@@startlink{#1}\@@href}%
\providecommand \@@href[1]{\endgroup#1\@@endlink}%
\providecommand \@sanitize@url [0]{\catcode `\\12\catcode `\$12\catcode `\&12\catcode `\#12\catcode `\^12\catcode `\_12\catcode `\%12\relax}%
\providecommand \@@startlink[1]{}%
\providecommand \@@endlink[0]{}%
\providecommand \url  [0]{\begingroup\@sanitize@url \@url }%
\providecommand \@url [1]{\endgroup\@href {#1}{\urlprefix }}%
\providecommand \urlprefix  [0]{URL }%
\providecommand \Eprint [0]{\href }%
\providecommand \doibase [0]{http://dx.doi.org/}%
\providecommand \selectlanguage [0]{\@gobble}%
\providecommand \bibinfo  [0]{\@secondoftwo}%
\providecommand \bibfield  [0]{\@secondoftwo}%
\providecommand \translation [1]{[#1]}%
\providecommand \BibitemOpen [0]{}%
\providecommand \bibitemStop [0]{}%
\providecommand \bibitemNoStop [0]{.\EOS\space}%
\providecommand \EOS [0]{\spacefactor3000\relax}%
\providecommand \BibitemShut  [1]{\csname bibitem#1\endcsname}%
\let\auto@bib@innerbib\@empty
%</preamble>
\bibitem [{\citenamefont {Bao}\ \emph {et~al.}(2022)\citenamefont {Bao}, \citenamefont {Deng}, \citenamefont {Ding}, \citenamefont {Gao}, \citenamefont {Gao}, \citenamefont {Huang}, \citenamefont {Jiang}, \citenamefont {Ku}, \citenamefont {Li}, \citenamefont {Ma}, \citenamefont {Ni}, \citenamefont {Qin}, \citenamefont {Song}, \citenamefont {Sun}, \citenamefont {Tang}, \citenamefont {Wang}, \citenamefont {Wu}, \citenamefont {Xia}, \citenamefont {Yu}, \citenamefont {Zhang}, \citenamefont {Zhang}, \citenamefont {Zhang}, \citenamefont {Zhou}, \citenamefont {Zhu}, \citenamefont {Shi}, \citenamefont {Chen}, \citenamefont {Zhao},\ and\ \citenamefont {Deng}}]{PhysRevLett.129.010502}%
  \BibitemOpen
  \bibfield  {author} {\bibinfo {author} {\bibfnamefont {F.}~\bibnamefont {Bao}}, \bibinfo {author} {\bibfnamefont {H.}~\bibnamefont {Deng}}, \bibinfo {author} {\bibfnamefont {D.}~\bibnamefont {Ding}}, \bibinfo {author} {\bibfnamefont {R.}~\bibnamefont {Gao}}, \bibinfo {author} {\bibfnamefont {X.}~\bibnamefont {Gao}}, \bibinfo {author} {\bibfnamefont {C.}~\bibnamefont {Huang}}, \bibinfo {author} {\bibfnamefont {X.}~\bibnamefont {Jiang}}, \bibinfo {author} {\bibfnamefont {H.-S.}\ \bibnamefont {Ku}}, \bibinfo {author} {\bibfnamefont {Z.}~\bibnamefont {Li}}, \bibinfo {author} {\bibfnamefont {X.}~\bibnamefont {Ma}}, \bibinfo {author} {\bibfnamefont {X.}~\bibnamefont {Ni}}, \bibinfo {author} {\bibfnamefont {J.}~\bibnamefont {Qin}}, \bibinfo {author} {\bibfnamefont {Z.}~\bibnamefont {Song}}, \bibinfo {author} {\bibfnamefont {H.}~\bibnamefont {Sun}}, \bibinfo {author} {\bibfnamefont {C.}~\bibnamefont {Tang}}, \bibinfo {author} {\bibfnamefont {T.}~\bibnamefont {Wang}}, \bibinfo {author} {\bibfnamefont
  {F.}~\bibnamefont {Wu}}, \bibinfo {author} {\bibfnamefont {T.}~\bibnamefont {Xia}}, \bibinfo {author} {\bibfnamefont {W.}~\bibnamefont {Yu}}, \bibinfo {author} {\bibfnamefont {F.}~\bibnamefont {Zhang}}, \bibinfo {author} {\bibfnamefont {G.}~\bibnamefont {Zhang}}, \bibinfo {author} {\bibfnamefont {X.}~\bibnamefont {Zhang}}, \bibinfo {author} {\bibfnamefont {J.}~\bibnamefont {Zhou}}, \bibinfo {author} {\bibfnamefont {X.}~\bibnamefont {Zhu}}, \bibinfo {author} {\bibfnamefont {Y.}~\bibnamefont {Shi}}, \bibinfo {author} {\bibfnamefont {J.}~\bibnamefont {Chen}}, \bibinfo {author} {\bibfnamefont {H.-H.}\ \bibnamefont {Zhao}}, \ and\ \bibinfo {author} {\bibfnamefont {C.}~\bibnamefont {Deng}},\ }\href {\doibase 10.1103/PhysRevLett.129.010502} {\bibfield  {journal} {\bibinfo  {journal} {Phys. Rev. Lett.}\ }\textbf {\bibinfo {volume} {129}},\ \bibinfo {pages} {010502} (\bibinfo {year} {2022})}\BibitemShut {NoStop}%
\bibitem [{\citenamefont {Ding}\ \emph {et~al.}(2023)\citenamefont {Ding}, \citenamefont {Hays}, \citenamefont {Sung}, \citenamefont {Kannan}, \citenamefont {An}, \citenamefont {Di~Paolo}, \citenamefont {Karamlou}, \citenamefont {Hazard}, \citenamefont {Azar}, \citenamefont {Kim}, \citenamefont {Niedzielski}, \citenamefont {Melville}, \citenamefont {Schwartz}, \citenamefont {Yoder}, \citenamefont {Orlando}, \citenamefont {Gustavsson}, \citenamefont {Grover}, \citenamefont {Serniak},\ and\ \citenamefont {Oliver}}]{PhysRevX.13.031035}%
  \BibitemOpen
  \bibfield  {author} {\bibinfo {author} {\bibfnamefont {L.}~\bibnamefont {Ding}}, \bibinfo {author} {\bibfnamefont {M.}~\bibnamefont {Hays}}, \bibinfo {author} {\bibfnamefont {Y.}~\bibnamefont {Sung}}, \bibinfo {author} {\bibfnamefont {B.}~\bibnamefont {Kannan}}, \bibinfo {author} {\bibfnamefont {J.}~\bibnamefont {An}}, \bibinfo {author} {\bibfnamefont {A.}~\bibnamefont {Di~Paolo}}, \bibinfo {author} {\bibfnamefont {A.~H.}\ \bibnamefont {Karamlou}}, \bibinfo {author} {\bibfnamefont {T.~M.}\ \bibnamefont {Hazard}}, \bibinfo {author} {\bibfnamefont {K.}~\bibnamefont {Azar}}, \bibinfo {author} {\bibfnamefont {D.~K.}\ \bibnamefont {Kim}}, \bibinfo {author} {\bibfnamefont {B.~M.}\ \bibnamefont {Niedzielski}}, \bibinfo {author} {\bibfnamefont {A.}~\bibnamefont {Melville}}, \bibinfo {author} {\bibfnamefont {M.~E.}\ \bibnamefont {Schwartz}}, \bibinfo {author} {\bibfnamefont {J.~L.}\ \bibnamefont {Yoder}}, \bibinfo {author} {\bibfnamefont {T.~P.}\ \bibnamefont {Orlando}}, \bibinfo {author} {\bibfnamefont
  {S.}~\bibnamefont {Gustavsson}}, \bibinfo {author} {\bibfnamefont {J.~A.}\ \bibnamefont {Grover}}, \bibinfo {author} {\bibfnamefont {K.}~\bibnamefont {Serniak}}, \ and\ \bibinfo {author} {\bibfnamefont {W.~D.}\ \bibnamefont {Oliver}},\ }\href {\doibase 10.1103/PhysRevX.13.031035} {\bibfield  {journal} {\bibinfo  {journal} {Phys. Rev. X}\ }\textbf {\bibinfo {volume} {13}},\ \bibinfo {pages} {031035} (\bibinfo {year} {2023})}\BibitemShut {NoStop}%
\bibitem [{\citenamefont {Sung}\ \emph {et~al.}(2021)\citenamefont {Sung}, \citenamefont {Ding}, \citenamefont {Braum{\"u}ller}, \citenamefont {Veps{\"a}l{\"a}inen}, \citenamefont {Kannan}, \citenamefont {Kjaergaard}, \citenamefont {Greene}, \citenamefont {Samach}, \citenamefont {McNally}, \citenamefont {Kim}, \citenamefont {Melville}, \citenamefont {Niedzielski}, \citenamefont {Schwartz}, \citenamefont {Yoder}, \citenamefont {Orlando}, \citenamefont {Gustavsson},\ and\ \citenamefont {Oliver}}]{sung_realization_2021}%
  \BibitemOpen
  \bibfield  {author} {\bibinfo {author} {\bibfnamefont {Y.}~\bibnamefont {Sung}}, \bibinfo {author} {\bibfnamefont {L.}~\bibnamefont {Ding}}, \bibinfo {author} {\bibfnamefont {J.}~\bibnamefont {Braum{\"u}ller}}, \bibinfo {author} {\bibfnamefont {A.}~\bibnamefont {Veps{\"a}l{\"a}inen}}, \bibinfo {author} {\bibfnamefont {B.}~\bibnamefont {Kannan}}, \bibinfo {author} {\bibfnamefont {M.}~\bibnamefont {Kjaergaard}}, \bibinfo {author} {\bibfnamefont {A.}~\bibnamefont {Greene}}, \bibinfo {author} {\bibfnamefont {G.~O.}\ \bibnamefont {Samach}}, \bibinfo {author} {\bibfnamefont {C.}~\bibnamefont {McNally}}, \bibinfo {author} {\bibfnamefont {D.}~\bibnamefont {Kim}}, \bibinfo {author} {\bibfnamefont {A.}~\bibnamefont {Melville}}, \bibinfo {author} {\bibfnamefont {B.~M.}\ \bibnamefont {Niedzielski}}, \bibinfo {author} {\bibfnamefont {M.~E.}\ \bibnamefont {Schwartz}}, \bibinfo {author} {\bibfnamefont {J.~L.}\ \bibnamefont {Yoder}}, \bibinfo {author} {\bibfnamefont {T.~P.}\ \bibnamefont {Orlando}}, \bibinfo {author}
  {\bibfnamefont {S.}~\bibnamefont {Gustavsson}}, \ and\ \bibinfo {author} {\bibfnamefont {W.~D.}\ \bibnamefont {Oliver}},\ }\href {\doibase 10.1103/PhysRevX.11.021058} {\bibfield  {journal} {\bibinfo  {journal} {Phys. Rev. X}\ }\textbf {\bibinfo {volume} {11}},\ \bibinfo {pages} {021058} (\bibinfo {year} {2021})}\BibitemShut {NoStop}%
\bibitem [{\citenamefont {Stehlik}\ \emph {et~al.}(2021)\citenamefont {Stehlik}, \citenamefont {Zajac}, \citenamefont {Underwood}, \citenamefont {Phung}, \citenamefont {Blair}, \citenamefont {Carnevale}, \citenamefont {Klaus}, \citenamefont {Keefe}, \citenamefont {Carniol}, \citenamefont {Kumph}, \citenamefont {Steffen},\ and\ \citenamefont {Dial}}]{PhysRevLett.127.080505}%
  \BibitemOpen
  \bibfield  {author} {\bibinfo {author} {\bibfnamefont {J.}~\bibnamefont {Stehlik}}, \bibinfo {author} {\bibfnamefont {D.~M.}\ \bibnamefont {Zajac}}, \bibinfo {author} {\bibfnamefont {D.~L.}\ \bibnamefont {Underwood}}, \bibinfo {author} {\bibfnamefont {T.}~\bibnamefont {Phung}}, \bibinfo {author} {\bibfnamefont {J.}~\bibnamefont {Blair}}, \bibinfo {author} {\bibfnamefont {S.}~\bibnamefont {Carnevale}}, \bibinfo {author} {\bibfnamefont {D.}~\bibnamefont {Klaus}}, \bibinfo {author} {\bibfnamefont {G.~A.}\ \bibnamefont {Keefe}}, \bibinfo {author} {\bibfnamefont {A.}~\bibnamefont {Carniol}}, \bibinfo {author} {\bibfnamefont {M.}~\bibnamefont {Kumph}}, \bibinfo {author} {\bibfnamefont {M.}~\bibnamefont {Steffen}}, \ and\ \bibinfo {author} {\bibfnamefont {O.~E.}\ \bibnamefont {Dial}},\ }\href {\doibase 10.1103/PhysRevLett.127.080505} {\bibfield  {journal} {\bibinfo  {journal} {Phys. Rev. Lett.}\ }\textbf {\bibinfo {volume} {127}},\ \bibinfo {pages} {080505} (\bibinfo {year} {2021})}\BibitemShut {NoStop}%
\bibitem [{\citenamefont {Zhang}\ \emph {et~al.}(2024)\citenamefont {Zhang}, \citenamefont {Ding}, \citenamefont {Weiss}, \citenamefont {Huang}, \citenamefont {Ma}, \citenamefont {Guinn}, \citenamefont {Sussman}, \citenamefont {Chitta}, \citenamefont {Chen}, \citenamefont {Houck}, \citenamefont {Koch},\ and\ \citenamefont {Schuster}}]{zhang2023tunable}%
  \BibitemOpen
  \bibfield  {author} {\bibinfo {author} {\bibfnamefont {H.}~\bibnamefont {Zhang}}, \bibinfo {author} {\bibfnamefont {C.}~\bibnamefont {Ding}}, \bibinfo {author} {\bibfnamefont {D.}~\bibnamefont {Weiss}}, \bibinfo {author} {\bibfnamefont {Z.}~\bibnamefont {Huang}}, \bibinfo {author} {\bibfnamefont {Y.}~\bibnamefont {Ma}}, \bibinfo {author} {\bibfnamefont {C.}~\bibnamefont {Guinn}}, \bibinfo {author} {\bibfnamefont {S.}~\bibnamefont {Sussman}}, \bibinfo {author} {\bibfnamefont {S.~P.}\ \bibnamefont {Chitta}}, \bibinfo {author} {\bibfnamefont {D.}~\bibnamefont {Chen}}, \bibinfo {author} {\bibfnamefont {A.~A.}\ \bibnamefont {Houck}}, \bibinfo {author} {\bibfnamefont {J.}~\bibnamefont {Koch}}, \ and\ \bibinfo {author} {\bibfnamefont {D.~I.}\ \bibnamefont {Schuster}},\ }\href {\doibase 10.1103/PRXQuantum.5.020326} {\bibfield  {journal} {\bibinfo  {journal} {PRX Quantum}\ }\textbf {\bibinfo {volume} {5}},\ \bibinfo {pages} {020326} (\bibinfo {year} {2024})}\BibitemShut {NoStop}%
\bibitem [{\citenamefont {Acharya}\ \emph {et~al.}(2023)\citenamefont {Acharya}, \citenamefont {Aleiner}, \citenamefont {Allen}, \citenamefont {Andersen}, \citenamefont {Ansmann}, \citenamefont {Arute}, \citenamefont {Arya}, \citenamefont {Asfaw}, \citenamefont {Atalaya}, \citenamefont {Babbush}, \citenamefont {Bacon}, \citenamefont {Bardin}, \citenamefont {Basso}, \citenamefont {Bengtsson}, \citenamefont {Boixo}, \citenamefont {Bortoli}, \citenamefont {Bourassa}, \citenamefont {Bovaird}, \citenamefont {Brill}, \citenamefont {Broughton}, \citenamefont {Buckley}, \citenamefont {Buell}, \citenamefont {Burger}, \citenamefont {Burkett}, \citenamefont {Bushnell}, \citenamefont {Chen}, \citenamefont {Chen}, \citenamefont {Chiaro}, \citenamefont {Cogan}, \citenamefont {Collins}, \citenamefont {Conner}, \citenamefont {Courtney}, \citenamefont {Crook}, \citenamefont {Curtin}, \citenamefont {Debroy}, \citenamefont {Del Toro~Barba}, \citenamefont {Demura}, \citenamefont {Dunsworth}, \citenamefont {Eppens}, \citenamefont
  {Erickson}, \citenamefont {Faoro}, \citenamefont {Farhi}, \citenamefont {Fatemi}, \citenamefont {Flores~Burgos}, \citenamefont {Forati}, \citenamefont {Fowler}, \citenamefont {Foxen}, \citenamefont {Giang}, \citenamefont {Gidney}, \citenamefont {Gilboa}, \citenamefont {Giustina}, \citenamefont {Grajales~Dau}, \citenamefont {Gross}, \citenamefont {Habegger}, \citenamefont {Hamilton}, \citenamefont {Harrigan}, \citenamefont {Harrington}, \citenamefont {Higgott}, \citenamefont {Hilton}, \citenamefont {Hoffmann}, \citenamefont {Hong}, \citenamefont {Huang}, \citenamefont {Huff}, \citenamefont {Huggins}, \citenamefont {Ioffe}, \citenamefont {Isakov}, \citenamefont {Iveland}, \citenamefont {Jeffrey}, \citenamefont {Jiang}, \citenamefont {Jones}, \citenamefont {Juhas}, \citenamefont {Kafri}, \citenamefont {Kechedzhi}, \citenamefont {Kelly}, \citenamefont {Khattar}, \citenamefont {Khezri}, \citenamefont {Kieferov{\'a}}, \citenamefont {Kim}, \citenamefont {Kitaev}, \citenamefont {Klimov}, \citenamefont {Klots},
  \citenamefont {Korotkov}, \citenamefont {Kostritsa}, \citenamefont {Kreikebaum}, \citenamefont {Landhuis}, \citenamefont {Laptev}, \citenamefont {Lau}, \citenamefont {Laws}, \citenamefont {Lee}, \citenamefont {Lee}, \citenamefont {Lester}, \citenamefont {Lill}, \citenamefont {Liu}, \citenamefont {Locharla}, \citenamefont {Lucero}, \citenamefont {Malone}, \citenamefont {Marshall}, \citenamefont {Martin}, \citenamefont {McClean}, \citenamefont {McCourt}, \citenamefont {McEwen}, \citenamefont {Megrant}, \citenamefont {Meurer~Costa}, \citenamefont {Mi}, \citenamefont {Miao}, \citenamefont {Mohseni}, \citenamefont {Montazeri}, \citenamefont {Morvan}, \citenamefont {Mount}, \citenamefont {Mruczkiewicz}, \citenamefont {Naaman}, \citenamefont {Neeley}, \citenamefont {Neill}, \citenamefont {Nersisyan}, \citenamefont {Neven}, \citenamefont {Newman}, \citenamefont {Ng}, \citenamefont {Nguyen}, \citenamefont {Nguyen}, \citenamefont {Niu}, \citenamefont {O'Brien}, \citenamefont {Opremcak}, \citenamefont {Platt},
  \citenamefont {Petukhov}, \citenamefont {Potter}, \citenamefont {Pryadko}, \citenamefont {Quintana}, \citenamefont {Roushan}, \citenamefont {Rubin}, \citenamefont {Saei}, \citenamefont {Sank}, \citenamefont {Sankaragomathi}, \citenamefont {Satzinger}, \citenamefont {Schurkus}, \citenamefont {Schuster}, \citenamefont {Shearn}, \citenamefont {Shorter}, \citenamefont {Shvarts}, \citenamefont {Skruzny}, \citenamefont {Smelyanskiy}, \citenamefont {Smith}, \citenamefont {Sterling}, \citenamefont {Strain}, \citenamefont {Szalay}, \citenamefont {Torres}, \citenamefont {Vidal}, \citenamefont {Villalonga}, \citenamefont {Vollgraff~Heidweiller}, \citenamefont {White}, \citenamefont {Xing}, \citenamefont {Yao}, \citenamefont {Yeh}, \citenamefont {Yoo}, \citenamefont {Young}, \citenamefont {Zalcman}, \citenamefont {Zhang}, \citenamefont {Zhu},\ and\ \citenamefont {{Google Quantum AI}}}]{acharyaSuppressingQuantumErrors2023}%
  \BibitemOpen
  \bibfield  {author} {\bibinfo {author} {\bibfnamefont {R.}~\bibnamefont {Acharya}}, \bibinfo {author} {\bibfnamefont {I.}~\bibnamefont {Aleiner}}, \bibinfo {author} {\bibfnamefont {R.}~\bibnamefont {Allen}}, \bibinfo {author} {\bibfnamefont {T.~I.}\ \bibnamefont {Andersen}}, \bibinfo {author} {\bibfnamefont {M.}~\bibnamefont {Ansmann}}, \bibinfo {author} {\bibfnamefont {F.}~\bibnamefont {Arute}}, \bibinfo {author} {\bibfnamefont {K.}~\bibnamefont {Arya}}, \bibinfo {author} {\bibfnamefont {A.}~\bibnamefont {Asfaw}}, \bibinfo {author} {\bibfnamefont {J.}~\bibnamefont {Atalaya}}, \bibinfo {author} {\bibfnamefont {R.}~\bibnamefont {Babbush}}, \bibinfo {author} {\bibfnamefont {D.}~\bibnamefont {Bacon}}, \bibinfo {author} {\bibfnamefont {J.~C.}\ \bibnamefont {Bardin}}, \bibinfo {author} {\bibfnamefont {J.}~\bibnamefont {Basso}}, \bibinfo {author} {\bibfnamefont {A.}~\bibnamefont {Bengtsson}}, \bibinfo {author} {\bibfnamefont {S.}~\bibnamefont {Boixo}}, \bibinfo {author} {\bibfnamefont {G.}~\bibnamefont
  {Bortoli}}, \bibinfo {author} {\bibfnamefont {A.}~\bibnamefont {Bourassa}}, \bibinfo {author} {\bibfnamefont {J.}~\bibnamefont {Bovaird}}, \bibinfo {author} {\bibfnamefont {L.}~\bibnamefont {Brill}}, \bibinfo {author} {\bibfnamefont {M.}~\bibnamefont {Broughton}}, \bibinfo {author} {\bibfnamefont {B.~B.}\ \bibnamefont {Buckley}}, \bibinfo {author} {\bibfnamefont {D.~A.}\ \bibnamefont {Buell}}, \bibinfo {author} {\bibfnamefont {T.}~\bibnamefont {Burger}}, \bibinfo {author} {\bibfnamefont {B.}~\bibnamefont {Burkett}}, \bibinfo {author} {\bibfnamefont {N.}~\bibnamefont {Bushnell}}, \bibinfo {author} {\bibfnamefont {Y.}~\bibnamefont {Chen}}, \bibinfo {author} {\bibfnamefont {Z.}~\bibnamefont {Chen}}, \bibinfo {author} {\bibfnamefont {B.}~\bibnamefont {Chiaro}}, \bibinfo {author} {\bibfnamefont {J.}~\bibnamefont {Cogan}}, \bibinfo {author} {\bibfnamefont {R.}~\bibnamefont {Collins}}, \bibinfo {author} {\bibfnamefont {P.}~\bibnamefont {Conner}}, \bibinfo {author} {\bibfnamefont {W.}~\bibnamefont {Courtney}},
  \bibinfo {author} {\bibfnamefont {A.~L.}\ \bibnamefont {Crook}}, \bibinfo {author} {\bibfnamefont {B.}~\bibnamefont {Curtin}}, \bibinfo {author} {\bibfnamefont {D.~M.}\ \bibnamefont {Debroy}}, \bibinfo {author} {\bibfnamefont {A.}~\bibnamefont {Del Toro~Barba}}, \bibinfo {author} {\bibfnamefont {S.}~\bibnamefont {Demura}}, \bibinfo {author} {\bibfnamefont {A.}~\bibnamefont {Dunsworth}}, \bibinfo {author} {\bibfnamefont {D.}~\bibnamefont {Eppens}}, \bibinfo {author} {\bibfnamefont {C.}~\bibnamefont {Erickson}}, \bibinfo {author} {\bibfnamefont {L.}~\bibnamefont {Faoro}}, \bibinfo {author} {\bibfnamefont {E.}~\bibnamefont {Farhi}}, \bibinfo {author} {\bibfnamefont {R.}~\bibnamefont {Fatemi}}, \bibinfo {author} {\bibfnamefont {L.}~\bibnamefont {Flores~Burgos}}, \bibinfo {author} {\bibfnamefont {E.}~\bibnamefont {Forati}}, \bibinfo {author} {\bibfnamefont {A.~G.}\ \bibnamefont {Fowler}}, \bibinfo {author} {\bibfnamefont {B.}~\bibnamefont {Foxen}}, \bibinfo {author} {\bibfnamefont {W.}~\bibnamefont {Giang}},
  \bibinfo {author} {\bibfnamefont {C.}~\bibnamefont {Gidney}}, \bibinfo {author} {\bibfnamefont {D.}~\bibnamefont {Gilboa}}, \bibinfo {author} {\bibfnamefont {M.}~\bibnamefont {Giustina}}, \bibinfo {author} {\bibfnamefont {A.}~\bibnamefont {Grajales~Dau}}, \bibinfo {author} {\bibfnamefont {J.~A.}\ \bibnamefont {Gross}}, \bibinfo {author} {\bibfnamefont {S.}~\bibnamefont {Habegger}}, \bibinfo {author} {\bibfnamefont {M.~C.}\ \bibnamefont {Hamilton}}, \bibinfo {author} {\bibfnamefont {M.~P.}\ \bibnamefont {Harrigan}}, \bibinfo {author} {\bibfnamefont {S.~D.}\ \bibnamefont {Harrington}}, \bibinfo {author} {\bibfnamefont {O.}~\bibnamefont {Higgott}}, \bibinfo {author} {\bibfnamefont {J.}~\bibnamefont {Hilton}}, \bibinfo {author} {\bibfnamefont {M.}~\bibnamefont {Hoffmann}}, \bibinfo {author} {\bibfnamefont {S.}~\bibnamefont {Hong}}, \bibinfo {author} {\bibfnamefont {T.}~\bibnamefont {Huang}}, \bibinfo {author} {\bibfnamefont {A.}~\bibnamefont {Huff}}, \bibinfo {author} {\bibfnamefont {W.~J.}\ \bibnamefont
  {Huggins}}, \bibinfo {author} {\bibfnamefont {L.~B.}\ \bibnamefont {Ioffe}}, \bibinfo {author} {\bibfnamefont {S.~V.}\ \bibnamefont {Isakov}}, \bibinfo {author} {\bibfnamefont {J.}~\bibnamefont {Iveland}}, \bibinfo {author} {\bibfnamefont {E.}~\bibnamefont {Jeffrey}}, \bibinfo {author} {\bibfnamefont {Z.}~\bibnamefont {Jiang}}, \bibinfo {author} {\bibfnamefont {C.}~\bibnamefont {Jones}}, \bibinfo {author} {\bibfnamefont {P.}~\bibnamefont {Juhas}}, \bibinfo {author} {\bibfnamefont {D.}~\bibnamefont {Kafri}}, \bibinfo {author} {\bibfnamefont {K.}~\bibnamefont {Kechedzhi}}, \bibinfo {author} {\bibfnamefont {J.}~\bibnamefont {Kelly}}, \bibinfo {author} {\bibfnamefont {T.}~\bibnamefont {Khattar}}, \bibinfo {author} {\bibfnamefont {M.}~\bibnamefont {Khezri}}, \bibinfo {author} {\bibfnamefont {M.}~\bibnamefont {Kieferov{\'a}}}, \bibinfo {author} {\bibfnamefont {S.}~\bibnamefont {Kim}}, \bibinfo {author} {\bibfnamefont {A.}~\bibnamefont {Kitaev}}, \bibinfo {author} {\bibfnamefont {P.~V.}\ \bibnamefont {Klimov}},
  \bibinfo {author} {\bibfnamefont {A.~R.}\ \bibnamefont {Klots}}, \bibinfo {author} {\bibfnamefont {A.~N.}\ \bibnamefont {Korotkov}}, \bibinfo {author} {\bibfnamefont {F.}~\bibnamefont {Kostritsa}}, \bibinfo {author} {\bibfnamefont {J.~M.}\ \bibnamefont {Kreikebaum}}, \bibinfo {author} {\bibfnamefont {D.}~\bibnamefont {Landhuis}}, \bibinfo {author} {\bibfnamefont {P.}~\bibnamefont {Laptev}}, \bibinfo {author} {\bibfnamefont {K.-M.}\ \bibnamefont {Lau}}, \bibinfo {author} {\bibfnamefont {L.}~\bibnamefont {Laws}}, \bibinfo {author} {\bibfnamefont {J.}~\bibnamefont {Lee}}, \bibinfo {author} {\bibfnamefont {K.}~\bibnamefont {Lee}}, \bibinfo {author} {\bibfnamefont {B.~J.}\ \bibnamefont {Lester}}, \bibinfo {author} {\bibfnamefont {A.}~\bibnamefont {Lill}}, \bibinfo {author} {\bibfnamefont {W.}~\bibnamefont {Liu}}, \bibinfo {author} {\bibfnamefont {A.}~\bibnamefont {Locharla}}, \bibinfo {author} {\bibfnamefont {E.}~\bibnamefont {Lucero}}, \bibinfo {author} {\bibfnamefont {F.~D.}\ \bibnamefont {Malone}}, \bibinfo
  {author} {\bibfnamefont {J.}~\bibnamefont {Marshall}}, \bibinfo {author} {\bibfnamefont {O.}~\bibnamefont {Martin}}, \bibinfo {author} {\bibfnamefont {J.~R.}\ \bibnamefont {McClean}}, \bibinfo {author} {\bibfnamefont {T.}~\bibnamefont {McCourt}}, \bibinfo {author} {\bibfnamefont {M.}~\bibnamefont {McEwen}}, \bibinfo {author} {\bibfnamefont {A.}~\bibnamefont {Megrant}}, \bibinfo {author} {\bibfnamefont {B.}~\bibnamefont {Meurer~Costa}}, \bibinfo {author} {\bibfnamefont {X.}~\bibnamefont {Mi}}, \bibinfo {author} {\bibfnamefont {K.~C.}\ \bibnamefont {Miao}}, \bibinfo {author} {\bibfnamefont {M.}~\bibnamefont {Mohseni}}, \bibinfo {author} {\bibfnamefont {S.}~\bibnamefont {Montazeri}}, \bibinfo {author} {\bibfnamefont {A.}~\bibnamefont {Morvan}}, \bibinfo {author} {\bibfnamefont {E.}~\bibnamefont {Mount}}, \bibinfo {author} {\bibfnamefont {W.}~\bibnamefont {Mruczkiewicz}}, \bibinfo {author} {\bibfnamefont {O.}~\bibnamefont {Naaman}}, \bibinfo {author} {\bibfnamefont {M.}~\bibnamefont {Neeley}}, \bibinfo {author}
  {\bibfnamefont {C.}~\bibnamefont {Neill}}, \bibinfo {author} {\bibfnamefont {A.}~\bibnamefont {Nersisyan}}, \bibinfo {author} {\bibfnamefont {H.}~\bibnamefont {Neven}}, \bibinfo {author} {\bibfnamefont {M.}~\bibnamefont {Newman}}, \bibinfo {author} {\bibfnamefont {J.~H.}\ \bibnamefont {Ng}}, \bibinfo {author} {\bibfnamefont {A.}~\bibnamefont {Nguyen}}, \bibinfo {author} {\bibfnamefont {M.}~\bibnamefont {Nguyen}}, \bibinfo {author} {\bibfnamefont {M.~Y.}\ \bibnamefont {Niu}}, \bibinfo {author} {\bibfnamefont {T.~E.}\ \bibnamefont {O'Brien}}, \bibinfo {author} {\bibfnamefont {A.}~\bibnamefont {Opremcak}}, \bibinfo {author} {\bibfnamefont {J.}~\bibnamefont {Platt}}, \bibinfo {author} {\bibfnamefont {A.}~\bibnamefont {Petukhov}}, \bibinfo {author} {\bibfnamefont {R.}~\bibnamefont {Potter}}, \bibinfo {author} {\bibfnamefont {L.~P.}\ \bibnamefont {Pryadko}}, \bibinfo {author} {\bibfnamefont {C.}~\bibnamefont {Quintana}}, \bibinfo {author} {\bibfnamefont {P.}~\bibnamefont {Roushan}}, \bibinfo {author}
  {\bibfnamefont {N.~C.}\ \bibnamefont {Rubin}}, \bibinfo {author} {\bibfnamefont {N.}~\bibnamefont {Saei}}, \bibinfo {author} {\bibfnamefont {D.}~\bibnamefont {Sank}}, \bibinfo {author} {\bibfnamefont {K.}~\bibnamefont {Sankaragomathi}}, \bibinfo {author} {\bibfnamefont {K.~J.}\ \bibnamefont {Satzinger}}, \bibinfo {author} {\bibfnamefont {H.~F.}\ \bibnamefont {Schurkus}}, \bibinfo {author} {\bibfnamefont {C.}~\bibnamefont {Schuster}}, \bibinfo {author} {\bibfnamefont {M.~J.}\ \bibnamefont {Shearn}}, \bibinfo {author} {\bibfnamefont {A.}~\bibnamefont {Shorter}}, \bibinfo {author} {\bibfnamefont {V.}~\bibnamefont {Shvarts}}, \bibinfo {author} {\bibfnamefont {J.}~\bibnamefont {Skruzny}}, \bibinfo {author} {\bibfnamefont {V.}~\bibnamefont {Smelyanskiy}}, \bibinfo {author} {\bibfnamefont {W.~C.}\ \bibnamefont {Smith}}, \bibinfo {author} {\bibfnamefont {G.}~\bibnamefont {Sterling}}, \bibinfo {author} {\bibfnamefont {D.}~\bibnamefont {Strain}}, \bibinfo {author} {\bibfnamefont {M.}~\bibnamefont {Szalay}}, \bibinfo
  {author} {\bibfnamefont {A.}~\bibnamefont {Torres}}, \bibinfo {author} {\bibfnamefont {G.}~\bibnamefont {Vidal}}, \bibinfo {author} {\bibfnamefont {B.}~\bibnamefont {Villalonga}}, \bibinfo {author} {\bibfnamefont {C.}~\bibnamefont {Vollgraff~Heidweiller}}, \bibinfo {author} {\bibfnamefont {T.}~\bibnamefont {White}}, \bibinfo {author} {\bibfnamefont {C.}~\bibnamefont {Xing}}, \bibinfo {author} {\bibfnamefont {Z.~J.}\ \bibnamefont {Yao}}, \bibinfo {author} {\bibfnamefont {P.}~\bibnamefont {Yeh}}, \bibinfo {author} {\bibfnamefont {J.}~\bibnamefont {Yoo}}, \bibinfo {author} {\bibfnamefont {G.}~\bibnamefont {Young}}, \bibinfo {author} {\bibfnamefont {A.}~\bibnamefont {Zalcman}}, \bibinfo {author} {\bibfnamefont {Y.}~\bibnamefont {Zhang}}, \bibinfo {author} {\bibfnamefont {N.}~\bibnamefont {Zhu}}, \ and\ \bibinfo {author} {\bibnamefont {{Google Quantum AI}}},\ }\href {\doibase 10.1038/s41586-022-05434-1} {\bibfield  {journal} {\bibinfo  {journal} {Nature}\ }\textbf {\bibinfo {volume} {614}},\ \bibinfo {pages}
  {676} (\bibinfo {year} {2023})}\BibitemShut {NoStop}%
\bibitem [{\citenamefont {Zhao}\ \emph {et~al.}(2022)\citenamefont {Zhao}, \citenamefont {Ye}, \citenamefont {Huang}, \citenamefont {Zhang}, \citenamefont {Wu}, \citenamefont {Guan}, \citenamefont {Zhu}, \citenamefont {Wei}, \citenamefont {He}, \citenamefont {Cao}, \citenamefont {Chen}, \citenamefont {Chung}, \citenamefont {Deng}, \citenamefont {Fan}, \citenamefont {Gong}, \citenamefont {Guo}, \citenamefont {Guo}, \citenamefont {Han}, \citenamefont {Li}, \citenamefont {Li}, \citenamefont {Li}, \citenamefont {Liang}, \citenamefont {Lin}, \citenamefont {Qian}, \citenamefont {Rong}, \citenamefont {Su}, \citenamefont {Sun}, \citenamefont {Wang}, \citenamefont {Wu}, \citenamefont {Xu}, \citenamefont {Ying}, \citenamefont {Yu}, \citenamefont {Zha}, \citenamefont {Zhang}, \citenamefont {Huo}, \citenamefont {Lu}, \citenamefont {Peng}, \citenamefont {Zhu},\ and\ \citenamefont {Pan}}]{zhaoRealizationErrorCorrectingSurface2022}%
  \BibitemOpen
  \bibfield  {author} {\bibinfo {author} {\bibfnamefont {Y.}~\bibnamefont {Zhao}}, \bibinfo {author} {\bibfnamefont {Y.}~\bibnamefont {Ye}}, \bibinfo {author} {\bibfnamefont {H.-L.}\ \bibnamefont {Huang}}, \bibinfo {author} {\bibfnamefont {Y.}~\bibnamefont {Zhang}}, \bibinfo {author} {\bibfnamefont {D.}~\bibnamefont {Wu}}, \bibinfo {author} {\bibfnamefont {H.}~\bibnamefont {Guan}}, \bibinfo {author} {\bibfnamefont {Q.}~\bibnamefont {Zhu}}, \bibinfo {author} {\bibfnamefont {Z.}~\bibnamefont {Wei}}, \bibinfo {author} {\bibfnamefont {T.}~\bibnamefont {He}}, \bibinfo {author} {\bibfnamefont {S.}~\bibnamefont {Cao}}, \bibinfo {author} {\bibfnamefont {F.}~\bibnamefont {Chen}}, \bibinfo {author} {\bibfnamefont {T.-H.}\ \bibnamefont {Chung}}, \bibinfo {author} {\bibfnamefont {H.}~\bibnamefont {Deng}}, \bibinfo {author} {\bibfnamefont {D.}~\bibnamefont {Fan}}, \bibinfo {author} {\bibfnamefont {M.}~\bibnamefont {Gong}}, \bibinfo {author} {\bibfnamefont {C.}~\bibnamefont {Guo}}, \bibinfo {author} {\bibfnamefont
  {S.}~\bibnamefont {Guo}}, \bibinfo {author} {\bibfnamefont {L.}~\bibnamefont {Han}}, \bibinfo {author} {\bibfnamefont {N.}~\bibnamefont {Li}}, \bibinfo {author} {\bibfnamefont {S.}~\bibnamefont {Li}}, \bibinfo {author} {\bibfnamefont {Y.}~\bibnamefont {Li}}, \bibinfo {author} {\bibfnamefont {F.}~\bibnamefont {Liang}}, \bibinfo {author} {\bibfnamefont {J.}~\bibnamefont {Lin}}, \bibinfo {author} {\bibfnamefont {H.}~\bibnamefont {Qian}}, \bibinfo {author} {\bibfnamefont {H.}~\bibnamefont {Rong}}, \bibinfo {author} {\bibfnamefont {H.}~\bibnamefont {Su}}, \bibinfo {author} {\bibfnamefont {L.}~\bibnamefont {Sun}}, \bibinfo {author} {\bibfnamefont {S.}~\bibnamefont {Wang}}, \bibinfo {author} {\bibfnamefont {Y.}~\bibnamefont {Wu}}, \bibinfo {author} {\bibfnamefont {Y.}~\bibnamefont {Xu}}, \bibinfo {author} {\bibfnamefont {C.}~\bibnamefont {Ying}}, \bibinfo {author} {\bibfnamefont {J.}~\bibnamefont {Yu}}, \bibinfo {author} {\bibfnamefont {C.}~\bibnamefont {Zha}}, \bibinfo {author} {\bibfnamefont {K.}~\bibnamefont
  {Zhang}}, \bibinfo {author} {\bibfnamefont {Y.-H.}\ \bibnamefont {Huo}}, \bibinfo {author} {\bibfnamefont {C.-Y.}\ \bibnamefont {Lu}}, \bibinfo {author} {\bibfnamefont {C.-Z.}\ \bibnamefont {Peng}}, \bibinfo {author} {\bibfnamefont {X.}~\bibnamefont {Zhu}}, \ and\ \bibinfo {author} {\bibfnamefont {J.-W.}\ \bibnamefont {Pan}},\ }\href {\doibase 10.1103/PhysRevLett.129.030501} {\bibfield  {journal} {\bibinfo  {journal} {Phys. Rev. Lett.}\ }\textbf {\bibinfo {volume} {129}},\ \bibinfo {pages} {030501} (\bibinfo {year} {2022})}\BibitemShut {NoStop}%
\bibitem [{\citenamefont {Krinner}\ \emph {et~al.}(2022)\citenamefont {Krinner}, \citenamefont {Lacroix}, \citenamefont {Remm}, \citenamefont {Di~Paolo}, \citenamefont {Genois}, \citenamefont {Leroux}, \citenamefont {Hellings}, \citenamefont {Lazar}, \citenamefont {Swiadek}, \citenamefont {Herrmann}, \citenamefont {Norris}, \citenamefont {Andersen}, \citenamefont {M{\"u}ller}, \citenamefont {Blais}, \citenamefont {Eichler},\ and\ \citenamefont {Wallraff}}]{krinnerRealizingRepeatedQuantum2022}%
  \BibitemOpen
  \bibfield  {author} {\bibinfo {author} {\bibfnamefont {S.}~\bibnamefont {Krinner}}, \bibinfo {author} {\bibfnamefont {N.}~\bibnamefont {Lacroix}}, \bibinfo {author} {\bibfnamefont {A.}~\bibnamefont {Remm}}, \bibinfo {author} {\bibfnamefont {A.}~\bibnamefont {Di~Paolo}}, \bibinfo {author} {\bibfnamefont {E.}~\bibnamefont {Genois}}, \bibinfo {author} {\bibfnamefont {C.}~\bibnamefont {Leroux}}, \bibinfo {author} {\bibfnamefont {C.}~\bibnamefont {Hellings}}, \bibinfo {author} {\bibfnamefont {S.}~\bibnamefont {Lazar}}, \bibinfo {author} {\bibfnamefont {F.}~\bibnamefont {Swiadek}}, \bibinfo {author} {\bibfnamefont {J.}~\bibnamefont {Herrmann}}, \bibinfo {author} {\bibfnamefont {G.~J.}\ \bibnamefont {Norris}}, \bibinfo {author} {\bibfnamefont {C.~K.}\ \bibnamefont {Andersen}}, \bibinfo {author} {\bibfnamefont {M.}~\bibnamefont {M{\"u}ller}}, \bibinfo {author} {\bibfnamefont {A.}~\bibnamefont {Blais}}, \bibinfo {author} {\bibfnamefont {C.}~\bibnamefont {Eichler}}, \ and\ \bibinfo {author} {\bibfnamefont
  {A.}~\bibnamefont {Wallraff}},\ }\href {\doibase 10.1038/s41586-022-04566-8} {\bibfield  {journal} {\bibinfo  {journal} {Nature}\ }\textbf {\bibinfo {volume} {605}},\ \bibinfo {pages} {669} (\bibinfo {year} {2022})}\BibitemShut {NoStop}%
\bibitem [{\citenamefont {Georgescu}\ \emph {et~al.}(2014)\citenamefont {Georgescu}, \citenamefont {Ashhab},\ and\ \citenamefont {Nori}}]{georgescu_quantum_2014}%
  \BibitemOpen
  \bibfield  {author} {\bibinfo {author} {\bibfnamefont {I.}~\bibnamefont {Georgescu}}, \bibinfo {author} {\bibfnamefont {S.}~\bibnamefont {Ashhab}}, \ and\ \bibinfo {author} {\bibfnamefont {F.}~\bibnamefont {Nori}},\ }\href {\doibase 10.1103/RevModPhys.86.153} {\bibfield  {journal} {\bibinfo  {journal} {Reviews of Modern Physics}\ }\textbf {\bibinfo {volume} {86}},\ \bibinfo {pages} {153} (\bibinfo {year} {2014})}\BibitemShut {NoStop}%
\bibitem [{\citenamefont {Zhu}\ \emph {et~al.}(2022)\citenamefont {Zhu}, \citenamefont {Sun}, \citenamefont {Gong}, \citenamefont {Chen}, \citenamefont {Zhang}, \citenamefont {Wu}, \citenamefont {Ye}, \citenamefont {Zha}, \citenamefont {Li}, \citenamefont {Guo}, \citenamefont {Qian}, \citenamefont {Huang}, \citenamefont {Yu}, \citenamefont {Deng}, \citenamefont {Rong}, \citenamefont {Lin}, \citenamefont {Xu}, \citenamefont {Sun}, \citenamefont {Guo}, \citenamefont {Li}, \citenamefont {Liang}, \citenamefont {Peng}, \citenamefont {Fan}, \citenamefont {Zhu},\ and\ \citenamefont {Pan}}]{zhu_observation_2022}%
  \BibitemOpen
  \bibfield  {author} {\bibinfo {author} {\bibfnamefont {Q.}~\bibnamefont {Zhu}}, \bibinfo {author} {\bibfnamefont {Z.-H.}\ \bibnamefont {Sun}}, \bibinfo {author} {\bibfnamefont {M.}~\bibnamefont {Gong}}, \bibinfo {author} {\bibfnamefont {F.}~\bibnamefont {Chen}}, \bibinfo {author} {\bibfnamefont {Y.-R.}\ \bibnamefont {Zhang}}, \bibinfo {author} {\bibfnamefont {Y.}~\bibnamefont {Wu}}, \bibinfo {author} {\bibfnamefont {Y.}~\bibnamefont {Ye}}, \bibinfo {author} {\bibfnamefont {C.}~\bibnamefont {Zha}}, \bibinfo {author} {\bibfnamefont {S.}~\bibnamefont {Li}}, \bibinfo {author} {\bibfnamefont {S.}~\bibnamefont {Guo}}, \bibinfo {author} {\bibfnamefont {H.}~\bibnamefont {Qian}}, \bibinfo {author} {\bibfnamefont {H.-L.}\ \bibnamefont {Huang}}, \bibinfo {author} {\bibfnamefont {J.}~\bibnamefont {Yu}}, \bibinfo {author} {\bibfnamefont {H.}~\bibnamefont {Deng}}, \bibinfo {author} {\bibfnamefont {H.}~\bibnamefont {Rong}}, \bibinfo {author} {\bibfnamefont {J.}~\bibnamefont {Lin}}, \bibinfo {author} {\bibfnamefont
  {Y.}~\bibnamefont {Xu}}, \bibinfo {author} {\bibfnamefont {L.}~\bibnamefont {Sun}}, \bibinfo {author} {\bibfnamefont {C.}~\bibnamefont {Guo}}, \bibinfo {author} {\bibfnamefont {N.}~\bibnamefont {Li}}, \bibinfo {author} {\bibfnamefont {F.}~\bibnamefont {Liang}}, \bibinfo {author} {\bibfnamefont {C.-Z.}\ \bibnamefont {Peng}}, \bibinfo {author} {\bibfnamefont {H.}~\bibnamefont {Fan}}, \bibinfo {author} {\bibfnamefont {X.}~\bibnamefont {Zhu}}, \ and\ \bibinfo {author} {\bibfnamefont {J.-W.}\ \bibnamefont {Pan}},\ }\href {\doibase 10.1103/PhysRevLett.128.160502} {\bibfield  {journal} {\bibinfo  {journal} {Phys. Rev. Lett.}\ }\textbf {\bibinfo {volume} {128}},\ \bibinfo {pages} {160502} (\bibinfo {year} {2022})}\BibitemShut {NoStop}%
\bibitem [{\citenamefont {{Google Quantum AI and Collaborators}}\ \emph {et~al.}(2023)\citenamefont {{Google Quantum AI and Collaborators}}, \citenamefont {Andersen}, \citenamefont {Lensky}, \citenamefont {Kechedzhi}, \citenamefont {Drozdov}, \citenamefont {Bengtsson}, \citenamefont {Hong}, \citenamefont {Morvan}, \citenamefont {Mi}, \citenamefont {Opremcak}, \citenamefont {Acharya}, \citenamefont {Allen}, \citenamefont {Ansmann}, \citenamefont {Arute}, \citenamefont {Arya}, \citenamefont {Asfaw}, \citenamefont {Atalaya}, \citenamefont {Babbush}, \citenamefont {Bacon}, \citenamefont {Bardin}, \citenamefont {Bortoli}, \citenamefont {Bourassa}, \citenamefont {Bovaird}, \citenamefont {Brill}, \citenamefont {Broughton}, \citenamefont {Buckley}, \citenamefont {Buell}, \citenamefont {Burger}, \citenamefont {Burkett}, \citenamefont {Bushnell}, \citenamefont {Chen}, \citenamefont {Chiaro}, \citenamefont {Chik}, \citenamefont {Chou}, \citenamefont {Cogan}, \citenamefont {Collins}, \citenamefont {Conner}, \citenamefont
  {Courtney}, \citenamefont {Crook}, \citenamefont {Curtin}, \citenamefont {Debroy}, \citenamefont {Del Toro~Barba}, \citenamefont {Demura}, \citenamefont {Dunsworth}, \citenamefont {Eppens}, \citenamefont {Erickson}, \citenamefont {Faoro}, \citenamefont {Farhi}, \citenamefont {Fatemi}, \citenamefont {Ferreira}, \citenamefont {Burgos}, \citenamefont {Forati}, \citenamefont {Fowler}, \citenamefont {Foxen}, \citenamefont {Giang}, \citenamefont {Gidney}, \citenamefont {Gilboa}, \citenamefont {Giustina}, \citenamefont {Gosula}, \citenamefont {Dau}, \citenamefont {Gross}, \citenamefont {Habegger}, \citenamefont {Hamilton}, \citenamefont {Hansen}, \citenamefont {Harrigan}, \citenamefont {Harrington}, \citenamefont {Heu}, \citenamefont {Hilton}, \citenamefont {Hoffmann}, \citenamefont {Huang}, \citenamefont {Huff}, \citenamefont {Huggins}, \citenamefont {Ioffe}, \citenamefont {Isakov}, \citenamefont {Iveland}, \citenamefont {Jeffrey}, \citenamefont {Jiang}, \citenamefont {Jones}, \citenamefont {Juhas}, \citenamefont
  {Kafri}, \citenamefont {Khattar}, \citenamefont {Khezri}, \citenamefont {Kieferov{\'a}}, \citenamefont {Kim}, \citenamefont {Kitaev}, \citenamefont {Klimov}, \citenamefont {Klots}, \citenamefont {Korotkov}, \citenamefont {Kostritsa}, \citenamefont {Kreikebaum}, \citenamefont {Landhuis}, \citenamefont {Laptev}, \citenamefont {Lau}, \citenamefont {Laws}, \citenamefont {Lee}, \citenamefont {Lee}, \citenamefont {Lester}, \citenamefont {Lill}, \citenamefont {Liu}, \citenamefont {Locharla}, \citenamefont {Lucero}, \citenamefont {Malone}, \citenamefont {Martin}, \citenamefont {McClean}, \citenamefont {McCourt}, \citenamefont {McEwen}, \citenamefont {Miao}, \citenamefont {Mieszala}, \citenamefont {Mohseni}, \citenamefont {Montazeri}, \citenamefont {Mount}, \citenamefont {Movassagh}, \citenamefont {Mruczkiewicz}, \citenamefont {Naaman}, \citenamefont {Neeley}, \citenamefont {Neill}, \citenamefont {Nersisyan}, \citenamefont {Newman}, \citenamefont {Ng}, \citenamefont {Nguyen}, \citenamefont {Nguyen}, \citenamefont
  {Niu}, \citenamefont {O'Brien}, \citenamefont {Omonije}, \citenamefont {Petukhov}, \citenamefont {Potter}, \citenamefont {Pryadko}, \citenamefont {Quintana}, \citenamefont {Rocque}, \citenamefont {Rubin}, \citenamefont {Saei}, \citenamefont {Sank}, \citenamefont {Sankaragomathi}, \citenamefont {Satzinger}, \citenamefont {Schurkus}, \citenamefont {Schuster}, \citenamefont {Shearn}, \citenamefont {Shorter}, \citenamefont {Shutty}, \citenamefont {Shvarts}, \citenamefont {Skruzny}, \citenamefont {Smith}, \citenamefont {Somma}, \citenamefont {Sterling}, \citenamefont {Strain}, \citenamefont {Szalay}, \citenamefont {Torres}, \citenamefont {Vidal}, \citenamefont {Villalonga}, \citenamefont {Heidweiller}, \citenamefont {White}, \citenamefont {Woo}, \citenamefont {Xing}, \citenamefont {Yao}, \citenamefont {Yeh}, \citenamefont {Yoo}, \citenamefont {Young}, \citenamefont {Zalcman}, \citenamefont {Zhang}, \citenamefont {Zhu}, \citenamefont {Zobrist}, \citenamefont {Neven}, \citenamefont {Boixo}, \citenamefont
  {Megrant}, \citenamefont {Kelly}, \citenamefont {Chen}, \citenamefont {Smelyanskiy}, \citenamefont {Kim}, \citenamefont {Aleiner},\ and\ \citenamefont {Roushan}}]{google_quantum_ai_and_collaborators_non-abelian_2023}%
  \BibitemOpen
  \bibfield  {author} {\bibinfo {author} {\bibnamefont {{Google Quantum AI and Collaborators}}}, \bibinfo {author} {\bibfnamefont {T.~I.}\ \bibnamefont {Andersen}}, \bibinfo {author} {\bibfnamefont {Y.~D.}\ \bibnamefont {Lensky}}, \bibinfo {author} {\bibfnamefont {K.}~\bibnamefont {Kechedzhi}}, \bibinfo {author} {\bibfnamefont {I.~K.}\ \bibnamefont {Drozdov}}, \bibinfo {author} {\bibfnamefont {A.}~\bibnamefont {Bengtsson}}, \bibinfo {author} {\bibfnamefont {S.}~\bibnamefont {Hong}}, \bibinfo {author} {\bibfnamefont {A.}~\bibnamefont {Morvan}}, \bibinfo {author} {\bibfnamefont {X.}~\bibnamefont {Mi}}, \bibinfo {author} {\bibfnamefont {A.}~\bibnamefont {Opremcak}}, \bibinfo {author} {\bibfnamefont {R.}~\bibnamefont {Acharya}}, \bibinfo {author} {\bibfnamefont {R.}~\bibnamefont {Allen}}, \bibinfo {author} {\bibfnamefont {M.}~\bibnamefont {Ansmann}}, \bibinfo {author} {\bibfnamefont {F.}~\bibnamefont {Arute}}, \bibinfo {author} {\bibfnamefont {K.}~\bibnamefont {Arya}}, \bibinfo {author} {\bibfnamefont
  {A.}~\bibnamefont {Asfaw}}, \bibinfo {author} {\bibfnamefont {J.}~\bibnamefont {Atalaya}}, \bibinfo {author} {\bibfnamefont {R.}~\bibnamefont {Babbush}}, \bibinfo {author} {\bibfnamefont {D.}~\bibnamefont {Bacon}}, \bibinfo {author} {\bibfnamefont {J.~C.}\ \bibnamefont {Bardin}}, \bibinfo {author} {\bibfnamefont {G.}~\bibnamefont {Bortoli}}, \bibinfo {author} {\bibfnamefont {A.}~\bibnamefont {Bourassa}}, \bibinfo {author} {\bibfnamefont {J.}~\bibnamefont {Bovaird}}, \bibinfo {author} {\bibfnamefont {L.}~\bibnamefont {Brill}}, \bibinfo {author} {\bibfnamefont {M.}~\bibnamefont {Broughton}}, \bibinfo {author} {\bibfnamefont {B.~B.}\ \bibnamefont {Buckley}}, \bibinfo {author} {\bibfnamefont {D.~A.}\ \bibnamefont {Buell}}, \bibinfo {author} {\bibfnamefont {T.}~\bibnamefont {Burger}}, \bibinfo {author} {\bibfnamefont {B.}~\bibnamefont {Burkett}}, \bibinfo {author} {\bibfnamefont {N.}~\bibnamefont {Bushnell}}, \bibinfo {author} {\bibfnamefont {Z.}~\bibnamefont {Chen}}, \bibinfo {author} {\bibfnamefont
  {B.}~\bibnamefont {Chiaro}}, \bibinfo {author} {\bibfnamefont {D.}~\bibnamefont {Chik}}, \bibinfo {author} {\bibfnamefont {C.}~\bibnamefont {Chou}}, \bibinfo {author} {\bibfnamefont {J.}~\bibnamefont {Cogan}}, \bibinfo {author} {\bibfnamefont {R.}~\bibnamefont {Collins}}, \bibinfo {author} {\bibfnamefont {P.}~\bibnamefont {Conner}}, \bibinfo {author} {\bibfnamefont {W.}~\bibnamefont {Courtney}}, \bibinfo {author} {\bibfnamefont {A.~L.}\ \bibnamefont {Crook}}, \bibinfo {author} {\bibfnamefont {B.}~\bibnamefont {Curtin}}, \bibinfo {author} {\bibfnamefont {D.~M.}\ \bibnamefont {Debroy}}, \bibinfo {author} {\bibfnamefont {A.}~\bibnamefont {Del Toro~Barba}}, \bibinfo {author} {\bibfnamefont {S.}~\bibnamefont {Demura}}, \bibinfo {author} {\bibfnamefont {A.}~\bibnamefont {Dunsworth}}, \bibinfo {author} {\bibfnamefont {D.}~\bibnamefont {Eppens}}, \bibinfo {author} {\bibfnamefont {C.}~\bibnamefont {Erickson}}, \bibinfo {author} {\bibfnamefont {L.}~\bibnamefont {Faoro}}, \bibinfo {author} {\bibfnamefont
  {E.}~\bibnamefont {Farhi}}, \bibinfo {author} {\bibfnamefont {R.}~\bibnamefont {Fatemi}}, \bibinfo {author} {\bibfnamefont {V.~S.}\ \bibnamefont {Ferreira}}, \bibinfo {author} {\bibfnamefont {L.~F.}\ \bibnamefont {Burgos}}, \bibinfo {author} {\bibfnamefont {E.}~\bibnamefont {Forati}}, \bibinfo {author} {\bibfnamefont {A.~G.}\ \bibnamefont {Fowler}}, \bibinfo {author} {\bibfnamefont {B.}~\bibnamefont {Foxen}}, \bibinfo {author} {\bibfnamefont {W.}~\bibnamefont {Giang}}, \bibinfo {author} {\bibfnamefont {C.}~\bibnamefont {Gidney}}, \bibinfo {author} {\bibfnamefont {D.}~\bibnamefont {Gilboa}}, \bibinfo {author} {\bibfnamefont {M.}~\bibnamefont {Giustina}}, \bibinfo {author} {\bibfnamefont {R.}~\bibnamefont {Gosula}}, \bibinfo {author} {\bibfnamefont {A.~G.}\ \bibnamefont {Dau}}, \bibinfo {author} {\bibfnamefont {J.~A.}\ \bibnamefont {Gross}}, \bibinfo {author} {\bibfnamefont {S.}~\bibnamefont {Habegger}}, \bibinfo {author} {\bibfnamefont {M.~C.}\ \bibnamefont {Hamilton}}, \bibinfo {author} {\bibfnamefont
  {M.}~\bibnamefont {Hansen}}, \bibinfo {author} {\bibfnamefont {M.~P.}\ \bibnamefont {Harrigan}}, \bibinfo {author} {\bibfnamefont {S.~D.}\ \bibnamefont {Harrington}}, \bibinfo {author} {\bibfnamefont {P.}~\bibnamefont {Heu}}, \bibinfo {author} {\bibfnamefont {J.}~\bibnamefont {Hilton}}, \bibinfo {author} {\bibfnamefont {M.~R.}\ \bibnamefont {Hoffmann}}, \bibinfo {author} {\bibfnamefont {T.}~\bibnamefont {Huang}}, \bibinfo {author} {\bibfnamefont {A.}~\bibnamefont {Huff}}, \bibinfo {author} {\bibfnamefont {W.~J.}\ \bibnamefont {Huggins}}, \bibinfo {author} {\bibfnamefont {L.~B.}\ \bibnamefont {Ioffe}}, \bibinfo {author} {\bibfnamefont {S.~V.}\ \bibnamefont {Isakov}}, \bibinfo {author} {\bibfnamefont {J.}~\bibnamefont {Iveland}}, \bibinfo {author} {\bibfnamefont {E.}~\bibnamefont {Jeffrey}}, \bibinfo {author} {\bibfnamefont {Z.}~\bibnamefont {Jiang}}, \bibinfo {author} {\bibfnamefont {C.}~\bibnamefont {Jones}}, \bibinfo {author} {\bibfnamefont {P.}~\bibnamefont {Juhas}}, \bibinfo {author} {\bibfnamefont
  {D.}~\bibnamefont {Kafri}}, \bibinfo {author} {\bibfnamefont {T.}~\bibnamefont {Khattar}}, \bibinfo {author} {\bibfnamefont {M.}~\bibnamefont {Khezri}}, \bibinfo {author} {\bibfnamefont {M.}~\bibnamefont {Kieferov{\'a}}}, \bibinfo {author} {\bibfnamefont {S.}~\bibnamefont {Kim}}, \bibinfo {author} {\bibfnamefont {A.}~\bibnamefont {Kitaev}}, \bibinfo {author} {\bibfnamefont {P.~V.}\ \bibnamefont {Klimov}}, \bibinfo {author} {\bibfnamefont {A.~R.}\ \bibnamefont {Klots}}, \bibinfo {author} {\bibfnamefont {A.~N.}\ \bibnamefont {Korotkov}}, \bibinfo {author} {\bibfnamefont {F.}~\bibnamefont {Kostritsa}}, \bibinfo {author} {\bibfnamefont {J.~M.}\ \bibnamefont {Kreikebaum}}, \bibinfo {author} {\bibfnamefont {D.}~\bibnamefont {Landhuis}}, \bibinfo {author} {\bibfnamefont {P.}~\bibnamefont {Laptev}}, \bibinfo {author} {\bibfnamefont {K.-M.}\ \bibnamefont {Lau}}, \bibinfo {author} {\bibfnamefont {L.}~\bibnamefont {Laws}}, \bibinfo {author} {\bibfnamefont {J.}~\bibnamefont {Lee}}, \bibinfo {author} {\bibfnamefont
  {K.~W.}\ \bibnamefont {Lee}}, \bibinfo {author} {\bibfnamefont {B.~J.}\ \bibnamefont {Lester}}, \bibinfo {author} {\bibfnamefont {A.~T.}\ \bibnamefont {Lill}}, \bibinfo {author} {\bibfnamefont {W.}~\bibnamefont {Liu}}, \bibinfo {author} {\bibfnamefont {A.}~\bibnamefont {Locharla}}, \bibinfo {author} {\bibfnamefont {E.}~\bibnamefont {Lucero}}, \bibinfo {author} {\bibfnamefont {F.~D.}\ \bibnamefont {Malone}}, \bibinfo {author} {\bibfnamefont {O.}~\bibnamefont {Martin}}, \bibinfo {author} {\bibfnamefont {J.~R.}\ \bibnamefont {McClean}}, \bibinfo {author} {\bibfnamefont {T.}~\bibnamefont {McCourt}}, \bibinfo {author} {\bibfnamefont {M.}~\bibnamefont {McEwen}}, \bibinfo {author} {\bibfnamefont {K.~C.}\ \bibnamefont {Miao}}, \bibinfo {author} {\bibfnamefont {A.}~\bibnamefont {Mieszala}}, \bibinfo {author} {\bibfnamefont {M.}~\bibnamefont {Mohseni}}, \bibinfo {author} {\bibfnamefont {S.}~\bibnamefont {Montazeri}}, \bibinfo {author} {\bibfnamefont {E.}~\bibnamefont {Mount}}, \bibinfo {author} {\bibfnamefont
  {R.}~\bibnamefont {Movassagh}}, \bibinfo {author} {\bibfnamefont {W.}~\bibnamefont {Mruczkiewicz}}, \bibinfo {author} {\bibfnamefont {O.}~\bibnamefont {Naaman}}, \bibinfo {author} {\bibfnamefont {M.}~\bibnamefont {Neeley}}, \bibinfo {author} {\bibfnamefont {C.}~\bibnamefont {Neill}}, \bibinfo {author} {\bibfnamefont {A.}~\bibnamefont {Nersisyan}}, \bibinfo {author} {\bibfnamefont {M.}~\bibnamefont {Newman}}, \bibinfo {author} {\bibfnamefont {J.~H.}\ \bibnamefont {Ng}}, \bibinfo {author} {\bibfnamefont {A.}~\bibnamefont {Nguyen}}, \bibinfo {author} {\bibfnamefont {M.}~\bibnamefont {Nguyen}}, \bibinfo {author} {\bibfnamefont {M.~Y.}\ \bibnamefont {Niu}}, \bibinfo {author} {\bibfnamefont {T.~E.}\ \bibnamefont {O'Brien}}, \bibinfo {author} {\bibfnamefont {S.}~\bibnamefont {Omonije}}, \bibinfo {author} {\bibfnamefont {A.}~\bibnamefont {Petukhov}}, \bibinfo {author} {\bibfnamefont {R.}~\bibnamefont {Potter}}, \bibinfo {author} {\bibfnamefont {L.~P.}\ \bibnamefont {Pryadko}}, \bibinfo {author} {\bibfnamefont
  {C.}~\bibnamefont {Quintana}}, \bibinfo {author} {\bibfnamefont {C.}~\bibnamefont {Rocque}}, \bibinfo {author} {\bibfnamefont {N.~C.}\ \bibnamefont {Rubin}}, \bibinfo {author} {\bibfnamefont {N.}~\bibnamefont {Saei}}, \bibinfo {author} {\bibfnamefont {D.}~\bibnamefont {Sank}}, \bibinfo {author} {\bibfnamefont {K.}~\bibnamefont {Sankaragomathi}}, \bibinfo {author} {\bibfnamefont {K.~J.}\ \bibnamefont {Satzinger}}, \bibinfo {author} {\bibfnamefont {H.~F.}\ \bibnamefont {Schurkus}}, \bibinfo {author} {\bibfnamefont {C.}~\bibnamefont {Schuster}}, \bibinfo {author} {\bibfnamefont {M.~J.}\ \bibnamefont {Shearn}}, \bibinfo {author} {\bibfnamefont {A.}~\bibnamefont {Shorter}}, \bibinfo {author} {\bibfnamefont {N.}~\bibnamefont {Shutty}}, \bibinfo {author} {\bibfnamefont {V.}~\bibnamefont {Shvarts}}, \bibinfo {author} {\bibfnamefont {J.}~\bibnamefont {Skruzny}}, \bibinfo {author} {\bibfnamefont {W.~C.}\ \bibnamefont {Smith}}, \bibinfo {author} {\bibfnamefont {R.}~\bibnamefont {Somma}}, \bibinfo {author}
  {\bibfnamefont {G.}~\bibnamefont {Sterling}}, \bibinfo {author} {\bibfnamefont {D.}~\bibnamefont {Strain}}, \bibinfo {author} {\bibfnamefont {M.}~\bibnamefont {Szalay}}, \bibinfo {author} {\bibfnamefont {A.}~\bibnamefont {Torres}}, \bibinfo {author} {\bibfnamefont {G.}~\bibnamefont {Vidal}}, \bibinfo {author} {\bibfnamefont {B.}~\bibnamefont {Villalonga}}, \bibinfo {author} {\bibfnamefont {C.~V.}\ \bibnamefont {Heidweiller}}, \bibinfo {author} {\bibfnamefont {T.}~\bibnamefont {White}}, \bibinfo {author} {\bibfnamefont {B.~W.~K.}\ \bibnamefont {Woo}}, \bibinfo {author} {\bibfnamefont {C.}~\bibnamefont {Xing}}, \bibinfo {author} {\bibfnamefont {Z.~J.}\ \bibnamefont {Yao}}, \bibinfo {author} {\bibfnamefont {P.}~\bibnamefont {Yeh}}, \bibinfo {author} {\bibfnamefont {J.}~\bibnamefont {Yoo}}, \bibinfo {author} {\bibfnamefont {G.}~\bibnamefont {Young}}, \bibinfo {author} {\bibfnamefont {A.}~\bibnamefont {Zalcman}}, \bibinfo {author} {\bibfnamefont {Y.}~\bibnamefont {Zhang}}, \bibinfo {author} {\bibfnamefont
  {N.}~\bibnamefont {Zhu}}, \bibinfo {author} {\bibfnamefont {N.}~\bibnamefont {Zobrist}}, \bibinfo {author} {\bibfnamefont {H.}~\bibnamefont {Neven}}, \bibinfo {author} {\bibfnamefont {S.}~\bibnamefont {Boixo}}, \bibinfo {author} {\bibfnamefont {A.}~\bibnamefont {Megrant}}, \bibinfo {author} {\bibfnamefont {J.}~\bibnamefont {Kelly}}, \bibinfo {author} {\bibfnamefont {Y.}~\bibnamefont {Chen}}, \bibinfo {author} {\bibfnamefont {V.}~\bibnamefont {Smelyanskiy}}, \bibinfo {author} {\bibfnamefont {E.-A.}\ \bibnamefont {Kim}}, \bibinfo {author} {\bibfnamefont {I.}~\bibnamefont {Aleiner}}, \ and\ \bibinfo {author} {\bibfnamefont {P.}~\bibnamefont {Roushan}},\ }\href {\doibase 10.1038/s41586-023-05954-4} {\bibfield  {journal} {\bibinfo  {journal} {Nature}\ }\textbf {\bibinfo {volume} {618}},\ \bibinfo {pages} {264} (\bibinfo {year} {2023})}\BibitemShut {NoStop}%
\bibitem [{\citenamefont {Shi}\ \emph {et~al.}(2023)\citenamefont {Shi}, \citenamefont {Yang}, \citenamefont {Xiang}, \citenamefont {Ge}, \citenamefont {Li}, \citenamefont {Wang}, \citenamefont {Huang}, \citenamefont {Tian}, \citenamefont {Song}, \citenamefont {Zheng}, \citenamefont {Xu}, \citenamefont {Cai},\ and\ \citenamefont {Fan}}]{shi_quantum_2023}%
  \BibitemOpen
  \bibfield  {author} {\bibinfo {author} {\bibfnamefont {Y.-H.}\ \bibnamefont {Shi}}, \bibinfo {author} {\bibfnamefont {R.-Q.}\ \bibnamefont {Yang}}, \bibinfo {author} {\bibfnamefont {Z.}~\bibnamefont {Xiang}}, \bibinfo {author} {\bibfnamefont {Z.-Y.}\ \bibnamefont {Ge}}, \bibinfo {author} {\bibfnamefont {H.}~\bibnamefont {Li}}, \bibinfo {author} {\bibfnamefont {Y.-Y.}\ \bibnamefont {Wang}}, \bibinfo {author} {\bibfnamefont {K.}~\bibnamefont {Huang}}, \bibinfo {author} {\bibfnamefont {Y.}~\bibnamefont {Tian}}, \bibinfo {author} {\bibfnamefont {X.}~\bibnamefont {Song}}, \bibinfo {author} {\bibfnamefont {D.}~\bibnamefont {Zheng}}, \bibinfo {author} {\bibfnamefont {K.}~\bibnamefont {Xu}}, \bibinfo {author} {\bibfnamefont {R.-G.}\ \bibnamefont {Cai}}, \ and\ \bibinfo {author} {\bibfnamefont {H.}~\bibnamefont {Fan}},\ }\href {\doibase 10.1038/s41467-023-39064-6} {\bibfield  {journal} {\bibinfo  {journal} {Nature Communications}\ }\textbf {\bibinfo {volume} {14}},\ \bibinfo {pages} {3263} (\bibinfo {year}
  {2023})}\BibitemShut {NoStop}%
\bibitem [{\citenamefont {Xiang}\ \emph {et~al.}(2023)\citenamefont {Xiang}, \citenamefont {Huang}, \citenamefont {Zhang}, \citenamefont {Liu}, \citenamefont {Shi}, \citenamefont {Deng}, \citenamefont {Liu}, \citenamefont {Li}, \citenamefont {Liang}, \citenamefont {Mei}, \citenamefont {Yu}, \citenamefont {Xue}, \citenamefont {Tian}, \citenamefont {Song}, \citenamefont {Liu}, \citenamefont {Xu}, \citenamefont {Zheng}, \citenamefont {Nori},\ and\ \citenamefont {Fan}}]{xiang_simulating_2023}%
  \BibitemOpen
  \bibfield  {author} {\bibinfo {author} {\bibfnamefont {Z.-C.}\ \bibnamefont {Xiang}}, \bibinfo {author} {\bibfnamefont {K.}~\bibnamefont {Huang}}, \bibinfo {author} {\bibfnamefont {Y.-R.}\ \bibnamefont {Zhang}}, \bibinfo {author} {\bibfnamefont {T.}~\bibnamefont {Liu}}, \bibinfo {author} {\bibfnamefont {Y.-H.}\ \bibnamefont {Shi}}, \bibinfo {author} {\bibfnamefont {C.-L.}\ \bibnamefont {Deng}}, \bibinfo {author} {\bibfnamefont {T.}~\bibnamefont {Liu}}, \bibinfo {author} {\bibfnamefont {H.}~\bibnamefont {Li}}, \bibinfo {author} {\bibfnamefont {G.-H.}\ \bibnamefont {Liang}}, \bibinfo {author} {\bibfnamefont {Z.-Y.}\ \bibnamefont {Mei}}, \bibinfo {author} {\bibfnamefont {H.}~\bibnamefont {Yu}}, \bibinfo {author} {\bibfnamefont {G.}~\bibnamefont {Xue}}, \bibinfo {author} {\bibfnamefont {Y.}~\bibnamefont {Tian}}, \bibinfo {author} {\bibfnamefont {X.}~\bibnamefont {Song}}, \bibinfo {author} {\bibfnamefont {Z.-B.}\ \bibnamefont {Liu}}, \bibinfo {author} {\bibfnamefont {K.}~\bibnamefont {Xu}}, \bibinfo {author}
  {\bibfnamefont {D.}~\bibnamefont {Zheng}}, \bibinfo {author} {\bibfnamefont {F.}~\bibnamefont {Nori}}, \ and\ \bibinfo {author} {\bibfnamefont {H.}~\bibnamefont {Fan}},\ }\href {\doibase 10.1038/s41467-023-41230-9} {\bibfield  {journal} {\bibinfo  {journal} {Nature Communications}\ }\textbf {\bibinfo {volume} {14}},\ \bibinfo {pages} {5433} (\bibinfo {year} {2023})}\BibitemShut {NoStop}%
\bibitem [{\citenamefont {Xu}\ \emph {et~al.}(2023)\citenamefont {Xu}, \citenamefont {Sun}, \citenamefont {Wang}, \citenamefont {Xiang}, \citenamefont {Bao}, \citenamefont {Zhu}, \citenamefont {Shen}, \citenamefont {Song}, \citenamefont {Zhang}, \citenamefont {Ren}, \citenamefont {Zhang}, \citenamefont {Dong}, \citenamefont {Deng}, \citenamefont {Chen}, \citenamefont {Wu}, \citenamefont {Tan}, \citenamefont {Gao}, \citenamefont {Jin}, \citenamefont {Zhu}, \citenamefont {Zhang}, \citenamefont {Wang}, \citenamefont {Zou}, \citenamefont {Zhong}, \citenamefont {Zhang}, \citenamefont {Li}, \citenamefont {Jiang}, \citenamefont {Yu}, \citenamefont {Yao}, \citenamefont {Wang}, \citenamefont {Li}, \citenamefont {Guo}, \citenamefont {Song}, \citenamefont {Wang},\ and\ \citenamefont {Deng}}]{xu_digital_2023}%
  \BibitemOpen
  \bibfield  {author} {\bibinfo {author} {\bibfnamefont {S.}~\bibnamefont {Xu}}, \bibinfo {author} {\bibfnamefont {Z.-Z.}\ \bibnamefont {Sun}}, \bibinfo {author} {\bibfnamefont {K.}~\bibnamefont {Wang}}, \bibinfo {author} {\bibfnamefont {L.}~\bibnamefont {Xiang}}, \bibinfo {author} {\bibfnamefont {Z.}~\bibnamefont {Bao}}, \bibinfo {author} {\bibfnamefont {Z.}~\bibnamefont {Zhu}}, \bibinfo {author} {\bibfnamefont {F.}~\bibnamefont {Shen}}, \bibinfo {author} {\bibfnamefont {Z.}~\bibnamefont {Song}}, \bibinfo {author} {\bibfnamefont {P.}~\bibnamefont {Zhang}}, \bibinfo {author} {\bibfnamefont {W.}~\bibnamefont {Ren}}, \bibinfo {author} {\bibfnamefont {X.}~\bibnamefont {Zhang}}, \bibinfo {author} {\bibfnamefont {H.}~\bibnamefont {Dong}}, \bibinfo {author} {\bibfnamefont {J.}~\bibnamefont {Deng}}, \bibinfo {author} {\bibfnamefont {J.}~\bibnamefont {Chen}}, \bibinfo {author} {\bibfnamefont {Y.}~\bibnamefont {Wu}}, \bibinfo {author} {\bibfnamefont {Z.}~\bibnamefont {Tan}}, \bibinfo {author} {\bibfnamefont
  {Y.}~\bibnamefont {Gao}}, \bibinfo {author} {\bibfnamefont {F.}~\bibnamefont {Jin}}, \bibinfo {author} {\bibfnamefont {X.}~\bibnamefont {Zhu}}, \bibinfo {author} {\bibfnamefont {C.}~\bibnamefont {Zhang}}, \bibinfo {author} {\bibfnamefont {N.}~\bibnamefont {Wang}}, \bibinfo {author} {\bibfnamefont {Y.}~\bibnamefont {Zou}}, \bibinfo {author} {\bibfnamefont {J.}~\bibnamefont {Zhong}}, \bibinfo {author} {\bibfnamefont {A.}~\bibnamefont {Zhang}}, \bibinfo {author} {\bibfnamefont {W.}~\bibnamefont {Li}}, \bibinfo {author} {\bibfnamefont {W.}~\bibnamefont {Jiang}}, \bibinfo {author} {\bibfnamefont {L.-W.}\ \bibnamefont {Yu}}, \bibinfo {author} {\bibfnamefont {Y.}~\bibnamefont {Yao}}, \bibinfo {author} {\bibfnamefont {Z.}~\bibnamefont {Wang}}, \bibinfo {author} {\bibfnamefont {H.}~\bibnamefont {Li}}, \bibinfo {author} {\bibfnamefont {Q.}~\bibnamefont {Guo}}, \bibinfo {author} {\bibfnamefont {C.}~\bibnamefont {Song}}, \bibinfo {author} {\bibfnamefont {H.}~\bibnamefont {Wang}}, \ and\ \bibinfo {author} {\bibfnamefont
  {D.-L.}\ \bibnamefont {Deng}},\ }\href {\doibase 10.1088/0256-307X/40/6/060301} {\bibfield  {journal} {\bibinfo  {journal} {Chinese Physics Letters}\ }\textbf {\bibinfo {volume} {40}},\ \bibinfo {pages} {060301} (\bibinfo {year} {2023})}\BibitemShut {NoStop}%
\bibitem [{\citenamefont {Rist{\`e}}\ \emph {et~al.}(2017)\citenamefont {Rist{\`e}}, \citenamefont {Da~Silva}, \citenamefont {Ryan}, \citenamefont {Cross}, \citenamefont {C{\'o}rcoles}, \citenamefont {Smolin}, \citenamefont {Gambetta}, \citenamefont {Chow},\ and\ \citenamefont {Johnson}}]{riste_demonstration_2017}%
  \BibitemOpen
  \bibfield  {author} {\bibinfo {author} {\bibfnamefont {D.}~\bibnamefont {Rist{\`e}}}, \bibinfo {author} {\bibfnamefont {M.~P.}\ \bibnamefont {Da~Silva}}, \bibinfo {author} {\bibfnamefont {C.~A.}\ \bibnamefont {Ryan}}, \bibinfo {author} {\bibfnamefont {A.~W.}\ \bibnamefont {Cross}}, \bibinfo {author} {\bibfnamefont {A.~D.}\ \bibnamefont {C{\'o}rcoles}}, \bibinfo {author} {\bibfnamefont {J.~A.}\ \bibnamefont {Smolin}}, \bibinfo {author} {\bibfnamefont {J.~M.}\ \bibnamefont {Gambetta}}, \bibinfo {author} {\bibfnamefont {J.~M.}\ \bibnamefont {Chow}}, \ and\ \bibinfo {author} {\bibfnamefont {B.~R.}\ \bibnamefont {Johnson}},\ }\href {\doibase 10.1038/s41534-017-0017-3} {\bibfield  {journal} {\bibinfo  {journal} {npj Quantum Inf.}\ }\textbf {\bibinfo {volume} {3}},\ \bibinfo {pages} {16} (\bibinfo {year} {2017})}\BibitemShut {NoStop}%
\bibitem [{\citenamefont {Gong}\ \emph {et~al.}(2021)\citenamefont {Gong}, \citenamefont {Wang}, \citenamefont {Zha}, \citenamefont {Chen}, \citenamefont {Huang}, \citenamefont {Wu}, \citenamefont {Zhu}, \citenamefont {Zhao}, \citenamefont {Li}, \citenamefont {Guo}, \citenamefont {Qian}, \citenamefont {Ye}, \citenamefont {Chen}, \citenamefont {Ying}, \citenamefont {Yu}, \citenamefont {Fan}, \citenamefont {Wu}, \citenamefont {Su}, \citenamefont {Deng}, \citenamefont {Rong}, \citenamefont {Zhang}, \citenamefont {Cao}, \citenamefont {Lin}, \citenamefont {Xu}, \citenamefont {Sun}, \citenamefont {Guo}, \citenamefont {Li}, \citenamefont {Liang}, \citenamefont {Bastidas}, \citenamefont {Nemoto}, \citenamefont {Munro}, \citenamefont {Huo}, \citenamefont {Lu}, \citenamefont {Peng}, \citenamefont {Zhu},\ and\ \citenamefont {Pan}}]{gong_quantum_2021}%
  \BibitemOpen
  \bibfield  {author} {\bibinfo {author} {\bibfnamefont {M.}~\bibnamefont {Gong}}, \bibinfo {author} {\bibfnamefont {S.}~\bibnamefont {Wang}}, \bibinfo {author} {\bibfnamefont {C.}~\bibnamefont {Zha}}, \bibinfo {author} {\bibfnamefont {M.-C.}\ \bibnamefont {Chen}}, \bibinfo {author} {\bibfnamefont {H.-L.}\ \bibnamefont {Huang}}, \bibinfo {author} {\bibfnamefont {Y.}~\bibnamefont {Wu}}, \bibinfo {author} {\bibfnamefont {Q.}~\bibnamefont {Zhu}}, \bibinfo {author} {\bibfnamefont {Y.}~\bibnamefont {Zhao}}, \bibinfo {author} {\bibfnamefont {S.}~\bibnamefont {Li}}, \bibinfo {author} {\bibfnamefont {S.}~\bibnamefont {Guo}}, \bibinfo {author} {\bibfnamefont {H.}~\bibnamefont {Qian}}, \bibinfo {author} {\bibfnamefont {Y.}~\bibnamefont {Ye}}, \bibinfo {author} {\bibfnamefont {F.}~\bibnamefont {Chen}}, \bibinfo {author} {\bibfnamefont {C.}~\bibnamefont {Ying}}, \bibinfo {author} {\bibfnamefont {J.}~\bibnamefont {Yu}}, \bibinfo {author} {\bibfnamefont {D.}~\bibnamefont {Fan}}, \bibinfo {author} {\bibfnamefont
  {D.}~\bibnamefont {Wu}}, \bibinfo {author} {\bibfnamefont {H.}~\bibnamefont {Su}}, \bibinfo {author} {\bibfnamefont {H.}~\bibnamefont {Deng}}, \bibinfo {author} {\bibfnamefont {H.}~\bibnamefont {Rong}}, \bibinfo {author} {\bibfnamefont {K.}~\bibnamefont {Zhang}}, \bibinfo {author} {\bibfnamefont {S.}~\bibnamefont {Cao}}, \bibinfo {author} {\bibfnamefont {J.}~\bibnamefont {Lin}}, \bibinfo {author} {\bibfnamefont {Y.}~\bibnamefont {Xu}}, \bibinfo {author} {\bibfnamefont {L.}~\bibnamefont {Sun}}, \bibinfo {author} {\bibfnamefont {C.}~\bibnamefont {Guo}}, \bibinfo {author} {\bibfnamefont {N.}~\bibnamefont {Li}}, \bibinfo {author} {\bibfnamefont {F.}~\bibnamefont {Liang}}, \bibinfo {author} {\bibfnamefont {V.~M.}\ \bibnamefont {Bastidas}}, \bibinfo {author} {\bibfnamefont {K.}~\bibnamefont {Nemoto}}, \bibinfo {author} {\bibfnamefont {W.~J.}\ \bibnamefont {Munro}}, \bibinfo {author} {\bibfnamefont {Y.-H.}\ \bibnamefont {Huo}}, \bibinfo {author} {\bibfnamefont {C.-Y.}\ \bibnamefont {Lu}}, \bibinfo {author}
  {\bibfnamefont {C.-Z.}\ \bibnamefont {Peng}}, \bibinfo {author} {\bibfnamefont {X.}~\bibnamefont {Zhu}}, \ and\ \bibinfo {author} {\bibfnamefont {J.-W.}\ \bibnamefont {Pan}},\ }\href {\doibase 10.1126/science.abg7812} {\bibfield  {journal} {\bibinfo  {journal} {Science}\ }\textbf {\bibinfo {volume} {372}},\ \bibinfo {pages} {948} (\bibinfo {year} {2021})}\BibitemShut {NoStop}%
\bibitem [{\citenamefont {Ni}\ \emph {et~al.}(2024)\citenamefont {Ni}, \citenamefont {Wang}, \citenamefont {Chao},\ and\ \citenamefont {Chen}}]{ni_superconducting_2023}%
  \BibitemOpen
  \bibfield  {author} {\bibinfo {author} {\bibfnamefont {X.}~\bibnamefont {Ni}}, \bibinfo {author} {\bibfnamefont {Z.}~\bibnamefont {Wang}}, \bibinfo {author} {\bibfnamefont {R.}~\bibnamefont {Chao}}, \ and\ \bibinfo {author} {\bibfnamefont {J.}~\bibnamefont {Chen}},\ }\href@noop {} {\  (\bibinfo {year} {2024})},\ \Eprint {http://arxiv.org/abs/2312.04186} {arXiv:2312.04186 [quant-ph]} \BibitemShut {NoStop}%
\bibitem [{\citenamefont {contributors}(2024)}]{supergrad2024github}%
  \BibitemOpen
  \bibfield  {author} {\bibinfo {author} {\bibfnamefont {S.}~\bibnamefont {contributors}},\ }\href {https://github.com/iqubit-org/supergrad} {\enquote {\bibinfo {title} {{SuperGrad}: Differentiable simulator for superconducting quantum processors},}\ } (\bibinfo {year} {2024})\BibitemShut {NoStop}%
\bibitem [{\citenamefont {Ni}\ \emph {et~al.}(2022)\citenamefont {Ni}, \citenamefont {Zhao}, \citenamefont {Wang}, \citenamefont {Wu},\ and\ \citenamefont {Chen}}]{ni_integrating_2022}%
  \BibitemOpen
  \bibfield  {author} {\bibinfo {author} {\bibfnamefont {X.}~\bibnamefont {Ni}}, \bibinfo {author} {\bibfnamefont {H.-H.}\ \bibnamefont {Zhao}}, \bibinfo {author} {\bibfnamefont {L.}~\bibnamefont {Wang}}, \bibinfo {author} {\bibfnamefont {F.}~\bibnamefont {Wu}}, \ and\ \bibinfo {author} {\bibfnamefont {J.}~\bibnamefont {Chen}},\ }\href {\doibase 10.1038/s41534-022-00614-3} {\bibfield  {journal} {\bibinfo  {journal} {npj Quantum Inf.}\ }\textbf {\bibinfo {volume} {8}},\ \bibinfo {pages} {106} (\bibinfo {year} {2022})}\BibitemShut {NoStop}%
\bibitem [{\citenamefont {Groszkowski}\ and\ \citenamefont {Koch}(2021)}]{groszkowski_scqubits_2021}%
  \BibitemOpen
  \bibfield  {author} {\bibinfo {author} {\bibfnamefont {P.}~\bibnamefont {Groszkowski}}\ and\ \bibinfo {author} {\bibfnamefont {J.}~\bibnamefont {Koch}},\ }\href {\doibase 10.22331/q-2021-11-17-583} {\bibfield  {journal} {\bibinfo  {journal} {Quantum}\ }\textbf {\bibinfo {volume} {5}},\ \bibinfo {pages} {583} (\bibinfo {year} {2021})}\BibitemShut {NoStop}%
\bibitem [{\citenamefont {Puzzuoli}\ \emph {et~al.}(2023{\natexlab{a}})\citenamefont {Puzzuoli}, \citenamefont {Wood}, \citenamefont {Egger}, \citenamefont {Rosand},\ and\ \citenamefont {Ueda}}]{puzzuoli_qiskit_2023}%
  \BibitemOpen
  \bibfield  {author} {\bibinfo {author} {\bibfnamefont {D.}~\bibnamefont {Puzzuoli}}, \bibinfo {author} {\bibfnamefont {C.~J.}\ \bibnamefont {Wood}}, \bibinfo {author} {\bibfnamefont {D.~J.}\ \bibnamefont {Egger}}, \bibinfo {author} {\bibfnamefont {B.}~\bibnamefont {Rosand}}, \ and\ \bibinfo {author} {\bibfnamefont {K.}~\bibnamefont {Ueda}},\ }\href {\doibase 10.21105/joss.05853} {\bibfield  {journal} {\bibinfo  {journal} {Journal of Open Source Software}\ }\textbf {\bibinfo {volume} {8}},\ \bibinfo {pages} {5853} (\bibinfo {year} {2023}{\natexlab{a}})}\BibitemShut {NoStop}%
\bibitem [{\citenamefont {Régent}\ \emph {et~al.}(2023)\citenamefont {Régent}, \citenamefont {Berdou}, \citenamefont {Leghtas}, \citenamefont {Guillaud},\ and\ \citenamefont {Mirrahimi}}]{regentHighperformanceRepetitionCat2023}%
  \BibitemOpen
  \bibfield  {author} {\bibinfo {author} {\bibfnamefont {F.-M.~L.}\ \bibnamefont {Régent}}, \bibinfo {author} {\bibfnamefont {C.}~\bibnamefont {Berdou}}, \bibinfo {author} {\bibfnamefont {Z.}~\bibnamefont {Leghtas}}, \bibinfo {author} {\bibfnamefont {J.}~\bibnamefont {Guillaud}}, \ and\ \bibinfo {author} {\bibfnamefont {M.}~\bibnamefont {Mirrahimi}},\ }\href {\doibase 10.22331/q-2023-12-06-1198} {\bibfield  {journal} {\bibinfo  {journal} {Quantum}\ }\textbf {\bibinfo {volume} {7}},\ \bibinfo {pages} {1198} (\bibinfo {year} {2023})}\BibitemShut {NoStop}%
\bibitem [{\citenamefont {Chamberland}\ \emph {et~al.}(2022)\citenamefont {Chamberland}, \citenamefont {Noh}, \citenamefont {Arrangoiz-Arriola}, \citenamefont {Campbell}, \citenamefont {Hann}, \citenamefont {Iverson}, \citenamefont {Putterman}, \citenamefont {Bohdanowicz}, \citenamefont {Flammia}, \citenamefont {Keller}, \citenamefont {Refael}, \citenamefont {Preskill}, \citenamefont {Jiang}, \citenamefont {Safavi-Naeini}, \citenamefont {Painter},\ and\ \citenamefont {Brandão}}]{chamberlandBuildingFaultTolerantQuantum2022}%
  \BibitemOpen
  \bibfield  {author} {\bibinfo {author} {\bibfnamefont {C.}~\bibnamefont {Chamberland}}, \bibinfo {author} {\bibfnamefont {K.}~\bibnamefont {Noh}}, \bibinfo {author} {\bibfnamefont {P.}~\bibnamefont {Arrangoiz-Arriola}}, \bibinfo {author} {\bibfnamefont {E.~T.}\ \bibnamefont {Campbell}}, \bibinfo {author} {\bibfnamefont {C.~T.}\ \bibnamefont {Hann}}, \bibinfo {author} {\bibfnamefont {J.}~\bibnamefont {Iverson}}, \bibinfo {author} {\bibfnamefont {H.}~\bibnamefont {Putterman}}, \bibinfo {author} {\bibfnamefont {T.~C.}\ \bibnamefont {Bohdanowicz}}, \bibinfo {author} {\bibfnamefont {S.~T.}\ \bibnamefont {Flammia}}, \bibinfo {author} {\bibfnamefont {A.}~\bibnamefont {Keller}}, \bibinfo {author} {\bibfnamefont {G.}~\bibnamefont {Refael}}, \bibinfo {author} {\bibfnamefont {J.}~\bibnamefont {Preskill}}, \bibinfo {author} {\bibfnamefont {L.}~\bibnamefont {Jiang}}, \bibinfo {author} {\bibfnamefont {A.~H.}\ \bibnamefont {Safavi-Naeini}}, \bibinfo {author} {\bibfnamefont {O.}~\bibnamefont {Painter}}, \ and\ \bibinfo
  {author} {\bibfnamefont {F.~G.}\ \bibnamefont {Brandão}},\ }\href {\doibase 10.1103/PRXQuantum.3.010329} {\bibfield  {journal} {\bibinfo  {journal} {PRX Quantum}\ }\textbf {\bibinfo {volume} {3}},\ \bibinfo {pages} {010329} (\bibinfo {year} {2022})}\BibitemShut {NoStop}%
\bibitem [{\citenamefont {Zhao}\ \emph {et~al.}(2024)\citenamefont {Zhao}, \citenamefont {Chen}, \citenamefont {Lyu},\ and\ \citenamefont {Koch}}]{qfit}%
  \BibitemOpen
  \bibfield  {author} {\bibinfo {author} {\bibfnamefont {T.}~\bibnamefont {Zhao}}, \bibinfo {author} {\bibfnamefont {D.}~\bibnamefont {Chen}}, \bibinfo {author} {\bibfnamefont {T.}~\bibnamefont {Lyu}}, \ and\ \bibinfo {author} {\bibfnamefont {J.}~\bibnamefont {Koch}},\ }\href {https://github.com/scqubits/qfit} {\enquote {\bibinfo {title} {Qfit: Interactive parameter fitting for superconducting circuits},}\ } (\bibinfo {year} {2024})\BibitemShut {NoStop}%
\bibitem [{\citenamefont {Krastanov}\ \emph {et~al.}(2019)\citenamefont {Krastanov}, \citenamefont {Zhou}, \citenamefont {Flammia},\ and\ \citenamefont {Jiang}}]{krastanov_stochastic_2019}%
  \BibitemOpen
  \bibfield  {author} {\bibinfo {author} {\bibfnamefont {S.}~\bibnamefont {Krastanov}}, \bibinfo {author} {\bibfnamefont {S.}~\bibnamefont {Zhou}}, \bibinfo {author} {\bibfnamefont {S.~T.}\ \bibnamefont {Flammia}}, \ and\ \bibinfo {author} {\bibfnamefont {L.}~\bibnamefont {Jiang}},\ }\href {\doibase 10.1088/2058-9565/ab18d5} {\bibfield  {journal} {\bibinfo  {journal} {Quantum Science and Technology}\ }\textbf {\bibinfo {volume} {4}},\ \bibinfo {pages} {035003} (\bibinfo {year} {2019})}\BibitemShut {NoStop}%
\bibitem [{\citenamefont {Bradbury}\ \emph {et~al.}(2018)\citenamefont {Bradbury}, \citenamefont {Frostig}, \citenamefont {Hawkins}, \citenamefont {Johnson}, \citenamefont {Leary}, \citenamefont {Maclaurin}, \citenamefont {Necula}, \citenamefont {Paszke}, \citenamefont {Vander{P}las}, \citenamefont {Wanderman-{M}ilne},\ and\ \citenamefont {Zhang}}]{jax2018github}%
  \BibitemOpen
  \bibfield  {author} {\bibinfo {author} {\bibfnamefont {J.}~\bibnamefont {Bradbury}}, \bibinfo {author} {\bibfnamefont {R.}~\bibnamefont {Frostig}}, \bibinfo {author} {\bibfnamefont {P.}~\bibnamefont {Hawkins}}, \bibinfo {author} {\bibfnamefont {M.~J.}\ \bibnamefont {Johnson}}, \bibinfo {author} {\bibfnamefont {C.}~\bibnamefont {Leary}}, \bibinfo {author} {\bibfnamefont {D.}~\bibnamefont {Maclaurin}}, \bibinfo {author} {\bibfnamefont {G.}~\bibnamefont {Necula}}, \bibinfo {author} {\bibfnamefont {A.}~\bibnamefont {Paszke}}, \bibinfo {author} {\bibfnamefont {J.}~\bibnamefont {Vander{P}las}}, \bibinfo {author} {\bibfnamefont {S.}~\bibnamefont {Wanderman-{M}ilne}}, \ and\ \bibinfo {author} {\bibfnamefont {Q.}~\bibnamefont {Zhang}},\ }\href {http://github.com/jax-ml/jax} {\enquote {\bibinfo {title} {{JAX}: composable transformations of {P}ython+{N}um{P}y programs},}\ } (\bibinfo {year} {2018})\BibitemShut {NoStop}%
\bibitem [{\citenamefont {Hennigan}\ \emph {et~al.}(2020)\citenamefont {Hennigan}, \citenamefont {Cai}, \citenamefont {Norman}, \citenamefont {Martens},\ and\ \citenamefont {Babuschkin}}]{haiku2020github}%
  \BibitemOpen
  \bibfield  {author} {\bibinfo {author} {\bibfnamefont {T.}~\bibnamefont {Hennigan}}, \bibinfo {author} {\bibfnamefont {T.}~\bibnamefont {Cai}}, \bibinfo {author} {\bibfnamefont {T.}~\bibnamefont {Norman}}, \bibinfo {author} {\bibfnamefont {L.}~\bibnamefont {Martens}}, \ and\ \bibinfo {author} {\bibfnamefont {I.}~\bibnamefont {Babuschkin}},\ }\href {http://github.com/google-deepmind/dm-haiku} {\enquote {\bibinfo {title} {{H}aiku: {S}onnet for {JAX}},}\ } (\bibinfo {year} {2020})\BibitemShut {NoStop}%
\bibitem [{\citenamefont {Trotter}(1959)}]{Trotter1959}%
  \BibitemOpen
  \bibfield  {author} {\bibinfo {author} {\bibfnamefont {H.~F.}\ \bibnamefont {Trotter}},\ }\href {\doibase 10.1090/S0002-9939-1959-0108732-6} {\bibfield  {journal} {\bibinfo  {journal} {Proceedings of the American Mathematical Society}\ }\textbf {\bibinfo {volume} {10}},\ \bibinfo {pages} {545} (\bibinfo {year} {1959})}\BibitemShut {NoStop}%
\bibitem [{\citenamefont {Suzuki}(1976)}]{Suzuki1976GeneralizedTF}%
  \BibitemOpen
  \bibfield  {author} {\bibinfo {author} {\bibfnamefont {M.}~\bibnamefont {Suzuki}},\ }\href {\doibase 10.1007/BF01609348} {\bibfield  {journal} {\bibinfo  {journal} {Communications in Mathematical Physics}\ }\textbf {\bibinfo {volume} {51}},\ \bibinfo {pages} {183} (\bibinfo {year} {1976})}\BibitemShut {NoStop}%
\bibitem [{\citenamefont {Suzuki}(1990)}]{suzuki_fractal_1990}%
  \BibitemOpen
  \bibfield  {author} {\bibinfo {author} {\bibfnamefont {M.}~\bibnamefont {Suzuki}},\ }\href {\doibase 10.1016/0375-9601(90)90962-N} {\bibfield  {journal} {\bibinfo  {journal} {Physics Letters A}\ }\textbf {\bibinfo {volume} {146}},\ \bibinfo {pages} {319} (\bibinfo {year} {1990})}\BibitemShut {NoStop}%
\bibitem [{\citenamefont {Pontryagin}(1985)}]{pontryagin_mathematical_1985}%
  \BibitemOpen
  \bibfield  {author} {\bibinfo {author} {\bibfnamefont {L.~S.}\ \bibnamefont {Pontryagin}},\ }\href@noop {} {\bibfield  {journal} {\bibinfo  {journal} {Trudy Mat. Inst. Steklov.}\ }\textbf {\bibinfo {volume} {169}},\ \bibinfo {pages} {119} (\bibinfo {year} {1985})}\BibitemShut {NoStop}%
\bibitem [{\citenamefont {Chen}\ \emph {et~al.}(2018)\citenamefont {Chen}, \citenamefont {Rubanova}, \citenamefont {Bettencourt},\ and\ \citenamefont {Duvenaud}}]{chen_neural_nodate}%
  \BibitemOpen
  \bibfield  {author} {\bibinfo {author} {\bibfnamefont {R.~T.~Q.}\ \bibnamefont {Chen}}, \bibinfo {author} {\bibfnamefont {Y.}~\bibnamefont {Rubanova}}, \bibinfo {author} {\bibfnamefont {J.}~\bibnamefont {Bettencourt}}, \ and\ \bibinfo {author} {\bibfnamefont {D.~K.}\ \bibnamefont {Duvenaud}},\ }in\ \href {https://proceedings.neurips.cc/paper_files/paper/2018/file/69386f6bb1dfed68692a24c8686939b9-Paper.pdf} {\emph {\bibinfo {booktitle} {Advances in Neural Information Processing Systems}}},\ Vol.~\bibinfo {volume} {31}\ (\bibinfo  {publisher} {Curran Associates, Inc.},\ \bibinfo {year} {2018})\BibitemShut {NoStop}%
\bibitem [{\citenamefont {Blondel}\ and\ \citenamefont {Roulet}(2024)}]{blondel_elements_2024}%
  \BibitemOpen
  \bibfield  {author} {\bibinfo {author} {\bibfnamefont {M.}~\bibnamefont {Blondel}}\ and\ \bibinfo {author} {\bibfnamefont {V.}~\bibnamefont {Roulet}},\ }\href@noop {} {\  (\bibinfo {year} {2024})},\ \Eprint {http://arxiv.org/abs/2403.14606} {arXiv:2403.14606 [cs.LG]} \BibitemShut {NoStop}%
\bibitem [{\citenamefont {Efthymiou}\ \emph {et~al.}(2022)\citenamefont {Efthymiou}, \citenamefont {Ramos-Calderer}, \citenamefont {Bravo-Prieto}, \citenamefont {P{\'e}rez-Salinas}, \citenamefont {Garc{\'\i}a-Mart{\'\i}n}, \citenamefont {Garcia-Saez}, \citenamefont {Latorre},\ and\ \citenamefont {Carrazza}}]{efthymiou_qibo_2022}%
  \BibitemOpen
  \bibfield  {author} {\bibinfo {author} {\bibfnamefont {S.}~\bibnamefont {Efthymiou}}, \bibinfo {author} {\bibfnamefont {S.}~\bibnamefont {Ramos-Calderer}}, \bibinfo {author} {\bibfnamefont {C.}~\bibnamefont {Bravo-Prieto}}, \bibinfo {author} {\bibfnamefont {A.}~\bibnamefont {P{\'e}rez-Salinas}}, \bibinfo {author} {\bibfnamefont {D.}~\bibnamefont {Garc{\'\i}a-Mart{\'\i}n}}, \bibinfo {author} {\bibfnamefont {A.}~\bibnamefont {Garcia-Saez}}, \bibinfo {author} {\bibfnamefont {J.~I.}\ \bibnamefont {Latorre}}, \ and\ \bibinfo {author} {\bibfnamefont {S.}~\bibnamefont {Carrazza}},\ }\href {\doibase 10.1088/2058-9565/ac39f5} {\bibfield  {journal} {\bibinfo  {journal} {Quantum Science and Technology}\ }\textbf {\bibinfo {volume} {7}},\ \bibinfo {pages} {015018} (\bibinfo {year} {2022})}\BibitemShut {NoStop}%
\bibitem [{\citenamefont {Johansson}\ \emph {et~al.}(2012)\citenamefont {Johansson}, \citenamefont {Nation},\ and\ \citenamefont {Nori}}]{JOHANSSON20121760}%
  \BibitemOpen
  \bibfield  {author} {\bibinfo {author} {\bibfnamefont {J.}~\bibnamefont {Johansson}}, \bibinfo {author} {\bibfnamefont {P.}~\bibnamefont {Nation}}, \ and\ \bibinfo {author} {\bibfnamefont {F.}~\bibnamefont {Nori}},\ }\href {\doibase 10.1016/j.cpc.2012.02.021} {\bibfield  {journal} {\bibinfo  {journal} {Computer Physics Communications}\ }\textbf {\bibinfo {volume} {183}},\ \bibinfo {pages} {1760} (\bibinfo {year} {2012})}\BibitemShut {NoStop}%
\bibitem [{\citenamefont {Baur}\ and\ \citenamefont {Strassen}(1983)}]{baurComplexityPartialDerivatives1983}%
  \BibitemOpen
  \bibfield  {author} {\bibinfo {author} {\bibfnamefont {W.}~\bibnamefont {Baur}}\ and\ \bibinfo {author} {\bibfnamefont {V.}~\bibnamefont {Strassen}},\ }\href {\doibase 10.1016/0304-3975(83)90110-X} {\bibfield  {journal} {\bibinfo  {journal} {Theor. Comput. Sci.}\ }\textbf {\bibinfo {volume} {22}},\ \bibinfo {pages} {317} (\bibinfo {year} {1983})}\BibitemShut {NoStop}%
\bibitem [{\citenamefont {Wiltschko}\ and\ \citenamefont {Johnson}(2018)}]{jaxautodiff}%
  \BibitemOpen
  \bibfield  {author} {\bibinfo {author} {\bibfnamefont {A.}~\bibnamefont {Wiltschko}}\ and\ \bibinfo {author} {\bibfnamefont {M.~J.}\ \bibnamefont {Johnson}},\ }\href {https://github.com/jax-ml/jax/blob/main/docs/notebooks/autodiff_cookbook.ipynb} {\enquote {\bibinfo {title} {{T}he {A}utodiff {C}ookbook},}\ } (\bibinfo {year} {2018})\BibitemShut {NoStop}%
\bibitem [{\citenamefont {Nguyen}\ \emph {et~al.}(2022)\citenamefont {Nguyen}, \citenamefont {Koolstra}, \citenamefont {Kim}, \citenamefont {Morvan}, \citenamefont {Chistolini}, \citenamefont {Singh}, \citenamefont {Nesterov}, \citenamefont {J\"unger}, \citenamefont {Chen}, \citenamefont {Pedramrazi}, \citenamefont {Mitchell}, \citenamefont {Kreikebaum}, \citenamefont {Puri}, \citenamefont {Santiago},\ and\ \citenamefont {Siddiqi}}]{nguyen_blueprint_2022}%
  \BibitemOpen
  \bibfield  {author} {\bibinfo {author} {\bibfnamefont {L.~B.}\ \bibnamefont {Nguyen}}, \bibinfo {author} {\bibfnamefont {G.}~\bibnamefont {Koolstra}}, \bibinfo {author} {\bibfnamefont {Y.}~\bibnamefont {Kim}}, \bibinfo {author} {\bibfnamefont {A.}~\bibnamefont {Morvan}}, \bibinfo {author} {\bibfnamefont {T.}~\bibnamefont {Chistolini}}, \bibinfo {author} {\bibfnamefont {S.}~\bibnamefont {Singh}}, \bibinfo {author} {\bibfnamefont {K.~N.}\ \bibnamefont {Nesterov}}, \bibinfo {author} {\bibfnamefont {C.}~\bibnamefont {J\"unger}}, \bibinfo {author} {\bibfnamefont {L.}~\bibnamefont {Chen}}, \bibinfo {author} {\bibfnamefont {Z.}~\bibnamefont {Pedramrazi}}, \bibinfo {author} {\bibfnamefont {B.~K.}\ \bibnamefont {Mitchell}}, \bibinfo {author} {\bibfnamefont {J.~M.}\ \bibnamefont {Kreikebaum}}, \bibinfo {author} {\bibfnamefont {S.}~\bibnamefont {Puri}}, \bibinfo {author} {\bibfnamefont {D.~I.}\ \bibnamefont {Santiago}}, \ and\ \bibinfo {author} {\bibfnamefont {I.}~\bibnamefont {Siddiqi}},\ }\href {\doibase 10.1103/PRXQuantum.3.037001} {\bibfield  {journal} {\bibinfo  {journal} {PRX Quantum}\ }\textbf {\bibinfo {volume} {3}},\ \bibinfo {pages} {037001} (\bibinfo {year} {2022})}\BibitemShut {NoStop}%
\bibitem [{\citenamefont {Dogan}\ \emph {et~al.}(2023)\citenamefont {Dogan}, \citenamefont {Rosenstock}, \citenamefont {Le~Guevel}, \citenamefont {Xiong}, \citenamefont {Mencia}, \citenamefont {Somoroff}, \citenamefont {Nesterov}, \citenamefont {Vavilov}, \citenamefont {Manucharyan},\ and\ \citenamefont {Wang}}]{dogan_two-fluxonium_2023}%
  \BibitemOpen
  \bibfield  {author} {\bibinfo {author} {\bibfnamefont {E.}~\bibnamefont {Dogan}}, \bibinfo {author} {\bibfnamefont {D.}~\bibnamefont {Rosenstock}}, \bibinfo {author} {\bibfnamefont {L.}~\bibnamefont {Le~Guevel}}, \bibinfo {author} {\bibfnamefont {H.}~\bibnamefont {Xiong}}, \bibinfo {author} {\bibfnamefont {R.~A.}\ \bibnamefont {Mencia}}, \bibinfo {author} {\bibfnamefont {A.}~\bibnamefont {Somoroff}}, \bibinfo {author} {\bibfnamefont {K.~N.}\ \bibnamefont {Nesterov}}, \bibinfo {author} {\bibfnamefont {M.~G.}\ \bibnamefont {Vavilov}}, \bibinfo {author} {\bibfnamefont {V.~E.}\ \bibnamefont {Manucharyan}}, \ and\ \bibinfo {author} {\bibfnamefont {C.}~\bibnamefont {Wang}},\ }\href {\doibase 10.1103/PhysRevApplied.20.024011} {\bibfield  {journal} {\bibinfo  {journal} {Phys. Rev. Applied}\ }\textbf {\bibinfo {volume} {20}},\ \bibinfo {pages} {024011} (\bibinfo {year} {2023})}\BibitemShut {NoStop}%
\bibitem [{\citenamefont {Lin}\ \emph {et~al.}(2025)\citenamefont {Lin}, \citenamefont {Cho}, \citenamefont {Chen}, \citenamefont {Vavilov}, \citenamefont {Wang},\ and\ \citenamefont {Manucharyan}}]{PRXQuantum.6.010349}%
  \BibitemOpen
  \bibfield  {author} {\bibinfo {author} {\bibfnamefont {W.-J.}\ \bibnamefont {Lin}}, \bibinfo {author} {\bibfnamefont {H.}~\bibnamefont {Cho}}, \bibinfo {author} {\bibfnamefont {Y.}~\bibnamefont {Chen}}, \bibinfo {author} {\bibfnamefont {M.~G.}\ \bibnamefont {Vavilov}}, \bibinfo {author} {\bibfnamefont {C.}~\bibnamefont {Wang}}, \ and\ \bibinfo {author} {\bibfnamefont {V.~E.}\ \bibnamefont {Manucharyan}},\ }\href {\doibase 10.1103/PRXQuantum.6.010349} {\bibfield  {journal} {\bibinfo  {journal} {PRX Quantum}\ }\textbf {\bibinfo {volume} {6}},\ \bibinfo {pages} {010349} (\bibinfo {year} {2025})}\BibitemShut {NoStop}%
\bibitem [{\citenamefont {Hagberg}\ \emph {et~al.}(2008)\citenamefont {Hagberg}, \citenamefont {Schult},\ and\ \citenamefont {Swart}}]{SciPyProceedings_11}%
  \BibitemOpen
  \bibfield  {author} {\bibinfo {author} {\bibfnamefont {A.~A.}\ \bibnamefont {Hagberg}}, \bibinfo {author} {\bibfnamefont {D.~A.}\ \bibnamefont {Schult}}, \ and\ \bibinfo {author} {\bibfnamefont {P.~J.}\ \bibnamefont {Swart}},\ }in\ \href {https://www.osti.gov/biblio/960616} {\emph {\bibinfo {booktitle} {Proceedings of the 7th Python in Science Conference}}}\ (\bibinfo {address} {Pasadena, CA USA},\ \bibinfo {year} {2008})\ pp.\ \bibinfo {pages} {11 -- 15}\BibitemShut {NoStop}%
\bibitem [{\citenamefont {Berke}\ \emph {et~al.}(2022)\citenamefont {Berke}, \citenamefont {Varvelis}, \citenamefont {Trebst}, \citenamefont {Altland},\ and\ \citenamefont {DiVincenzo}}]{berke_transmon_2022}%
  \BibitemOpen
  \bibfield  {author} {\bibinfo {author} {\bibfnamefont {C.}~\bibnamefont {Berke}}, \bibinfo {author} {\bibfnamefont {E.}~\bibnamefont {Varvelis}}, \bibinfo {author} {\bibfnamefont {S.}~\bibnamefont {Trebst}}, \bibinfo {author} {\bibfnamefont {A.}~\bibnamefont {Altland}}, \ and\ \bibinfo {author} {\bibfnamefont {D.~P.}\ \bibnamefont {DiVincenzo}},\ }\href {\doibase 10.1038/s41467-022-29940-y} {\bibfield  {journal} {\bibinfo  {journal} {Nature Communications}\ }\textbf {\bibinfo {volume} {13}},\ \bibinfo {pages} {2495} (\bibinfo {year} {2022})}\BibitemShut {NoStop}%
\bibitem [{\citenamefont {Byrd}\ \emph {et~al.}(1995)\citenamefont {Byrd}, \citenamefont {Lu}, \citenamefont {Nocedal},\ and\ \citenamefont {Zhu}}]{doi:10.1137/0916069}%
  \BibitemOpen
  \bibfield  {author} {\bibinfo {author} {\bibfnamefont {R.~H.}\ \bibnamefont {Byrd}}, \bibinfo {author} {\bibfnamefont {P.}~\bibnamefont {Lu}}, \bibinfo {author} {\bibfnamefont {J.}~\bibnamefont {Nocedal}}, \ and\ \bibinfo {author} {\bibfnamefont {C.}~\bibnamefont {Zhu}},\ }\href {\doibase 10.1137/0916069} {\bibfield  {journal} {\bibinfo  {journal} {SIAM Journal on Scientific Computing}\ }\textbf {\bibinfo {volume} {16}},\ \bibinfo {pages} {1190} (\bibinfo {year} {1995})}\BibitemShut {NoStop}%
\bibitem [{\citenamefont {Lei}(2023)}]{private_communication_lei_wang}%
  \BibitemOpen
  \bibfield  {author} {\bibinfo {author} {\bibfnamefont {W.}~\bibnamefont {Lei}},\ }\href@noop {} {\bibfield  {journal} {\bibinfo  {journal} {Private communication}\ } (\bibinfo {year} {2023})}\BibitemShut {NoStop}%
\bibitem [{\citenamefont {Leung}\ \emph {et~al.}(2017)\citenamefont {Leung}, \citenamefont {Abdelhafez}, \citenamefont {Koch},\ and\ \citenamefont {Schuster}}]{PhysRevA.95.042318}%
  \BibitemOpen
  \bibfield  {author} {\bibinfo {author} {\bibfnamefont {N.}~\bibnamefont {Leung}}, \bibinfo {author} {\bibfnamefont {M.}~\bibnamefont {Abdelhafez}}, \bibinfo {author} {\bibfnamefont {J.}~\bibnamefont {Koch}}, \ and\ \bibinfo {author} {\bibfnamefont {D.}~\bibnamefont {Schuster}},\ }\href {\doibase 10.1103/PhysRevA.95.042318} {\bibfield  {journal} {\bibinfo  {journal} {Phys. Rev. A}\ }\textbf {\bibinfo {volume} {95}},\ \bibinfo {pages} {042318} (\bibinfo {year} {2017})}\BibitemShut {NoStop}%
\bibitem [{\citenamefont {Wang}\ and\ \citenamefont {Ni}(2024)}]{ziang_2024_11192761}%
  \BibitemOpen
  \bibfield  {author} {\bibinfo {author} {\bibfnamefont {Z.}~\bibnamefont {Wang}}\ and\ \bibinfo {author} {\bibfnamefont {X.}~\bibnamefont {Ni}},\ }\href {https://doi.org/10.5281/zenodo.11192761} {\bibfield  {journal} {\bibinfo  {journal} {Zenodo}\ } (\bibinfo {year} {2024})}\BibitemShut {NoStop}%
\bibitem [{\citenamefont {Puzzuoli}\ \emph {et~al.}(2023{\natexlab{b}})\citenamefont {Puzzuoli}, \citenamefont {Lin}, \citenamefont {Malekakhlagh}, \citenamefont {Pritchett}, \citenamefont {Rosand},\ and\ \citenamefont {Wood}}]{PUZZUOLI2023112262}%
  \BibitemOpen
  \bibfield  {author} {\bibinfo {author} {\bibfnamefont {D.}~\bibnamefont {Puzzuoli}}, \bibinfo {author} {\bibfnamefont {S.~F.}\ \bibnamefont {Lin}}, \bibinfo {author} {\bibfnamefont {M.}~\bibnamefont {Malekakhlagh}}, \bibinfo {author} {\bibfnamefont {E.}~\bibnamefont {Pritchett}}, \bibinfo {author} {\bibfnamefont {B.}~\bibnamefont {Rosand}}, \ and\ \bibinfo {author} {\bibfnamefont {C.~J.}\ \bibnamefont {Wood}},\ }\href {\doibase 10.1016/j.jcp.2023.112262} {\bibfield  {journal} {\bibinfo  {journal} {Journal of Computational Physics}\ }\textbf {\bibinfo {volume} {489}},\ \bibinfo {pages} {112262} (\bibinfo {year} {2023}{\natexlab{b}})}\BibitemShut {NoStop}%
\bibitem [{\citenamefont {Heek}\ \emph {et~al.}(2023)\citenamefont {Heek}, \citenamefont {Levskaya}, \citenamefont {Oliver}, \citenamefont {Ritter}, \citenamefont {Rondepierre}, \citenamefont {Steiner},\ and\ \citenamefont {van {Z}ee}}]{flax2020github}%
  \BibitemOpen
  \bibfield  {author} {\bibinfo {author} {\bibfnamefont {J.}~\bibnamefont {Heek}}, \bibinfo {author} {\bibfnamefont {A.}~\bibnamefont {Levskaya}}, \bibinfo {author} {\bibfnamefont {A.}~\bibnamefont {Oliver}}, \bibinfo {author} {\bibfnamefont {M.}~\bibnamefont {Ritter}}, \bibinfo {author} {\bibfnamefont {B.}~\bibnamefont {Rondepierre}}, \bibinfo {author} {\bibfnamefont {A.}~\bibnamefont {Steiner}}, \ and\ \bibinfo {author} {\bibfnamefont {M.}~\bibnamefont {van {Z}ee}},\ }\href {http://github.com/google/flax} {\enquote {\bibinfo {title} {{F}lax: A neural network library and ecosystem for {JAX}},}\ } (\bibinfo {year} {2023})\BibitemShut {NoStop}%
\bibitem [{\citenamefont {Xu}\ \emph {et~al.}(2020)\citenamefont {Xu}, \citenamefont {Chu}, \citenamefont {Yuan}, \citenamefont {Qiu}, \citenamefont {Zhou}, \citenamefont {Zhang}, \citenamefont {Tan}, \citenamefont {Yu}, \citenamefont {Liu}, \citenamefont {Li}, \citenamefont {Yan},\ and\ \citenamefont {Yu}}]{xu_high-fidelity_2020}%
  \BibitemOpen
  \bibfield  {author} {\bibinfo {author} {\bibfnamefont {Y.}~\bibnamefont {Xu}}, \bibinfo {author} {\bibfnamefont {J.}~\bibnamefont {Chu}}, \bibinfo {author} {\bibfnamefont {J.}~\bibnamefont {Yuan}}, \bibinfo {author} {\bibfnamefont {J.}~\bibnamefont {Qiu}}, \bibinfo {author} {\bibfnamefont {Y.}~\bibnamefont {Zhou}}, \bibinfo {author} {\bibfnamefont {L.}~\bibnamefont {Zhang}}, \bibinfo {author} {\bibfnamefont {X.}~\bibnamefont {Tan}}, \bibinfo {author} {\bibfnamefont {Y.}~\bibnamefont {Yu}}, \bibinfo {author} {\bibfnamefont {S.}~\bibnamefont {Liu}}, \bibinfo {author} {\bibfnamefont {J.}~\bibnamefont {Li}}, \bibinfo {author} {\bibfnamefont {F.}~\bibnamefont {Yan}}, \ and\ \bibinfo {author} {\bibfnamefont {D.}~\bibnamefont {Yu}},\ }\href {\doibase 10.1103/PhysRevLett.125.240503} {\bibfield  {journal} {\bibinfo  {journal} {Phys. Rev. Lett.}\ }\textbf {\bibinfo {volume} {125}},\ \bibinfo {pages} {240503} (\bibinfo {year} {2020})}\BibitemShut {NoStop}%
\bibitem [{\citenamefont {Vidal}(2007)}]{PhysRevLett.98.070201}%
  \BibitemOpen
  \bibfield  {author} {\bibinfo {author} {\bibfnamefont {G.}~\bibnamefont {Vidal}},\ }\href {\doibase 10.1103/PhysRevLett.98.070201} {\bibfield  {journal} {\bibinfo  {journal} {Phys. Rev. Lett.}\ }\textbf {\bibinfo {volume} {98}},\ \bibinfo {pages} {070201} (\bibinfo {year} {2007})}\BibitemShut {NoStop}%
\bibitem [{\citenamefont {a.~Smith}\ and\ \citenamefont {Gray}(2018)}]{g_a_smith_opt_einsum_2018}%
  \BibitemOpen
  \bibfield  {author} {\bibinfo {author} {\bibfnamefont {D.~G.}\ \bibnamefont {a.~Smith}}\ and\ \bibinfo {author} {\bibfnamefont {J.}~\bibnamefont {Gray}},\ }\href {\doibase 10.21105/joss.00753} {\bibfield  {journal} {\bibinfo  {journal} {Journal of Open Source Software}\ }\textbf {\bibinfo {volume} {3}},\ \bibinfo {pages} {753} (\bibinfo {year} {2018})}\BibitemShut {NoStop}%
\bibitem [{\citenamefont {Gray}\ and\ \citenamefont {Kourtis}(2021)}]{Gray2021hyperoptimized}%
  \BibitemOpen
  \bibfield  {author} {\bibinfo {author} {\bibfnamefont {J.}~\bibnamefont {Gray}}\ and\ \bibinfo {author} {\bibfnamefont {S.}~\bibnamefont {Kourtis}},\ }\href {\doibase 10.22331/q-2021-03-15-410} {\bibfield  {journal} {\bibinfo  {journal} {{Quantum}}\ }\textbf {\bibinfo {volume} {5}},\ \bibinfo {pages} {410} (\bibinfo {year} {2021})}\BibitemShut {NoStop}%
\bibitem [{\citenamefont {O'Gorman}(2019)}]{ogorman_parameterization_2019}%
  \BibitemOpen
  \bibfield  {author} {\bibinfo {author} {\bibfnamefont {B.}~\bibnamefont {O'Gorman}},\ }in\ \href {\doibase 10.4230/LIPIcs.TQC.2019.10} {\emph {\bibinfo {booktitle} {14th Conference on the Theory of Quantum Computation, Communication and Cryptography (TQC 2019)}}},\ Vol.\ \bibinfo {volume} {135}\ (\bibinfo {year} {2019})\ pp.\ \bibinfo {pages} {10:1--10:19}\BibitemShut {NoStop}%
\bibitem [{\citenamefont {Kourtis}\ \emph {et~al.}(2019)\citenamefont {Kourtis}, \citenamefont {Chamon}, \citenamefont {Mucciolo},\ and\ \citenamefont {Ruckenstein}}]{kourtis_fast_2019}%
  \BibitemOpen
  \bibfield  {author} {\bibinfo {author} {\bibfnamefont {S.}~\bibnamefont {Kourtis}}, \bibinfo {author} {\bibfnamefont {C.}~\bibnamefont {Chamon}}, \bibinfo {author} {\bibfnamefont {E.}~\bibnamefont {Mucciolo}}, \ and\ \bibinfo {author} {\bibfnamefont {A.}~\bibnamefont {Ruckenstein}},\ }\href {\doibase 10.21468/SciPostPhys.7.5.060} {\bibfield  {journal} {\bibinfo  {journal} {SciPost Physics}\ }\textbf {\bibinfo {volume} {7}},\ \bibinfo {pages} {060} (\bibinfo {year} {2019})}\BibitemShut {NoStop}%
\bibitem [{\citenamefont {McKay}\ \emph {et~al.}(2017)\citenamefont {McKay}, \citenamefont {Wood}, \citenamefont {Sheldon}, \citenamefont {Chow},\ and\ \citenamefont {Gambetta}}]{mckayEfficientGatesQuantum2017}%
  \BibitemOpen
  \bibfield  {author} {\bibinfo {author} {\bibfnamefont {D.~C.}\ \bibnamefont {McKay}}, \bibinfo {author} {\bibfnamefont {C.~J.}\ \bibnamefont {Wood}}, \bibinfo {author} {\bibfnamefont {S.}~\bibnamefont {Sheldon}}, \bibinfo {author} {\bibfnamefont {J.~M.}\ \bibnamefont {Chow}}, \ and\ \bibinfo {author} {\bibfnamefont {J.~M.}\ \bibnamefont {Gambetta}},\ }\href {\doibase 10.1103/PhysRevA.96.022330} {\bibfield  {journal} {\bibinfo  {journal} {Phys. Rev. A}\ }\textbf {\bibinfo {volume} {96}},\ \bibinfo {pages} {022330} (\bibinfo {year} {2017})}\BibitemShut {NoStop}%
\bibitem [{\citenamefont {Rigetti}\ and\ \citenamefont {Devoret}(2010)}]{rigettiFullyMicrowavetunableUniversal2010}%
  \BibitemOpen
  \bibfield  {author} {\bibinfo {author} {\bibfnamefont {C.}~\bibnamefont {Rigetti}}\ and\ \bibinfo {author} {\bibfnamefont {M.}~\bibnamefont {Devoret}},\ }\href {\doibase 10.1103/PhysRevB.81.134507} {\bibfield  {journal} {\bibinfo  {journal} {Phys. Rev. B}\ }\textbf {\bibinfo {volume} {81}},\ \bibinfo {pages} {134507} (\bibinfo {year} {2010})}\BibitemShut {NoStop}%
\bibitem [{\citenamefont {{de Groot}}\ \emph {et~al.}(2010)\citenamefont {{de Groot}}, \citenamefont {Lisenfeld}, \citenamefont {Schouten}, \citenamefont {Ashhab}, \citenamefont {Lupa{\c s}cu}, \citenamefont {Harmans},\ and\ \citenamefont {Mooij}}]{degrootSelectiveDarkeningDegenerate2010}%
  \BibitemOpen
  \bibfield  {author} {\bibinfo {author} {\bibfnamefont {P.~C.}\ \bibnamefont {{de Groot}}}, \bibinfo {author} {\bibfnamefont {J.}~\bibnamefont {Lisenfeld}}, \bibinfo {author} {\bibfnamefont {R.~N.}\ \bibnamefont {Schouten}}, \bibinfo {author} {\bibfnamefont {S.}~\bibnamefont {Ashhab}}, \bibinfo {author} {\bibfnamefont {A.}~\bibnamefont {Lupa{\c s}cu}}, \bibinfo {author} {\bibfnamefont {C.~J. P.~M.}\ \bibnamefont {Harmans}}, \ and\ \bibinfo {author} {\bibfnamefont {J.~E.}\ \bibnamefont {Mooij}},\ }\href {\doibase 10.1038/nphys1733} {\bibfield  {journal} {\bibinfo  {journal} {Nature Physics}\ }\textbf {\bibinfo {volume} {6}},\ \bibinfo {pages} {763} (\bibinfo {year} {2010})}\BibitemShut {NoStop}%
\bibitem [{\citenamefont {Pommerening}\ and\ \citenamefont {DiVincenzo}(2020)}]{PhysRevA.102.032623}%
  \BibitemOpen
  \bibfield  {author} {\bibinfo {author} {\bibfnamefont {J.~C.}\ \bibnamefont {Pommerening}}\ and\ \bibinfo {author} {\bibfnamefont {D.~P.}\ \bibnamefont {DiVincenzo}},\ }\href {\doibase 10.1103/PhysRevA.102.032623} {\bibfield  {journal} {\bibinfo  {journal} {Phys. Rev. A}\ }\textbf {\bibinfo {volume} {102}},\ \bibinfo {pages} {032623} (\bibinfo {year} {2020})}\BibitemShut {NoStop}%
\bibitem [{\citenamefont {Nielsen}(2002)}]{nielsen_simple_2002}%
  \BibitemOpen
  \bibfield  {author} {\bibinfo {author} {\bibfnamefont {M.~A.}\ \bibnamefont {Nielsen}},\ }\href {\doibase 10.1016/S0375-9601(02)01272-0} {\bibfield  {journal} {\bibinfo  {journal} {Physics Letters A}\ }\textbf {\bibinfo {volume} {303}},\ \bibinfo {pages} {249} (\bibinfo {year} {2002})}\BibitemShut {NoStop}%
\bibitem [{\citenamefont {Havel}(2003)}]{havel_robust_2003}%
  \BibitemOpen
  \bibfield  {author} {\bibinfo {author} {\bibfnamefont {T.~F.}\ \bibnamefont {Havel}},\ }\href {\doibase 10.1063/1.1518555} {\bibfield  {journal} {\bibinfo  {journal} {Journal of Mathematical Physics}\ }\textbf {\bibinfo {volume} {44}},\ \bibinfo {pages} {534} (\bibinfo {year} {2003})}\BibitemShut {NoStop}%
\end{thebibliography}%
%%%%%%%%%%%%%%%%%%%%% APPENDIX %%%%%%%%%%%%%%%%%%%%%%%%%%%%%%%%
\clearpage
\appendix
\onecolumn

\section{Higher-order Suzuki-Trotter decomposition}
\label{appen:higher_order_std}

    Consider a Hamiltonian represented as $\mathcal{H}=\sum_{k=1}^{N_H} H_k$, where each term $H_k$ denotes a local Hamiltonian that acts non-trivially on the subsystem $k$, $N_H$ is the number of local Hamiltonian terms. The first-order decomposition employs the product of local time evolution operators $e^{-i H_k \delta t}$, which can be expressed as:
\begin{equation}
    Q_1(\delta t) = \prod_{k=1}^{N_H} e^{-i H_k \delta t},
\end{equation}
    where each $k$ corresponds to a specific subset of the quantum system.

    The second-order decomposition, derived from a symmetric (palindromic) sequence of $Q_1(\delta t)$ is given by
\begin{equation}
    Q_2(\delta t) = \prod_{k=1}^{N_H} e^{-i H_k \frac{\delta t}{2}} \prod_{k=N_H}^{1} e^{-i H_k \frac{\delta t}{2}}.
\end{equation}

    The complex fourth-order Suzuki-Trotter decomposition, which satisfies~\autoref{eq:trotter_4j} with $p=(3 - \sqrt{3}i/6)$.
\begin{equation}
    \label{eq:trotter_4j}
    Q_{4j}(\delta t) = \prod_{k=1}^{N_H} e^{-i pH_k \frac{\delta t}{2}} \prod_{k=N_H}^{1} e^{-i pH_k \frac{\delta t}{2}}\prod_{k=1}^{N_H} e^{-i p^* H_k \frac{\delta t}{2}} \prod_{k=N_H}^{1} e^{-i p^* H_k \frac{\delta t}{2}},
\end{equation}
    where the star operator $*$ indicates the complex conjugate. The real fourth-order Suzuki-Trotter decomposition, which includes more terms compared to the complex version, is formulated as~\autoref{eq:trotter_4} with $p=1/(2-\sqrt[3]{2})$.
\begin{equation}
    \label{eq:trotter_4}
    Q_4(\delta t) = \prod_{k=1}^{N_H} e^{-i pH_k \frac{\delta t}{2}} \prod_{k=N_H}^{1} e^{-i pH_k \frac{\delta t}{2}}\prod_{k=1}^{N_H} e^{-i (1-2p)H_k \frac{\delta t}{2}} \prod_{k=N_H}^{1} e^{-i (1-2p)H_k \frac{\delta t}{2}}\prod_{k=1}^{N_H} e^{-i pH_k \frac{\delta t}{2}} \prod_{k=N_H}^{1} e^{-i pH_k \frac{\delta t}{2}}
\end{equation}

    These decompositions offer varying levels of accuracy and computational efficiency, catering to the precision requirements and the specific property of the quantum system being modeled.

\section{Tensor Network Contraction}
\label{appen:tn_contraction}
    We represent the quantum state of the composite Hilbert space as a rank-$N$ tensor, denoted by $\psi_{1,2,3,...,N}$. To address the dynamics of this state, we consider the product of the time evolution operator and the tensor state. Specifically, we focus on the contraction between the first-order STD decomposition $Q_1(\delta t)$ and $\psi_{1,2,3,...,N}$ in the context of time evolution. In the simplest case that only involves a single two-body local time evolution operator, we define the operator as $u_{ij,i^\prime j^\prime} = \exp\{-i H_{ij,i^\prime j^\prime} \delta t\}$, where $\delta t$ represents a small time interval. The resulting tensor contraction can be expressed as
\begin{equation}
    \psi^\prime_{1,2,3,\cdots, i^\prime, j^\prime, \cdots,N} = \sum_{i,j} u_{ij,i^\prime j^\prime} \psi_{1,2,3,\cdots, i,j, \cdots,N}.
\end{equation}

    This computation is performed using tensor network contraction along an optimized path, performed by efficient packages~\cite{g_a_smith_opt_einsum_2018, Gray2021hyperoptimized}. In this section, we specifically consider the first-order Suzuki-Trotter decomposition to illustrate the integration of the Suzuki-Trotter decomposition with the tensor network contraction. It is important to note that the computation methodology for higher-order Suzuki-Trotter decompositions is analogous.

    Assuming a quantum processor with $N$ qudits and each qudit subsystem having a dimension $d$, the computational cost of tensor contraction is determined by the size of the $k$-body evolution operator $dim(U_X) = d^{2k}$ and the dimension of the tensor state $dim(\psi) = d^N$. This cost is represented as $\frac{|dim(\psi)| |dim(U_X)|}{|dim(\psi) \cap dim(U_X)|}$, where $\mathrm{X}$ labels subsets of the local quantum system. The time complexity for evolving the quantum state with a $k$-body evolution operator is $O(d^{N+k})$. This is lower than the time complexity of matrix-vector multiplication in the Hilbert space, which is $O(d^{2N})$. Thus, using tensor contraction can significantly enhance computational efficiency, especially when leveraging GPU with cuTENSOR or hipTensor.

    After introducing the Suzuki-Trotter decomposition, the time evolution over a given interval can be represented as a tensor network contraction problem. This requires identifying an efficient contraction path to minimize the size of intermediate tensors and the computational cost, measured in floating-point operations per second (FLOPS). Optimally determining such a path is an NP-hard problem~\cite{ogorman_parameterization_2019}. Nonetheless, the Greedy algorithm provides an effective method for finding contraction paths, especially when dealing with a multitude of low-rank tensors~\cite{Gray2021hyperoptimized, kourtis_fast_2019}. The algorithm follows these principal steps:

\begin{enumerate}
    \item Compute Hadamard products of tensors in an arbitrary order.
    \item Contract pairs of remaining tensors, specifically choosing those pairs that minimize the resultant tensor's size and computational cost.
    \item Compute any necessary pairwise outer products.
\end{enumerate}

    Once established, this contraction path can be repeatedly used for both the evolution of quantum states and their adjoint states, thereby justifying a higher initial computational investment to determine the most optimal path possible.

\section{Simultaneous Gate Simulation Details}
\label{appen:gate_simulation}

    A commonly implemented single-qubit gate is the Rabi $\pi$ rotation. To execute a single-qubit gate, a drive pulse is applied through the diplexed flux control line of the qubit $i$. The associated driving Hamiltonian is given by
\begin{equation}
    H_\epsilon(t) = \mathcal{E}(t) \cos(\omega_d t+\phi) \opv_i, 
    \label{eq:microwave_control_hamiltonian}
\end{equation}
    where the pulse envelope $\mathcal{E}(t)$ is specifically shaped as
\begin{equation}
    \mathcal{E}(t) = \frac{\epsilon_d}{2} \left[ 1 - \cos\left(\frac{2\pi t}{\tau_{\mathrm{gate}}}\right) \right].
    \label{eq:single_qubit_cosine_envelope}
\end{equation}

    Considering the presence of higher energy levels and adjacent qubits, the optimal frequency $\omega_d$ is derived by analyzing the composited quantum system. Furthermore, arbitrary $Z$-rotations are incorporated both before and after the gate operations. We also allow arbitrary $Z$-rotations before and after the gates, since they can be essentially free~\cite{mckayEfficientGatesQuantum2017}.

    For two-qubit operations, we employ the Cross-Resonance ($\mathrm{CR}$) pulse~\cite{rigettiFullyMicrowavetunableUniversal2010,degrootSelectiveDarkeningDegenerate2010}. There is one of the two-qubit gate schemes discussed in~\cite{nguyen_blueprint_2022}. The $\mathrm{CR}$ pulse requires the application of a microwave tone to the control qubit, the microwave tuned approximately to the frequency $\omega_{10}$ of the target qubit. The drive employs the same form as the Hamiltonian defined in~\autoref{eq:microwave_control_hamiltonian}, with the envelope $\mathcal{E}(t)$ designed to have a ramp time $\tau_{\mathrm{ramp}}$ equal to half the gate duration $\tau_{\mathrm{gate}} / 2$.

    By tuning $\omega_d$ to approximate the $0-1$ transition frequency of the neighboring qubit, the resultant two-qubit gate $U_\mathrm{CR}$ can effectively emulate a $\mathrm{CNOT}$ gate, with the addition of some single-qubit unitary transformations. To simplify the simulation of $U_\mathrm{CNOT}$ based on $U_\mathrm{CR}$, we incorporate ideal single-qubit rotations before and after the driving of simultaneous pulses.

\section{Details of time evolution computation}
    \label{appen:basis_transform}

    The commonly used basis for multi-qubit processors is the product basis (the tensor product of single-qubit eigenbasis) or the multi-qubit eigenbasis. This latter basis is derived by diagonalizing the idling Hamiltonian matrix in the product basis and then adjusting the phases of its eigenstates~\cite{ni_superconducting_2023}. The product basis offers the advantage of storing the local Hamiltonian in the qubit subspace, facilitating efficient time evolution through the use of a series of 2-body operators. However, the local Hamiltonian in the multi-qubit eigenbasis tends to occupy the full Hilbert space, which significantly degrades the performance during time evolution.
 
    Computing in a different basis is almost equivalent, but some information will be lost when we truncate the matrices to the computational subspace. The multi-qubit eigenbasis is considered to be a good approach for the experimental basis~\cite{PhysRevA.102.032623}. We suggest a method that utilizes the multi-qubit eigenbasis to represent the final state while employing the product basis for time evolution calculations. The primary computational cost here involves the diagonalization of the idling Hamiltonian. The time complexity for diagonalization is $O(d^{3N})$, where $N$ represents the number of qudits and $d$ denotes the dimension of the qudit subsystem. This complexity becomes prohibitive for large quantum processors. Therefore, finding an efficient method to approximate the eigenstates of the idling Hamiltonian remains a question for future study. Alternatively, employing the product basis to represent the final state may reduce the diagonalization overhead but increase the error.

    We begin the process by diagonalizing the idling Hamiltonian and using the eigenbasis for the time evolution unitary matrices. We denote the eigenstates by the matrix $V$, which is written in the product basis $B_{\textrm{product}}$. 
    In this work, the Hamiltonians are always written in the product basis since we want to use Trotter decomposition.
    Therefore, the time evolution unitary in the eigenbasis is
\begin{equation}
    \label{eq:basis_transform}
    U(t_0, t_g) \mapsto V^{\dagger} \left( \mathcal{T} \exp \left\{-i \int_{t_0}^{t_g} \mathcal{H}(t) dt \right\} \right) V.
\end{equation}

    We can absorb $V^{\dagger}$ and $V$ into the quantum state by evolving the initial state according to $|\psi^\prime_{\textrm{initial}}\rangle \mapsto V |\psi_{\textrm{initial}}\rangle $ in the product basis. Subsequently, we transform the final state back to the eigenbasis via $|\psi_{\textrm{final}}\rangle \mapsto V^{\dagger} |\psi^\prime_{\textrm{final}}\rangle $. The method for transforming the basis and calculating the objective function comprises the following steps:

\begin{enumerate}
    \item For each idling qubit, compute its single-qubit eigenbasis and represent the subsystem Hamiltonian, $\opn$, and $\opv$ operators within this eigenbasis. This method achieves faster convergence with an increasing number of subsystem levels due to the more information captured in the low-energy spectrum.
    \item Determine the eigenstate of the idling system's Hamiltonian.
    \item To model the unitary for simultaneous driving, simulate the evolution of all possible initial states in the computational basis in parallel.
    \item Employ a Trotterization solver to compute the time evolution, using a time step size of $\delta t = 0.05\ $ns. We used a fourth-order complex Suzuki-Trotter decomposition to simulate simultaneous two-qubit gates.
    \item Truncate the final state to the computational basis and assemble the time evolution operator $U_{\textrm{sim}}$. We assume that the leakage error is negligible, which allows the unitary to be written in $B_{\textrm{eigen}}$ as $U_{\textrm{sim}}$.
    \item Apply $\mathrm{SU}(4)$ unitary compensation to $U_{\textrm{sim}}$ by incorporating errorless single-qubit gates before and after the simultaneous pulses, thus aligning $U_{\textrm{sim}}$ as closely as possible with the target unitary $U_{\textrm{target}}$.
    \item Evaluate the fidelity metric and use it as the objective function for optimization. The formula is detailed in~\autoref{eq:average_fidelity}.
    \item Optimize using the L-BFGS-B method, leveraging backpropagation to compute the gradient.
\end{enumerate}

    The objective function is defined as $\mathcal{L}=1-\mathcal{F}$, and the formula for the fidelity metric~\cite{nielsen_simple_2002} satisfies
\begin{equation}
    \mathcal{F} (\mathcal{E},I) = \frac{\sum_j \operatorname{tr} (U_j^\dagger \mathcal{E}(U_j)) + D^2}{D^2(D+1)}, 
    \label{eq:average_fidelity}
\end{equation}
    where $D$ is the dimension of the system and the unitary operators $U_j$ form an orthonormal operator basis.

    It's valuable to note that the dynamics of a density matrix could be characterized using a superoperator, which is computed by solving the vectorized Lindblad master equation~\cite{havel_robust_2003}. In this way, SuperGrad migrated the Trotterization solver and backpropagation method to the open quantum systems.

\section{Miscellaneous Data}
\label{appen:data_details}

\renewcommand{\arraystretch}{1.5}%
\begin{table*}
    \small
    \centering
    \resizebox{0.9\textwidth}{!}{%
    % \begin{sidewaystable}%
    \begin{tabular}{| l | r | r | r || l | r | r | r |}%
    \hline%
    \multicolumn{4}{|c||}{Device parameters}&\multicolumn{4}{c|}{Control parameters}\\%
    \hline%
    Parameter&Value (GHz)&Gradient [TAD]&Relative Error [LCAM]&Parameter&Value&Gradient [TAD]&Relative Error [LCAM]\\%
    \hline%
    $E_{C, 0}$&9.945e{-}01&{-}5.253e{-}04&5.498e{-}11&$\epsilon_{{d}}$ ($\mathrm{Q}_{0}$)&8.995e{-}03 GHz&{-}2.443e{-}04 &{-}6.275e{-}12\\%
    $E_{J, 0}$&3.978e+00&1.636e{-}04&5.939e{-}11&$t_{\text{single}}$ ($\mathrm{Q}_{0}$)&5.000e+01 ns&{-}1.138e{-}03 &{-}1.755e{-}08\\%
    $E_{L, 0}$&8.950e{-}01&{-}4.703e{-}04&9.996e{-}11&$\omega_{{d}}/2 \pi$ ($\mathrm{Q}_{0}$)&4.859e{-}01 GHz&{-}3.821e{-}02 &{-}9.866e{-}13\\%
    $E_{C, 1}$&9.967e{-}01&3.492e{-}04&{-}2.101e{-}11&$\epsilon_{{d}}$ ($\mathrm{Q}_{1}$)&9.433e{-}03 GHz&{-}7.167e{-}04 &{-}1.786e{-}12\\%
    $E_{J, 1}$&3.987e+00&{-}1.106e{-}04&{-}2.148e{-}11&$t_{\text{single}}$ ($\mathrm{Q}_{1}$)&5.000e+01 ns&{-}1.121e{-}03 &{-}1.781e{-}08\\%
    $E_{L, 1}$&9.967e{-}01&3.064e{-}04&{-}8.477e{-}11&$\omega_{{d}}/2 \pi$ ($\mathrm{Q}_{1}$)&5.913e{-}01 GHz&1.019e{-}02 &{-}1.503e{-}12\\%
    $E_{C, 2}$&1.005e+00&1.929e{-}03&{-}6.928e{-}12&$\epsilon_{{d}}$ ($\mathrm{Q}_{2}$)&9.745e{-}03 GHz&{-}1.840e{-}04 &{-}9.815e{-}12\\%
    $E_{J, 2}$&4.021e+00&{-}6.279e{-}04&{-}6.326e{-}12&$t_{\text{single}}$ ($\mathrm{Q}_{2}$)&5.000e+01 ns&{-}1.125e{-}03 &{-}1.775e{-}08\\%
    $E_{L, 2}$&1.106e+00&1.654e{-}03&{-}1.781e{-}11&$\omega_{{d}}/2 \pi$ ($\mathrm{Q}_{2}$)&6.710e{-}01 GHz&6.646e{-}03 &7.352e{-}13\\%
    $E_{C, 3}$&1.014e+00&{-}1.672e{-}03&4.218e{-}13&$\epsilon_{{d}}$ ($\mathrm{Q}_{3}$)&9.119e{-}03 GHz&{-}1.113e{-}03 &1.287e{-}11\\%
    $E_{J, 3}$&4.056e+00&5.166e{-}04&1.422e{-}12&$t_{\text{single}}$ ($\mathrm{Q}_{3}$)&5.000e+01 ns&{-}1.091e{-}03 &{-}1.830e{-}08\\%
    $E_{L, 3}$&9.126e{-}01&{-}1.521e{-}03&1.420e{-}11&$\omega_{{d}}/2 \pi$ ($\mathrm{Q}_{3}$)&5.348e{-}01 GHz&4.803e{-}02 &{-}3.982e{-}13\\%
    $E_{C, 4}$&9.863e{-}01&1.089e{-}03&{-}1.270e{-}11&$\epsilon_{{d}}$ ($\mathrm{Q}_{4}$)&9.446e{-}03 GHz&{-}1.628e{-}03 &9.624e{-}12\\%
    $E_{J, 4}$&3.945e+00&{-}3.492e{-}04&{-}1.272e{-}11&$t_{\text{single}}$ ($\mathrm{Q}_{4}$)&5.000e+01 ns&{-}1.115e{-}03 &{-}1.790e{-}08\\%
    $E_{L, 4}$&9.863e{-}01&9.692e{-}04&{-}3.276e{-}11&$\omega_{{d}}/2 \pi$ ($\mathrm{Q}_{4}$)&5.843e{-}01 GHz&{-}6.521e{-}02 &2.924e{-}13\\%
    $E_{C, 5}$&1.010e+00&1.786e{-}03&{-}9.624e{-}12&$\epsilon_{{d}}$ ($\mathrm{Q}_{5}$)&9.694e{-}03 GHz&{-}2.304e{-}03 &4.820e{-}12\\%
    $E_{J, 5}$&4.041e+00&{-}5.796e{-}04&{-}8.902e{-}12&$t_{\text{single}}$ ($\mathrm{Q}_{5}$)&5.000e+01 ns&{-}1.036e{-}03 &{-}1.927e{-}08\\%
    $E_{L, 5}$&1.111e+00&1.519e{-}03&{-}2.144e{-}11&$\omega_{{d}}/2 \pi$ ($\mathrm{Q}_{5}$)&6.637e{-}01 GHz&{-}8.102e{-}02 &4.253e{-}14\\%
    $E_{C, 6}$&1.001e+00&{-}8.966e{-}04&{-}3.022e{-}12&$\epsilon_{{d}}$ ($\mathrm{Q}_{6}$)&9.007e{-}03 GHz&{-}4.521e{-}04 &2.623e{-}11\\%
    $E_{J, 6}$&4.003e+00&2.779e{-}04&{-}5.447e{-}13&$t_{\text{single}}$ ($\mathrm{Q}_{6}$)&5.000e+01 ns&{-}1.059e{-}03 &{-}1.885e{-}08\\%
    $E_{L, 6}$&9.006e{-}01&{-}8.016e{-}04&2.326e{-}11&$\omega_{{d}}/2 \pi$ ($\mathrm{Q}_{6}$)&4.958e{-}01 GHz&6.750e{-}02 &{-}5.386e{-}13\\%
    $E_{C, 7}$&1.010e+00&1.071e{-}03&4.939e{-}12&$\epsilon_{{d}}$ ($\mathrm{Q}_{7}$)&9.396e{-}03 GHz&{-}5.403e{-}04 &{-}6.577e{-}12\\%
    $E_{J, 7}$&4.041e+00&{-}3.382e{-}04&4.827e{-}12&$t_{\text{single}}$ ($\mathrm{Q}_{7}$)&5.000e+01 ns&{-}1.153e{-}03 &{-}1.731e{-}08\\%
    $E_{L, 7}$&1.010e+00&9.346e{-}04&{-}1.536e{-}11&$\omega_{{d}}/2 \pi$ ($\mathrm{Q}_{7}$)&5.947e{-}01 GHz&4.621e{-}02 &{-}7.426e{-}13\\%
    $J_{C,(0, 1)}$&2.000e{-}02&{-}1.321e{-}05&6.382e{-}12&&&&\\%
    $J_{L,(0, 1)}$&{-}2.000e{-}03&{-}2.940e{-}03&5.496e{-}12&&&&\\%
    $J_{C,(1, 2)}$&2.000e{-}02&{-}3.149e{-}05&{-}1.993e{-}12&&&&\\%
    $J_{L,(1, 2)}$&{-}2.000e{-}03&{-}5.407e{-}03&{-}1.771e{-}12&&&&\\%
    $J_{C,(2, 3)}$&2.000e{-}02&{-}2.055e{-}05&{-}2.378e{-}12&&&&\\%
    $J_{L,(2, 3)}$&{-}2.000e{-}03&{-}3.753e{-}03&{-}1.977e{-}12&&&&\\%
    $J_{C,(3, 4)}$&2.000e{-}02&{-}2.748e{-}05&2.024e{-}12&&&&\\%
    $J_{L,(3, 4)}$&{-}2.000e{-}03&{-}5.455e{-}03&2.385e{-}12&&&&\\%
    $J_{C,(4, 5)}$&2.000e{-}02&{-}5.965e{-}05&{-}1.502e{-}12&&&&\\%
    $J_{L,(4, 5)}$&{-}2.000e{-}03&{-}1.019e{-}02&{-}1.508e{-}12&&&&\\%
    $J_{C,(5, 6)}$&2.000e{-}02&3.242e{-}06&1.378e{-}11&&&&\\%
    $J_{L,(5, 6)}$&{-}2.000e{-}03&1.521e{-}04&4.129e{-}11&&&&\\%
    $J_{C,(6, 7)}$&2.000e{-}02&{-}3.169e{-}05&{-}8.270e{-}13&&&&\\%
    $J_{L,(6, 7)}$&{-}2.000e{-}03&{-}7.191e{-}03&5.958e{-}14&&&&\\%
    \hline%
    \end{tabular}%
    }
    % \end{sidewaystable}%
    \caption{\label{table:parameters_relative_error} The table contains device and control parameters for an 8-qubit chain conducted in~\autoref{fig:benchmark_grad_method}. The gradients listed are $\nabla \mathcal{L} (\hamparam)$ calculated through the auto-diff of Trotterization (TAD), where $\mathcal{L}$ is the objective function containing simultaneous $\mathrm{X}^{\otimes 8}$ gate fidelity defined in~\autoref{eq:average_fidelity}. The units of gradients are the reciprocal of the units of their corresponding values. The relative errors listed are determined using gradients computed by the local continuous adjoint method (LCAM) for each parameter component.}
\end{table*}

\end{document}